\pdfoutput=1

\documentclass[11pt,nofootinbib]{article}
\usepackage{amsmath}
\usepackage{amssymb}
\usepackage{graphicx}
\usepackage[colorlinks=true,linkcolor=blue,citecolor=blue]{hyperref}

\numberwithin{equation}{section}

\begin{document}

\begin{titlepage}

\setcounter{page}{1} \baselineskip=15.5pt \thispagestyle{empty}

\begin{flushright}
RESCEU-44/12
\end{flushright}
\vfil

\bigskip\
\begin{center}
{\LARGE \textbf{Primordial Spikes from Wrapped Brane Inflation}}
\vskip 15pt
\end{center}

\vspace{0.5cm}
\begin{center}
{\Large 
Takeshi Kobayashi$^{\star, \dagger}$\footnote{takeshi@cita.utoronto.ca}
and
Jun'ichi Yokoyama$^{\ast, \ddagger}$\footnote{yokoyama@resceu.s.u-tokyo.ac.jp}
}
\end{center}

\vspace{0.3cm}

\begin{center}
\textit{$^{\star}$ Canadian Institute for Theoretical Astrophysics,
 University of Toronto, \\ 60 St. George Street, Toronto, Ontario M5S
 3H8, Canada}\\ 

\vskip 4pt
\textit{$^{\dagger}$ Perimeter Institute for Theoretical Physics, \\ 
 31 Caroline Street North, Waterloo, Ontario N2L 2Y5, Canada}\\ 

\vskip 4pt
\textit{$^{\ast}$ Research Center for the Early Universe, School of
 Science, The University of Tokyo, \\ 7-3-1 Hongo, Bunkyo-ku, Tokyo
 113-0033, Japan}\\ 
 
\vskip 4pt
\textit{$^{\ddagger}$ Kavli Institute for the Physics and Mathematics of the
 Universe, The University of Tokyo, \\ 5-1-5 Kashiwanoha,
 Kashiwa, Chiba 277-8582, Japan}

\end{center} \vfil

\vspace{0.8cm}

\noindent
Cosmic inflation driven by branes wrapping the extra dimensions involves
 Kaluza-Klein (KK) degrees of freedom in addition to the zero-mode position
 of the brane which plays the role of the inflaton. As the wrapped brane
 passes by localized sources or features along its inflationary
 trajectory in the extra dimensional space, the KK modes along the
 wrapped direction are excited and start to oscillate during
 inflation. We show that the oscillating KK modes induce parametric
 resonance for the curvature perturbations, generating sharp signals in the
 perturbation spectrum. The effective four dimensional picture is a
 theory where the inflaton couples to the heavy KK modes. The Nambu-Goto
 action of the brane sources couplings between the inflaton kinetic
 terms and the KK modes, which trigger significant resonant amplification
 of the curvature perturbations. We find that the strong resonant
 effects are localized to narrow wave number ranges, producing spikes in
 the perturbation spectrum. Investigation of such resonant signals opens
 up the possibility of probing the extra dimensional space through
 cosmological observations.  

\vfil

\end{titlepage}

\newpage
\tableofcontents

\section{Introduction}
\label{sec:intro}

Cosmic inflation~\cite{Starobinsky:1980te,Sato:1980yn,Guth:1980zm} not
only sets the initial conditions for the Hot Big Bang cosmology, but
also seeds structure formation in the universe by generating the
primordial density perturbations.
Conversely, the primordial perturbations are a powerful probe of
inflationary cosmology.
Upcoming precision cosmological measurements will further reveal 
details of the perturbation spectrum, which contains
significant information about the underlying physics of the inflationary
era. Since inflation is sensitive to Planck-scale physics, this
provides a fascinating opportunity of experimentally studying physics at
energy scales far beyond the reach of current terrestrial accelerators. 

In this paper, we explore the possibility that theories with extra
dimensions can produce specific signals such as spikes and/or
oscillations in the primordial perturbation spectrum. 
Inflation driven by branes wrapping the extra dimensions is
investigated, and we will show that in the presence of the excited
Kaluza-Klein (KK) modes (i.e. oscillation modes) of the brane along the wrapped
directions, sharp signals in the perturbation spectrum are generated
due to parametric resonance.
The excitation of the KK modes can be triggered during inflation as the
inflaton brane passes by localized sources or features along the
wrapped inflationary trajectory of the extra dimensions. 
Since the mass of the KK modes are
determined by the size of the wrapped cycles, for small extra dimensions
the excited KK modes are heavy and thus oscillate, leaving imprints on
the density perturbations. 
In this sense the wrapped inflaton brane scans the extra dimensions,
whose output is, as we will show in this paper, sharp signals
in the primordial perturbation spectrum. 
Moreover, the signals produced from the KK tower have a periodic nature
in $k$-space.
Thus all these features of wrapped brane inflation offer us an opportunity
of extracting information about the extra dimensions through
cosmological observations.

Our work is motivated by a class of models in string theory where
inflation is driven by D-branes wrapped on cycles of the internal
geometry, e.g.,
\cite{Kobayashi:2007hm,Becker:2007ui,Silverstein:2008sg,McAllister:2008hb}.  
We especially have in mind the construction
of~\cite{Silverstein:2008sg} where the tension of the wrapped D-brane
sources the potential energy driving inflation, and the monodromy
of the wrapped brane extends the field space to yield
large-field inflation. 
However, let us stress that our work is
not limited to such explicit construction, but discusses rather generic
phenomenon that arise when inflation is driven by physically extended objects.

From the effective four-dimensional point of view, the zero mode
position of the brane plays the role of the inflaton, while the excited
KK modes are heavy oscillating fields coupled to the inflaton. 
Perturbations with wave modes which resonate with the heavy fields' oscillation
frequencies are enhanced/suppressed while inside the Hubble horizon,
generating features in the resulting perturbation
spectrum. Effects of heavy fields during inflation have much in common with
inflaton potentials with sharp and/or repeated structures, and have been
the subject of extensive study, 
e.g., \cite{Starobinsky:1992ts,Adams:2001vc,Cline:2003gq,Martin:2003kp,Chen:2006xjb,Joy:2007na,Chen:2008wn,Bean:2008na,Flauger:2009ab,Flauger:2010ja,Nakashima:2010sa,Kawasaki:2010ux,Achucarro:2010da,Chen:2011zf,Chen:2011tu,Behbahani:2011it,Gao:2012uq,Chen:2012ja,Saito:2012pd}.
(See also~\cite{Nagata:2008zj,Ichiki:2009xs} for observational
analyses of spiky modulations in the primordial spectrum.)
Normally one finds that such effects are spread over a rather wide
$k$~range in the perturbation spectrum, since it takes a few e-foldings
during inflation for the inflaton to come back to its attractor
trajectory, or for the heavy field oscillations to damp away. However in
this paper, we show that parametric resonance due to the oscillating KK
modes can source sharp features that are localized to narrow wave number
ranges of the density perturbations. 
This is due to the Nambu-Goto action of the brane (or
the Dirac-Born-Infeld (DBI) action for D-branes) yielding nontrivial couplings
between the heavy KK modes and the inflaton, including those with the
inflaton kinetic term. Such kinetic couplings will turn out to be
extremely efficient in producing strongly resonant features. 
Furthermore, resonant amplification is highly sensitive to the
oscillating amplitude of the KK modes.
Since the excited KK mode oscillations are damped by the
inflationary expansion, strong resonance happens only for a narrow wave
number range, thus results in producing sharp spikes in the perturbation
spectrum. 
In this sense, our work is not only a wrapped brane realization of the
well-studied heavy field physics during inflation, but explores a new
possibility for generating extremely sharp features in the primordial
perturbation spectrum.

This paper is organized as follows: First we explain the setup for the
wrapped brane inflation we consider, and derive its effective
four-dimensional action in Section~\ref{sec:sec2}. 
The dynamics of the brane, i.e. of the zero mode inflaton and KK modes,
are discussed in Section~\ref{sec:dyna}.
Then in Section~\ref{sec:CPcalc}, we derive the evolution equation of
the inflaton field fluctuations, or equivalently, of the curvature
perturbations. The resonant signals in the perturbation spectrum
have different nature, depending on the amplitude of the excited KK
modes. KK modes with small amplitudes give rise to oscillations with
unique envelopes on the perturbation spectrum. This case, which we refer
to as weak resonance, is analyzed mostly analytically in
Section~\ref{sec:smallKK}. For largely excited KK modes, the parametric
resonance becomes extremely efficient and gives rise to spiky features in
the perturbation spectrum. Such strong resonance is investigated in
Section~\ref{sec:largeKK}. Finally we present our conclusions in
Section~\ref{sec:Conc}. 
In order to analyze effects from KK modes in the Nambu-Goto (or DBI)
action for wrapped branes, in Appendix~\ref{sec:app}, we derive the
second order action for field fluctuations in general multi-field
inflation with various forms of kinetic terms.
Explicit forms of the functions that show up in the evolution equation
for field fluctuations in wrapped brane inflation are laid out in
Appendix~\ref{app:B}. 
In Appendix~\ref{app:c}, we give detailed analyses of the evolution
equation for perturbations from weak parametric resonance.

\section{Effective Action of a Wrapped 4-Brane}
\label{sec:sec2}

In this paper we would like to address effects of heavily
oscillating KK modes on inflation driven by wrapped branes. 
For that purpose, let us consider the simple case where the inflaton
brane is a 4-brane that stretches along the external directions, while
wrapped on an internal 1-cycle. Inflation is considered to happen while
the 4-brane is moving along another direction of the internal manifold.
We will suppose the six-dimensional part of the metric relevant
to us to take the following diagonal form:
\begin{equation}
 ds^2 = g_{\mu\nu} \,  dx^\mu dx^\nu + g_{rr}\,  dr^2 +
  g_{\lambda\lambda}  d\lambda^2.
 \label{A1}
\end{equation}
Here $x^\mu$ ($\mu = 0,1,2,3$) are the external four-dimensional
coordinates, $r$ denotes the internal direction along which the inflaton
brane moves, and $\lambda$ is another internal direction wrapped by the
brane, which we consider to be compactified by 
$ \lambda \simeq \lambda + 2 \pi$.
For an explicit realization of wrapped brane inflation, we have in
mind the construction of~\cite{Silverstein:2008sg} which considers wrapped
D4-branes in ten-dimensional type IIA string theory. 
The detailed form of the internal metric will also be chosen based
on~\cite{Silverstein:2008sg} later on, 
however we should also stress that the phenomenon with KK modes that we
will discuss are not restricted to the explicit string construction, but
are rather generic features of wrapped brane inflation.
(See also \cite{Kobayashi:2010fm} which discusses wrapped brane
inflation with light KK modes along the wrapped directions.)

We start by considering the Nambu-Goto action of the wrapped
brane. Taking the brane coordinates~$\xi^M$ ($M = 0, \cdots , 4$) to
coincide with $x^\mu$ and $\lambda $, then the Nambu-Goto
action of the wrapped 4-brane with winding number~$p$ is\footnote{Though
we also denote the brane coordinate along the wrapped direction
by~$\lambda$ as in (\ref{nambugoto}), it should be noted that the
brane coordinate~$\lambda$ has periodicity $\lambda \simeq \lambda + 2 \pi p$.}
\begin{equation}
 S = -T_4 \int d^5 \xi \sqrt{-\det (G_{MN})} = 
-T_4 \int d^4 x \int^{2 \pi p }_0 d \lambda 
\sqrt{ -\det (G_{MN} ) },
 \label{nambugoto}
\end{equation}
where $T_4$ is the 4-brane tension and $G_{MN}$ is the induced metric on
the brane world-volume. $G_{MN}$ is given by the metric and the brane
position~$r(x^\mu, \lambda)$ as 
\begin{equation}
 G_{MN} d\xi^M d\xi^N = 
 g_{\mu\nu} dx^\mu dx^\nu + g_{\lambda\lambda} d\lambda^2 + 
 g_{rr} \left\{\partial_\mu r \partial_\nu r \, dx^\mu dx^\nu + 2
	 \partial_\mu r \partial_\lambda r \, dx^\mu d\lambda +
	 (\partial_\lambda r)^2 d\lambda^2  \right\}.
\end{equation}
One can check that its determinant takes the form
\begin{equation}
 \det (G_{MN} )  =  g \left\{ g_{\lambda \lambda } +
       g_{\lambda  \lambda } g_{rr} g^{\mu\nu}\partial_{\mu}r
       \partial_{\nu}r 
       + g_{rr} (\partial_{\lambda} r)^2 \right\},
\end{equation}
where $g = \det (g_{\mu\nu})$. Thus far, we have been including
derivatives of~$r$ up to all orders in the Nambu-Goto action. 

To make our discussion concrete, throughout this paper we suppose the
Nambu-Goto action to be the main contribution to the inflationary
Lagrangian, in other words, that inflation to be driven by the tension
of the wrapped brane. We further assume the
four-dimensional metric to depend only on the external coordinates,
i.e. $ g_{\mu\nu} =  g_{\mu\nu} (x)$, and consider the internal metric
to take the following form during inflation,
\begin{equation}
 g_{rr} = A^2, \qquad 
 g_{\lambda\lambda} = B^2 r^2,
\label{internalmetric}
\end{equation}
where $A^2$ and $B^2$ are positive constants. 
Thus the basic picture is that the wrapped 4-brane starts off at some
nonzero value of~$r$, and drives cosmic inflation as it moves towards
smaller~$|r|$ due to the wrapped brane tension. 
(\ref{internalmetric}) can be considered as an approximation of the
internal metric during inflation, i.e., for the region of the internal
space where the inflaton brane moves along.
We should also note that this is equivalent to the case constructed
in~\cite{Silverstein:2008sg} where the extra six-dimensional space of
type IIA string theory is compactified on nil
manifolds~\cite{Silverstein:2007ac}. (Though (\ref{internalmetric}) is
not exactly the geometry of the compactified nil manifold, 
monodromy of suitably wrapped D-branes extends the effective field
range, i.e. allows large ranges for~$r$, and realizes an effective metric
of the form~(\ref{internalmetric}) in the large field limit.)
The internal metric (\ref{internalmetric}) allows large-field inflation
for the canonically normalized zero mode position of the brane~$\phi$
with the effective potential $V \propto \phi^{2/3}$, as we will soon see.

In order to obtain the effective four-dimensional action, we now expand
the 4-brane position in the internal space as
\begin{equation}
 r(x^\mu, \lambda) = \sum_{n=-\infty}^{\infty} r_n (x^\mu)\, 
  e^{i n \lambda / p}
\end{equation}
where $\bar{r}_n = r_{-n}$. In the following we use the abbreviations
\begin{equation}
 \partial f \cdot \partial g \equiv
 g^{\mu\nu} \partial_\mu f \partial_\nu g , 
 \qquad
 (\partial f)^2 \equiv 
  g^{\mu\nu} \partial_\mu f \partial_\nu f,
\qquad 
 | \partial f |^2 \equiv g^{\mu\nu} \partial_\mu f \partial_\nu \bar{f},
\end{equation}
and we also assume
\begin{equation}
 r, r_0 > 0,  \qquad  1 + A^2 (\partial r_0)^2 > 0. 
\label{rgtr0}
\end{equation}
Then, expanding the action~(\ref{nambugoto}) in terms of the (nonzero) KK
modes, one obtains the effective four-dimensional Lagrangian 
($ S  = \int d^4 x \mathcal{L}$) as
\begin{multline}
 \mathcal{L} = -\frac{2 \pi p B T_4 r_0}{\gamma}\sqrt{-g} 
 \Biggl[ 1 
+ \sum_{n \neq 0 } 
 \biggl\{
 \gamma^2 A^2 \frac{\bar{r}_n}{r_0} (\partial r_0 \cdot \partial r_n) +
 \frac{A^2}{2} \gamma^2
 |\partial r_n |^2
 + \frac{\gamma^2}{2} \frac{A^2}{B^2} \frac{n^2}{p^2}
 \frac{|r_n|^2}{r_0^2}
\\
- \frac{\gamma^4}{2} A^4 (\partial r_0 \cdot \partial r_n)
 (\partial r_0 \cdot \partial \bar{r}_n)
\biggr\}
 + (\mathrm{cubic\, \,  or\, \,  higher\, \, in\, \, KK\, \, modes})
\Biggr],
 \label{L8}
\end{multline}
where $\sum_{n\neq 0 }$ denotes the sum over all nonzero~$n$, i.e. $
\sum_{n = -\infty, n\neq 0}^{\infty}$,  and $\gamma$ is defined as
\begin{equation}
 \gamma \equiv \frac{1}{\sqrt{1 + A^2 (\partial r_0)^2}}.
\end{equation}
The wrapped brane inflation under consideration is basically
a slow-roll one, and the inflaton brane does not go into the so-called
DBI regime $\gamma \gg 1$~\cite{Silverstein:2003hf,Alishahiha:2004eh},
however it is important to keep the $\gamma$ factor since it gives rise
to kinetic couplings between the zero mode and KK modes, as can be seen
in for e.g. the $A^2 \gamma^2 |\partial r_n|^2$ term. 

We have skipped terms including cubic or higher
order in terms of the (nonzero) KK modes~$r_n$ and/or their
derivatives~$\partial r_n$. 
It can be checked that such higher order terms in the Lagrangian are smaller than the
quadratic ones shown in the $\left\{\right\}$ parentheses in (\ref{L8})
under the following conditions:
\begin{equation}
 \left| \frac{r_n}{r_0}\right|, \, \, 
\left| A^2 (\partial r_0)^2 \right|, \, \, 
\left| A^2 (\partial r_0 \cdot \partial r_n)\right|, \, \, 
\left| A^2 (\partial r_n )^2 \right|, \, \, 
\left| \frac{A^2}{B^2} \frac{n^2}{p^2} \frac{r_n^2}{r_0^2} \right|
 \ll 1,
 \label{maru1}
\end{equation}
where $n$ in (\ref{maru1}) represents any nonzero KK
mode. Hereafter we restrict ourselves to this condition and consider
interactions with the KK modes up to the quadratic order.
Under (\ref{maru1}), one may consider the term $\gamma^4 A^4 (\partial
r_0 \cdot \partial r_n) (\partial r_0 \cdot \partial \bar{r}_n)$ in the
second line of (\ref{L8}) to be negligible compared to $A^2 \gamma^2 |\partial
r_n|$ in the first line, however it should be noted that when focusing
on their kinetic couplings to $\partial r_0$, they both are of the same strength.
(Actually, we will see that these terms both give important contributions
to the resulting KK-mode effects.)\footnote{To be precise, (\ref{maru1})
is the sufficient condition 
for the quadratic terms of the most excited KK modes to be larger than
any other cubic or higher KK terms. We also note that terms in the
Lagrangian being small does not necessarily guarantee their effects on
the equations of motion to be negligibly tiny. However, in this paper we 
simply drop the higher order KK terms based on the
condition~(\ref{maru1}).}

Let us now redefine the fields as follows,
\begin{equation}
\begin{split}
 \phi & \equiv \left(\frac{8}{9} \pi p T_4 A^2 B  \right)^{1/2} r_0^{3/2}
 \\
  \psi_n & \equiv 
  \left(4 \pi p T_4 A^2 B\right)^{1/2} r_0^{1/2} \times
 \left\{
   \begin{array}{cl}
     \mathrm{Re}(r_n) & \quad (\mbox{for}\, \, \, n > 0) \\    
     \mathrm{Im}(r_n) & \quad (\mbox{for}\, \, \, n < 0)  
   \end{array}
\right.
\end{split}
\end{equation}
so that the action contains only real fields, and the zero mode~$\phi$
becomes canonical in the slow-roll limit in the absence of the KK modes.
Then the Lagrangian is rewritten as
\begin{multline}
 \frac{\mathcal{L}}{\sqrt{-g}} = -V \left(\frac{1}{\gamma }
 + 2 \gamma \sum_{n \neq 0} \alpha_n^2 
 \frac{\psi_n^2}{\phi^2} \right)
 - \gamma \sum_{n \neq 0} 
 \left\{ \frac{1}{2} (\partial \psi_n )^2 - \frac{1}{6} \frac{
  \psi_n^2}{\phi^2} (\partial \phi)^2
 + \frac{1}{3} \frac{\psi_n}{\phi } ( \partial \phi \cdot
 \partial \psi_n )
 \right\}
\\
+ \frac{\gamma^3}{2 V}\sum_{n \neq 0} 
\left\{
\left( \partial \phi \cdot \partial \psi_n  \right)^2 
+ \frac{1}{9} \frac{ \psi_n^2}{\phi^2} 
\left( (\partial \phi)^2 \right)^2
- \frac{2}{3} \frac{\psi_n}{\phi } (\partial \phi)^2 (\partial \phi
 \cdot \partial \psi_n) 
\right\},
\label{Lphi}
\end{multline}
where again $\sum_{n\neq 0 } = \sum_{n = -\infty, n\neq 0}^{\infty}$.
We have defined an effective potential
\begin{equation}
 V(\phi) = \mu^{10/3} \phi^{2/3}
 \qquad \mathrm{with} \qquad
  \mu  \equiv \left(\frac{3 \pi p T_4 B}{A} \right)  ^{1/5},
\label{Vphi}
\end{equation}
and a dimensionless constant
\begin{equation}
 \alpha_n^2  \equiv \frac{1}{9} \frac{A^2 }{B^2 } \frac{n^2}{ p^2}.
\label{alpha_n}
\end{equation}
Note that $\alpha_n^2 = \alpha_{-n}^2$, and $\alpha_n$ denotes the 
positive root of (\ref{alpha_n}).
The $\gamma$ factor is now expressed as
\begin{equation}
 \gamma = \left(1 +\frac{ (\partial \phi)^2}{V}\right)^{-1/2},
\end{equation}
and the condition (\ref{maru1}) for neglecting the cubic or higher order
terms of the KK modes is transformed to
\begin{equation}
 \left|\frac{\psi_n}{\phi}  \right|, \, \, 
 \left| \frac{(\partial \phi)^2}{V}  \right|, \, \,
 \left| \frac{\partial \phi \cdot \partial \psi_n}{ V}  \right|, \, \,
 \left|\frac{(\partial \psi_n )^2 }{V}\right| , \, \, 
 \left| \alpha_n^2  \frac{\psi_n^2}{\phi^2} \right|
 \ll 1.
 \label{maru1dash}
\end{equation}

In the absence of the KK modes, the Lagrangian~(\ref{Lphi}) in the
slow-roll limit reduces to the canonical form of 
$\mathcal{L} / \sqrt{-g} \simeq -(\partial \phi)^2 / 2 - V(\phi)$,
which realizes large-field inflation.

\section{Dynamics of Wrapped Brane Inflation}
\label{sec:dyna}

Now that we have the effective four-dimensional action~(\ref{Lphi}), let
us study the inflationary dynamics of the wrapped 4-brane. 
We start by discussing the homogeneous background.

\subsection{Homogeneous Background}

Upon discussing the homogeneous equations of motion, we fix the
background metric to the flat FRW:  
\begin{equation}
 ds^2 = -dt^2 + a^2 (t) d \boldsymbol{x}^2.
\end{equation}
Throughout this paper an overdot is used to denote derivatives in terms
of the time~$t$, and the Hubble parameter is defined by $H = \dot{a} /
a$. Then the Einstein equation of the action~(\ref{Lphi}) gives the
Friedmann equation
\begin{multline}
 3 M_p^2 H^2 = 
\gamma V \left(1 - 2 \gamma^2 \sum_{n \neq 0} \alpha_n^2
 \frac{\psi_n^2}{\phi^2}  \right) 
 + \gamma^3 \sum_{n \neq 0}
\left( \frac{1}{2} \dot{\psi}_n^2 - \frac{1}{6}
  \frac{\dot{\phi}^2   }{\phi^2} \psi_n^2+ 
 \frac{1}{3} \frac{\dot{\phi}}{\phi} \psi_n \dot{\psi}_n
 \right) 
\\
 + 3 \gamma^3 (\gamma^2 - 1)
\sum_{n \neq 0}
\left( \frac{1}{2} \dot{\psi}_n^2 + \frac{1}{18}
  \frac{\dot{\phi}^2   }{\phi^2} \psi_n^2 -
 \frac{1}{3} \frac{\dot{\phi}}{\phi} \psi_n \dot{\psi}_n
 \right) 
,
\label{Fvarphi}
\end{multline}
as well as the evolution equation
\begin{multline}
 - 2 M_p^2 \dot{H} 
 = \gamma \dot{\phi}^2 
 \left( 1 -  2 \gamma^2 \sum_{n \neq 0}\alpha_n^2 
 \frac{\psi_n^2}{ \phi^2} 
   \right)
 + \gamma (\gamma^2 + 1)
\sum_{n \neq 0} 
\left( \frac{1}{2} \dot{\psi}_n^2 - \frac{1}{6}
  \frac{\dot{\phi}^2   }{\phi^2} \psi_n^2+ 
 \frac{1}{3} \frac{\dot{\phi}}{\phi} \psi_n \dot{\psi}_n
 \right) 
\\
 + \gamma ( 3 \gamma^2 + 1) (\gamma^2 - 1)
\sum_{n \neq 0}
\left( \frac{1}{2} \dot{\psi}_n^2 + \frac{1}{18}
  \frac{\dot{\phi}^2   }{\phi^2} \psi_n^2 -
 \frac{1}{3} \frac{\dot{\phi}}{\phi} \psi_n \dot{\psi}_n
 \right) .
\label{dotH}
\end{multline}

The full form of the equations of motion of the brane positions
$\phi$ and $\psi_n$ are rather complicated, so let us partially write
down the equations. 
The equation of motion of the zero mode~$\phi$ is 
\begin{equation}
\label{EoMphi}
 0 = \gamma^2 \ddot{\phi} + 3 H  \dot{\phi} +
 \frac{3 - \gamma^2}{2}  \frac{dV}{d\phi } + \cdots,
\end{equation}
where $\cdots$ denotes terms containing the KK modes $\psi_n$ and their
time derivatives. 
This expression suffices for discussing the slow-roll dynamics of the
inflaton~$\phi$. The interactions between $\phi$ and $\psi_n$ will be
analyzed in detail in the next section when we study perturbations.

As for the KK mode~$\psi_n$, the term $-2 V\gamma \alpha_n^2 \psi_n^2 /
\phi^2 $ in the Lagrangian~(\ref{Lphi}) sources the effective mass 
that is set by the length of the wrapped cycle,
\begin{equation}
m_{\mathrm{KK}}^2 
\simeq \frac{4  \alpha_n^2 V}{\phi^2 }
=  \frac{n^2}{p^2 B^2 r_0^2},
\end{equation}
where we have ignored the $\gamma $ factor. 
Only writing down terms that are most relevant for us, the equation of
motion of the KK mode $\psi_{n}$ is  
\begin{equation}
\label{EoMvarphi}
 0 =  \ddot{\psi}_n    +  3 H  \dot{\psi}_n   + 4  V \alpha_n^2
 \frac{\psi_n}{\phi^2}   + \cdots .
\end{equation}

\subsection{Slow-Roll and Heavy-Field Approximations}
\label{subsec:approx}

For super-Planckian field values $\phi > M_p$, the action (\ref{Lphi})
can realize large field inflation with the potential $V(\phi)$
(\ref{Vphi}) where $\phi$ is a nearly canonical inflaton field. 
If the KK modes~$\psi_n$ are excited during inflation, then
given that their effective masses $\sim V \alpha_n^2 / \phi^2$ are larger
than $H^2$, the KK modes would oscillate and leave resonant imprints on the
curvature perturbation spectrum. 

We examine such case in this subsection and analyze the fields' dynamics. 
Specifically, we consider the case where the equations of motion 
(\ref{Fvarphi}), (\ref{EoMphi}), and (\ref{EoMvarphi}) are 
well approximated by, respectively, the slow-roll approximations
\begin{equation}
 3 M_p^2 H^2 \simeq V ,
 \label{star2}
\end{equation}
\begin{equation}
 3 H \dot{\phi} \simeq -\frac{dV}{d\phi} ,
\label{star1}
\end{equation}
and the heavy-field approximation
\begin{equation}
 \ddot{\psi}_n \simeq - 4 V \alpha_n^2 \frac{\psi_n}{\phi^2}.
 \label{star3}
\end{equation}
We further suppose that the order of magnitude of the $\psi_n$
velocity is given by
\begin{equation}
 \dot{\psi_n}  \sim 
V^{1/2} \alpha_n \frac{\psi_n}{\phi  },
 \label{star4}
\end{equation}
when averaged over the oscillation period.
We should remark that throughout this paper, we use ``$\sim$'' to denote that the
orders of magnitude of both sides of the equation are the same. For more precise
approximations, we use ``$\simeq$''. 
Effects of KK modes on the curvature perturbations are discussed in
Section~\ref{sec:smallKK} based on 
the above approximations, then in Section~\ref{sec:largeKK} we go
beyond this case.

\vspace{\baselineskip}

Let us now show that under the approximations (\ref{star2}) -
(\ref{star4}), the amplitudes of the following three parameters need to
be sufficiently smaller than unity
(We note that throughout the discussions we do not consider miraculous
cancellations among terms in the equations.):
\begin{equation}
 \epsilon \equiv \frac{M_p^2}{ \phi^2} , \quad
 \eta_n \equiv \alpha_n^3 \frac{M_p \psi_n^2}{\phi^3} , \quad
 \xi_n \equiv \frac{\phi }{\alpha_n M_p}.
\label{3srpar}
\end{equation}
Note that these are all positive parameters, since
we are considering positive~$\phi$, cf.~(\ref{rgtr0}).

Comparing with the full equations of motion, one can show that the
sufficient condition for (\ref{star2}) and (\ref{star3}) to hold is 
\begin{equation}
 \epsilon, \, \, \xi_n, \, \, \eta_n \xi_n  , \,  \, 
\left| \epsilon^{5/6}  \xi_n \frac{\ddot{\phi} M_p^{1/3}}{\mu^{10/3}} 
\right|
 \ll 1.
 \label{4cond}
\end{equation}
Here we remark that when we simply write, for e.g., $\eta_n$ as in
(\ref{4cond}), then $\eta_n$ represents the parameter for all nonzero~$n$.
Then, assuming $\epsilon, \, \xi_n, \, \eta_n \xi_n  \ll 1 $, one can
show from (\ref{dotH}) that\footnote{Expressions such as $y =
\mathcal{O} (x)$ are used to denote that $|y| \lesssim |x|$.}   
\begin{equation}
 \frac{\dot{H}}{H^2} = \mathcal{O}(\epsilon) + \mathcal{O} (\eta_n \xi_n).
\end{equation}
It should be noted that 
differentiating both sides of approximate relations does not necessarily
provide similarly good approximations.
Hence, in order to estimate the amplitude of~$\ddot{\phi}$, let us
introduce a parameter 
\begin{equation}
 \kappa \equiv \frac{3 H \dot{\phi}}{dV/d\phi} + 1.
\label{kappadef}
\end{equation}
From the equation of motion of~$\phi$, one can obtain
\begin{equation}
 \frac{\ddot{\phi}}{dV/d\phi} + \kappa = \mathcal{O} (\epsilon) + \mathcal{O} (\eta_n).
\end{equation}
Combining this with the time-derivative of (\ref{kappadef}), one arrives at
\begin{equation}
 \frac{\dot{\kappa}}{H}  = 
-3 \kappa + \mathcal{O} (\epsilon) + \mathcal{O} (\eta_n),
\end{equation}
which shows that $|\kappa| $ damps as the universe expands while
$|\kappa| \gg \epsilon,\, \eta_n$. Hence one can conclude that 
the amplitude of $\kappa$ soon approaches 
\begin{equation}
 \kappa = \mathcal{O} (\epsilon) + \mathcal{O} (\eta_n),
\end{equation}
and thus
\begin{equation}
 \frac{\ddot{\phi}}{H \dot{\phi}} = \mathcal{O} (\epsilon) + \mathcal{O}
  (\eta_n). 
\label{ddphiamp}
\end{equation}
Therefore the term $ \epsilon^{5/6}  \xi_n \ddot{\phi} M_p^{1/3} \mu^{-10/3}$
in (\ref{4cond}) is estimated to be of size $\mathcal{O} (\epsilon^2
\xi_n) + \mathcal{O} (\epsilon \eta_n \xi_n)$.
One can also show using (\ref{ddphiamp}) that 
the approximation (\ref{star1}) further requires $\eta_n$ to be small. 

Moreover, since
\begin{equation}
 \frac{\dot{\phi}}{H \phi }\sim \epsilon, \qquad
 \frac{\dot{\psi}_n}{H \psi_n} \sim \frac{1}{\xi_n} ,
\end{equation}
one can neglect the time variation of~$\phi$ in the approximation
(\ref{star3}) and see that $\psi_n$ at the leading order harmonically
oscillates as $\cos (2 V^{1/2} \alpha_n t / \phi )$.
This validates the order of magnitude estimation of $\dot{\psi}_n$ in
(\ref{star4}). (See also discussions around (\ref{anvarphi}).) 

\vspace{\baselineskip}

In summary, we have seen that when the approximations (\ref{star2}) -
(\ref{star4}) hold, and given that there is no miraculous cancellation
among the terms, then the following condition is satisfied,
\begin{equation}
  \epsilon, \, \, \eta_n, \, \, \xi_n \ll 1.
\label{smallpar}
\end{equation}
This guarantees $\alpha_n \gg  \phi / M_p \gg 1
$. It should also be noted that the smallness of $\xi_n$ implies that the
effective mass of the KK mode $\psi_n$ is sufficiently larger than the Hubble
parameter during inflation. 

For later convenience, here we lay out the order-of-magnitude estimations of
various quantities:
\begin{gather}
\gamma \simeq 1, \qquad
 \frac{\dot{\gamma}}{H \gamma } = \mathcal{O}  (\epsilon^2) +
 \mathcal{O}  (\epsilon \eta_n), \qquad
\frac{\dot{H}}{H^2 } = \mathcal{O} (\epsilon) + \mathcal{O}  (\eta_n
\xi_n), \qquad
\label{3737}
\\
 \frac{\dot{\phi}^2}{V} \sim \epsilon, \qquad
 \frac{\dot{\phi}}{H \phi} \sim \frac{\dot{V}}{H V} \sim \epsilon, \qquad
 \frac{\ddot{\phi}}{H \dot{\phi}} = \mathcal{O}  (\epsilon) +
 \mathcal{O}  (\eta_n), 
 \label{3838}
\\
 \frac{\dot{\psi}_n}{H \psi_n} \sim \frac{\ddot{\psi}_n}{H
 \dot{\psi}_n} \sim \frac{1}{ \xi_n}.
 \label{3939}
\end{gather}

We also note that the condition (\ref{maru1dash}) that was assumed in the
previous section is rewritten in terms of the small parameters as
\begin{equation}
 (\epsilon \eta_n \xi_n^3)^{1/2}, \, \, 
\epsilon, \, \, 
(\epsilon \eta_n \xi_n)^{1/2}, \, \, 
\eta_n \xi_n, \,  \, 
\eta_n \xi_n
 \ll 1,
\label{3.23}
\end{equation}
where we have neglected the spatial derivatives.

\vspace{\baselineskip}

We end this section by writing down an approximate solution for $\psi_n$
beyond the leading order equation~(\ref{star3}), which will be used upon
discussing curvature perturbations in the next section.  
One can check that the ansatz of the form\footnote{This ansatz can
easily be guessed from the next-to-leading order
approximation of the equation of motion of $\psi_{n}$,
\begin{equation}
 \ddot{\psi}_n + 4 V \alpha_n^2 \frac{\psi_n}{\phi^2} + 3 H \dot{\psi}_n 
 - \frac{\dot{\phi}^2}{V} \ddot{\psi}_n - 8 \alpha_n^2
 \frac{\dot{\phi}^2}{\phi^2} \psi_n \simeq 0,
\end{equation}
where the third term in the left hand side is suppressed by $\sim
\xi_n$ compared to the leading (i.e. first and second) terms, while
the forth and fifth terms are suppressed by $\sim \epsilon$. 
We note that (\ref{anvarphi}) is a good ansatz independently 
of $ \xi_n  \lessgtr \epsilon$.}
\begin{equation}
 \psi_n =    \frac{C}{a^{3/2}} 
 \cos \left\{ 2 \alpha_n \frac{V^{1/2}}{\phi} (t-t_*) + \theta  \right\}
 \label{anvarphi}
\end{equation}
(here $C$, $t_*$, and $\theta$ are constants) satisfies the
full equation of motion of $\psi_n$
within errors of $\mathcal{O}
(\epsilon, \, \xi_n^2)$, given that $H (t-t_*) \lesssim 1$, i.e.,
the time scale of interest is not much greater than the Hubble time.
Here we note that upon estimating the size of the error, 
we have considered cosines and sines to be $\sim 1$, and made
use of (\ref{star2}), (\ref{3737}), and (\ref{3838}), but instead of (\ref{3939}) we
have used the ansatz~(\ref{anvarphi}).
In a similar fashion, one can also check that during $H (t-t_*) \lesssim 1$, the
ansatz (\ref{anvarphi}) satisfies (\ref{star4}), and also
$\dot{\psi}_n / H \psi_n \sim
\ddot{\psi}_n / H \dot{\psi}_n \sim 1 / \xi_n$ as in~(\ref{3939}).
Since the oscillations of the KK modes are quickly damped, the above approximate
solution which is valid for a time range of $\sim H^{-1}$ will be useful
upon discussing effects from the KK modes.

\section{Effects of KK Modes on the Curvature Perturbations}
\label{sec:CPcalc}

The KK modes of the wrapped brane can be excited during inflation when the
brane passes by localized defects along the wrapped direction, such as
other branes and ``bumps'' in the internal manifold.
In this section we compute the curvature perturbations from wrapped
brane inflation, considering the KK mode oscillations. 
We start by applying the general discussions in Appendix~\ref{sec:app} to our
action~(\ref{Lphi}) and calculate the field fluctuations of
the inflaton~$\phi$, which will be transformed into the curvature
perturbations via the $\delta \mathcal{N}$-formalism.

\subsection{Fluctuation Equation}
\label{subsec:FE}

We are interested in the case where the nonzero KK modes $\psi_n$ are
heavy (i.e. $m_{\mathrm{KK}}^2 \gg H^2$), hence we neglect their field
fluctuations and only consider fluctuations of the zero mode inflaton, 
\begin{equation}
  \phi = \phi_0 + \frac{q}{a},
\end{equation}
where $\phi_0$ is the homogeneous classical background.
We work on flat spatial hypersurfaces, and hereafter we 
drop the subscript~$0$ denoting the background. 
Then the second order action for the field fluctuations~(\ref{S2Q})
is 
(see Appendix~\ref{sec:app} for the detailed
derivation\footnote{Discussions in Appendix~\ref{sec:app} are applied to
our case by substituting
\begin{equation}
\begin{split}
 \varphi^I \, & \longrightarrow \, \phi, \quad  \psi_n,  \\
 X^A \, & \longrightarrow \, -(\partial \phi)^2, \quad
 -\sum_{n \neq 0} (\partial \psi_n)^2, \quad 
 -(\partial \phi \cdot \partial \psi_n ) .
\end{split}
\end{equation}
}) 
\begin{equation}
 S_2 = \int d\tau  d^3 x 
 \left\{
 -L_A G^A_{\phi \phi} (\delta^{ij} \partial_i q \partial_j q) +
 \frac{a^2}{2} \left( M_{\phi 
     \phi} -      \dot{C}_{\phi \phi} - 3 \frac{a'}{a^2}C_{\phi \phi} 
\right) q^2 + \frac{1}{2} B_{\phi \phi} \left(
 q' - \frac{a'}{a} q \right)^2
\right\},
 \label{S2q}
\end{equation}
where $\tau$ is the conformal time $dt = a \, d\tau $, a prime denotes a
$\tau$-derivative, and the quantities except for $q$ are those of the
homogeneous background.
The definition of the functions $B_{\phi \phi}$, $C_{\phi\phi}$,
etc. are laid out in Appendix~\ref{app:B}. 
The second order action gives a linear equation of motion for~$q$,
which, after Fourier expansion
\begin{equation}
 q (\tau , \boldsymbol{x}) = \frac{1}{ (2 \pi)^3} 
 \int d^3 k\,  e^{-i \boldsymbol{k \cdot x}}
 \,  q_{\boldsymbol{k}} (\tau), 
\end{equation}
takes the form (cf.~(\ref{EoMq}))
\begin{equation}
 0 = q_{\boldsymbol{k}}'' +  \frac{\dot{B}_{\phi \phi}}{B_{\phi \phi} }
 a  q_{\boldsymbol{k}}' + 
\left\{ \frac{2 L_A G_{\phi \phi}^A}{B_{\phi \phi} } k^2 -  \frac{a''}{a}
 + \frac{-M_{\phi \phi} + \dot{C}_{\phi \phi}}{B_{\phi \phi} } a^2  + 
 \frac{ -\dot{B}_{\phi \phi} + 3 C_{\phi \phi} }{B_{\phi \phi} } a'\right\}
q_{\boldsymbol{k}}. 
 \label{4747}
\end{equation}
Here, $k = |\boldsymbol{k} |$.
We remark that $B_{\phi \phi}$ is positive throughout the cases
discussed in this paper.

\subsection{Analytic Study of Field Fluctuations}

In the rest of this section we carry out analytic calculations of the field
fluctuations by restricting ourselves to the case studied in 
Section~\ref{subsec:approx}, where the excited KK mode amplitudes are
small (cf. (\ref{smallpar})) and the background field dynamics are
approximated by (\ref{star2}) - (\ref{star4}).
We will compute the leading order effects from the KK
modes on the curvature perturbation spectrum.

\subsubsection{Evolution Equation for Small KK Excitations}

Under the condition~(\ref{smallpar}),
the homogeneous functions in (\ref{4747}) (cf. (\ref{eqB4}) -
(\ref{eqB7})) can be evaluated as 
\begin{gather}
 B_{\phi \phi} = 1 - 2 \sum_{n \neq 0} \alpha_n^2
 \frac{\psi_n^2}{\phi^2} + \frac{3}{2} \sum_{n \neq 0} \frac{\dot{\psi}_n^2}{V}
 + \mathcal{O} (\epsilon), 
 \label{Bphiphi}
\\
 \frac{\dot{B}_{\phi \phi}}{H}  = -\frac{4}{H} \sum_{n \neq 0}
  \alpha_n^2 \frac{\psi_n \dot{\psi}_n}{\phi^2} + \frac{3}{VH} \sum_{n
  \neq 0} \dot{\psi}_n \ddot{\psi}_n + 
\mathcal{O} (\epsilon \eta_n) + \mathcal{O} (\epsilon^2), 
\label{4949}
\\
L_{A}G^A_{\phi \phi} = \frac{1}{2} 
-  \sum_{n \neq 0} \alpha_n^2
 \frac{\psi_n^2}{\phi^2} + \frac{1}{4} \sum_{n \neq 0} \frac{\dot{\psi}_n^2}{V}
 + \mathcal{O} (\epsilon), 
\label{LAGAphiphi}
\\
 \frac{C_{\phi \phi}}{H} = \mathcal{O} (\epsilon^2) + \mathcal{O}
  (\epsilon \eta_n \xi_n),
\qquad
 \frac{\dot{C}_{\phi \phi}}{H^2} = \mathcal{O} (\epsilon^3) +
  \mathcal{O} (\epsilon \eta_n),
\qquad 
 \frac{M_{\phi \phi}}{H^2} = \mathcal{O} (\epsilon ),
\end{gather}
where we have used (\ref{3737}), (\ref{3838}), and (\ref{3939}) upon
estimating the amplitude of the dropped terms.
For the term $\dot{B}_{\phi \phi} / H $ (\ref{4949}), the first two
terms in the right hand side each has amplitude $ \sim \eta_n$.
We also note that for $B_{\phi \phi}$ and $L_A G^A_{\phi \phi}$, the
second and third terms on the right hand sides are of order $\eta_n
\xi_n$, and the other dropped KK terms (terms depending on $\psi_n$ and
$\dot{\psi}_n$) in (\ref{Bphiphi}) and (\ref{LAGAphiphi}) are
$\mathcal{O} (\epsilon \eta_n \xi_n)$.  

The conformal time can be computed by integrating
\begin{equation}
 d \left(- \frac{1}{aH}\right) = \left(1 + \frac{\dot{H}}{H^2}\right) d
  \tau .
\end{equation}
Hence one finds
\begin{equation}
 \tau = -\frac{1}{aH} \left( 1 + \mathcal{O} (\epsilon) + \mathcal{O}
      (\eta_n \xi_n) \right),
 \label{esttau}
\end{equation}
where we have took $\tau$ to (at the leading order) approach $0$ from
the negative side as $a \to \infty$.\footnote{Strictly speaking, the
$\mathcal{O} (\epsilon) $ and $ \mathcal{O} (\eta_n \xi_n) $ in
(\ref{esttau}) are integrated values of the parameters over a finite
period of $\tau$, however for simplicity we treat them as 
$\mathcal{O} (\epsilon, \eta_n \xi_n) $.\label{foot:appint}}
Then one can further show 
\begin{equation}
 \frac{ a'}{a} =  - \frac{1}{\tau} 
 \left( 1 + \mathcal{O} (\epsilon) + \mathcal{O} (\eta_n \xi_n) \right),
 \qquad
 \frac{a''}{a} = \frac{2}{ \tau^2} \left( 1 + \mathcal{O} (\epsilon) +
				    \mathcal{O} (\eta_n \xi_n) \right) .
 \label{adash}
\end{equation}

Thus by combining the above estimations, one can rewrite (\ref{4747}),
keeping only the leading order contributions from the KK modes as
\begin{multline}
 0 =  q_{\boldsymbol{k}}''  
+   \biggl\{
\frac{4}{H} \sum_{n \neq 0}
  \alpha_n^2 \frac{\psi_n \dot{\psi}_n}{\phi^2} -\frac{3}{VH} \sum_{n
  \neq 0} \dot{\psi}_n \ddot{\psi}_n 
+ \mathcal{O} (\epsilon^2) + \mathcal{O} (\epsilon \eta_n)  + \mathcal{O}
(\eta_n^2 \xi_n)
\biggr\} \frac{q_{\boldsymbol{k}}'}{ \tau}
 \\
 +  \Biggl[
 \biggl\{ 1 
 - \frac{1}{V} \sum_{n \neq 0} \dot{\psi}_n^2
+ \mathcal{O} (\epsilon) + 
 \mathcal{O} (\eta_n^2 \xi_n^2)  \biggr\}
 k^2 \tau^2
 \qquad \qquad  \qquad \qquad \qquad
 \\
   - 2 + 
\frac{4}{H} \sum_{n \neq 0}
  \alpha_n^2 \frac{\psi_n \dot{\psi}_n}{\phi^2} -\frac{3}{VH} \sum_{n
  \neq 0} \dot{\psi}_n \ddot{\psi}_n 
+ \mathcal{O} (\epsilon)  + \mathcal{O} (\eta_n \xi_n)
\Biggr] \frac{q_{\boldsymbol{k}}}{\tau^2}.
 \label{eq:qk}
\end{multline}
Let us repeat that the $\psi_n$ terms explicitly written inside the
parentheses on the first and third lines have amplitude of $\sim
\eta_n$, while that in the second line is $\sim \eta_n
\xi_n$.\footnote{We also note that the term $\dot{\psi}_n^2 / V$ in 
the second line of (\ref{eq:qk})
is the leading KK mode
contribution to the coefficient of $k^2 \tau^2$, since the
next-to-leading KK terms are of $\mathcal{O} (\epsilon \eta_n \xi_n,
\eta_n^2 \xi_n^2)$.}
One may except the term $\dot{\psi}_n^2 / V $ in the second line to have
much smaller contribution than the other KK terms, however this is not
the case for the following reason: The field fluctuation~$q_k$
experiences parametric resonance with the KK mode oscillations when
their frequencies synchronize. This happens when $ k / a \sim
m_{\mathrm{KK}} \sim \alpha_n
V^{1/2} / \phi$, i.e., when $k^2 \tau^2 \sim \xi_n^{-2}$. (We will see
this explicitly in Section~\ref{sec:smallKK}.) 
Thus for $q_{\boldsymbol{k}}$ that experience parametric resonance,
the term $k^2 \tau^2 \dot{\psi}_n^2 / V$ is of order $\eta_n / \xi_n$
until the wave mode passes the resonance band. 
Thus in the end, this term gives contributions comparable to those from
the KK terms of order~$ \eta_n$ in the first line of
(\ref{eq:qk}).\footnote{One may then naively think that 
the term $k^2 \tau^2 \dot{\psi}_n^2 / V$ should have much larger
effects than the other KK terms, however their contributions turn
out to be comparable when taking into account that it is actually
the absolute value of $q_{\boldsymbol{k}}$ that matters for physical
observables.}  
We also note that the KK terms in the third line turn out to be irrelevant inside the
horizon, as we will soon see.

We stress that the leading KK terms shown explicitly in (\ref{eq:qk})
(or in (\ref{Bphiphi}), (\ref{4949}), and (\ref{LAGAphiphi})) 
arise from the following three terms in the
Lagrangian~(\ref{Lphi}) through their kinetic couplings with the inflaton,
\begin{equation}
 - 2 \gamma V  \sum_{n \neq 0} \alpha_n^2 \frac{\psi_n^2}{\phi^2},
\qquad
-\frac{\gamma}{2} \sum_{n \neq 0}
 (\partial \psi_n)^2,
\qquad
 \frac{\gamma^3}{2 V} \sum_{n \neq 0}
 (\partial \phi \cdot \partial \psi_n)^2.
\label{threekings}
\end{equation}
These kinetic coupling terms play important roles even when we go beyond
the small KK excitations and study strong resonance in
Section~\ref{sec:largeKK}. 

\vspace{\baselineskip}

We expand the field fluctuation~$q_{\boldsymbol{k}}$ in terms of
$\eta_n$ as
\begin{equation}
 q_{\boldsymbol{k}} = q_{\boldsymbol{k}}^{(0)} +
  q_{\boldsymbol{k}}^{(1)} + \cdots,
\end{equation}
where $q_{\boldsymbol{k}}^{(m)} = \mathcal{O} (\eta_n^m)$, and solve
(\ref{eq:qk}) at each order.
(We remark that here we are implicitly assuming a hierarchy between
$\epsilon$ and $\eta_n$ (i.e., either $\epsilon \ll \eta_n $ or
$\epsilon \gg \eta_n $), otherwise we may have to analyze the
$\mathcal{O}(\epsilon)$ correction at each order in~$\eta_n$.) 
Then the $\eta_n^0$~order (\ref{eq:qk}) is, at zeroth order
of~$\epsilon$, 
\begin{equation}
 q_{\boldsymbol{k}}''^{(0)} + \left( k^2 - \frac{2}{\tau^2} \right) 
 q_{\boldsymbol{k}}^{(0)} \simeq 0.
\label{qk0eq}
\end{equation}
The $\eta_n^1$~order (\ref{eq:qk}) is, at zeroth order
of~$\epsilon$ and $\xi_n$, 
\begin{equation}
  q_{\boldsymbol{k}}''^{(1)} + \left(k^2 - \frac{2}{\tau^2}\right)
   q_{\boldsymbol{k}}^{(1)} \simeq 
 - 
 \left( \frac{q_{\boldsymbol{k}}^{(0)}}{\tau^2}
+ \frac{q_{\boldsymbol{k}}'^{(0)}}{\tau} 
\right) 
\frac{16}{H}
 \sum_{n \neq 0} \alpha_n^2 \frac{\psi_n
 \dot{\psi}_n}{\phi^2} 
+ k^2 q_{\boldsymbol{k}}^{(0)} \sum_{n \neq 0}  \frac{\dot{\psi}_n^2}{V},
 \label{maruA}
\end{equation}
where in the right hand side we have used (\ref{star3}). 

\subsubsection{Zeroth Order in $\eta_n$}

Let us first focus on the $\eta_n^0$-order
equation~(\ref{qk0eq}). The general solution to this equation is given
by a linear combination of the Hankel function
\begin{equation}
 \sqrt{-\tau} H_{3/2}^{(1)} (-k \tau) = 
		\sqrt{\frac{2 }{\pi k}}	
 \left( \frac{i}{k \tau} - 1\right) e^{-i k \tau}
 \label{Hankel}
\end{equation}
and its complex conjugate. The explicit form of
$q_{\boldsymbol{k}}^{(0)}$ is determined through quantizing the field
fluctuations. 
We stress here that the KK modes are assumed to be excited during
inflation, hence in the early stage of inflation their effects
are absent, i.e. $q_{\boldsymbol{k}} = q_{\boldsymbol{k}}^{(0)}$.

From the action $S_2 = \int d \tau d^3 x \, \mathcal{L}_2$
(\ref{S2q}), one obtains the conjugate momentum of $q$, 
\begin{equation}
 \Pi = \frac{\partial \mathcal{L}_2}{\partial q'} = B_{\phi \phi} \left(q' -
  \frac{a'}{a} q \right).
\end{equation}
We expand $q$ in terms of the mode functions~$q_{\boldsymbol{k}}$ and further
assign annihilation and creation operators 
($a_{\boldsymbol{k}}$ and $a_{\boldsymbol{k}}^{\dagger}$ respectively),
\begin{equation}
 q = \int d^3 k
 \left\{ q_{\boldsymbol{k}} e^{-i \boldsymbol{k \cdot x}}
  a_{\boldsymbol{k}}
+ \left(q_{\boldsymbol{k}} e^{-i \boldsymbol{k \cdot x}} \right)^*
  a_{\boldsymbol{k}}^{\dagger}
\right\},
\end{equation}
and impose the following commutation relations
\begin{equation}
\begin{split}
 [ a_{\boldsymbol{k}},\,  a_{\boldsymbol{q}}^{\dagger} ] &= (2 \pi)^3 \delta^{(3)}
 (\boldsymbol{k} - \boldsymbol{q}) , \\
 [ a_{\boldsymbol{k}},\,  a_{\boldsymbol{q}} ] &=
 [ a_{\boldsymbol{k}}^{\dagger},\,  a_{\boldsymbol{q}}^{\dagger} ] = 0.
 \label{eq:commu}
\end{split}
\end{equation}
as well as
\begin{equation}
\begin{split}
 \left[ q(\tau, \boldsymbol{x}),\,  \Pi (\tau, \boldsymbol{y})
 \right] &= i \delta^{(3)}
  (\boldsymbol{x} - \boldsymbol{y}), \\
 \left[ q(\tau, \boldsymbol{x}),\,  q (\tau, \boldsymbol{y}) \right] &= 
 \left[ \Pi(\tau, \boldsymbol{x}),\,  \Pi (\tau, \boldsymbol{y}) \right]
 = 0.
\label{eq:commu2}
\end{split}
\end{equation}
Requiring $q_{\boldsymbol{k}}$ to realize the positive frequency mode of
the plane wave solution $q_{\boldsymbol{k}} \propto e^{-i k \tau}$ in
the early times when the mode is well inside the horizon, and 
taking into account that $q_{\boldsymbol{k}}$ is initially equivalent to
$q_{\boldsymbol{k}}^{(0)}$, then we can choose the
solution~(\ref{Hankel}) as the mode function, i.e. 
$ q_{\boldsymbol{k}} = q_{\boldsymbol{k}}^{(0)} = \widetilde{C} \sqrt{-\tau}
H_{3/2}^{(1)} (-k \tau)$,  
at times before the KK modes are excited.
We choose the normalization~$\widetilde{C}$ 
(which is a constant at zeroth order in~$\epsilon$),
such that the commutation relations (\ref{eq:commu}) and
(\ref{eq:commu2}) are satisfied.
Hence $\widetilde{C}$ can be taken as
\begin{equation}
 \widetilde{C}  = \frac{\pi^{1/2}}{2 (2 \pi)^3}  
 \left( 1 + \mathcal{O} (\epsilon)  \right) ,
\label{widetildeC}
\end{equation}
where the estimated error $\mathcal{O} (\epsilon)$
arise also from $B_{\phi\phi}$~(\ref{Bphiphi}) and $a' /
a$~(\ref{adash}).\footnote{Strictly speaking, the error $\mathcal{O}
(\epsilon)$ in (\ref{widetildeC}) is 
actually $\epsilon$ further multiplied by factors such as $k$ and
$\tau$. However, here for simplicity we ignore such factors.} 
(The error $\mathcal{O} (\eta_n \xi_n) $ is absent here since we are
discussing times before the KK excitations.)

After the KK modes are excited, the solution~(\ref{Hankel}) of
$q_{\boldsymbol{k}}$ is succeeded to $q_{\boldsymbol{k}}^{(0)}$, and KK
mode contributions $q_{\boldsymbol{k}}^{(1)}, \, 
q_{\boldsymbol{k}}^{(2)}, \, \cdots $ arise. Here, let us note that when
starting with an initial condition that is independent of the direction
of~$\boldsymbol{k}$ (such as~(\ref{Hankel})) and satisfies the commutation
relations~(\ref{eq:commu2}), then one can show that (\ref{eq:commu2}) is
satisfied as long as the mode function~$q_{\boldsymbol{k}}$ follows the
equation of motion~(\ref{4747}).

Therefore, within errors of $\mathcal{O} (\epsilon)$,
we have obtained $q_{\boldsymbol{k}}^{(0)}$ as
\begin{equation}
 q_{\boldsymbol{k}}^{(0)} \simeq \frac{1}{(2 \pi)^3}  \frac{1}{ (2
  k)^{1/2}}
 \left( \frac{i}{k \tau} - 1 \right) e^{-i k \tau}.
 \label{qkzero}
\end{equation}

\subsubsection{First Order in $\eta_n$}

Let us express the source terms in the right hand side of (\ref{maruA})
as functions of the conformal time.
In order to analyze the leading effects from the KK modes, it basically
suffices to obtain the leading order expressions. By solving
(\ref{star2}) and (\ref{star1}) using $\tau \simeq -1 / aH$, one finds
\begin{equation}
 \phi^2 \simeq \phi_*^2 + \frac{4}{3} M_p^2 \ln \frac{\tau}{\tau_*},
 \label{phitau}
\end{equation}
where the subscript~$*$ denotes values at some fixed time~$\tau_*$. 
For the KK mode we use the expression (\ref{anvarphi}), giving
\begin{equation}
 \psi_n \simeq  \widetilde{\psi}_{n*} \left(\frac{a_*}{a}\right)^{3/2}
 \cos \left\{2 \alpha_n \frac{V^{1/2}}{\phi } (t-t_*) + \theta_n
      \right\},
\label{6666}
\end{equation}
where $\widetilde{\psi}_{n*}$, $a_*$, $t_*$, and $\theta_n$ are
constants.\footnote{Here we use the approximate solution~(\ref{6666}) beyond
the leading order equation~(\ref{star3}), since the damping of the
oscillations $\psi_n \propto a^{-3/2}$ need to be taken into account for
understanding the behavior of the KK effects.} 
We obtain the expression for $\dot{\psi}_n$ by differentiating the
ansatz (\ref{6666}), which is at the leading 
order (for $ H (t-t_*) \lesssim 1$),
\begin{equation}
\dot{ \psi}_n \simeq  - 
2 \alpha_n \frac{V^{1/2}}{\phi }
\widetilde{\psi}_{n*} \left(\frac{a_*}{a}\right)^{3/2}
 \sin \left\{2 \alpha_n \frac{V^{1/2}}{\phi } (t-t_*) + \theta_n
      \right\}.
\label{666n}
\end{equation}
Moreover, given that $|\frac{\dot{H}}{H^2} \ln (\frac{a_*}{a})| \ll 1$,
one can check that 
\begin{equation}
 dt \simeq d \left\{ -\frac{1}{H} \ln (-a_* \tau H) \right\}.
\label{dtd1H}
\end{equation}
Integrating (\ref{dtd1H})\footnote{Here we suppose that the integrated
errors stay small, as in Footnote~\ref{foot:appint}.} and further using
(\ref{star2}), we obtain
\begin{equation}
 t  - t_*  \simeq - \frac{3^{1/2} M_p}{\mu^{5/3} \phi^{1/3}} \ln
 \left\{
\frac{\tau }{\tau_*} \left( \frac{\phi
			       }{\phi_*}\right)^{1/3}\right\}.
\end{equation}

Hence by combining the results, one can write the source terms of
(\ref{maruA}) as functions of the conformal time,
\begin{multline}
  \frac{16}{H} \sum_{n \neq 0} \alpha_n^2 \frac{\psi_n
 \dot{\psi}_n}{\phi^2} 
\\
\simeq 
 - 16 \cdot 3^{1/2} \left(
					   \frac{\tau}{\tau_*}\right)^3
 \sum_{n \neq 0} \alpha_n^3 
\frac{ M_p \widetilde{\psi}_{n*}^2 }{\phi^2 \phi_*}
 \sin \left\{
-4 \cdot 3^{1/2} \alpha_n \frac{M_p}{ \phi} 
\ln \left( \frac{\tau}{\tau_*} \left( \frac{\phi}{\phi_*}\right)^{1/3}
    \right) + 2 \theta_n
\right\},
\label{8psipsi}
\end{multline}
\begin{equation}
 \sum_{n \neq 0} \frac{\dot{\psi}_n^2 }{V} \simeq
 2 \left(\frac{\tau}{\tau_*} \right)^3 \sum_{n \neq 0} \alpha_n^2
 \frac{\widetilde{\psi}_{n*}^2}{\phi \phi_*} 
 \left[
1 -  \cos \left\{
-4 \cdot 3^{1/2} \alpha_n \frac{M_p}{ \phi} 
\ln \left( \frac{\tau}{\tau_*} \left( \frac{\phi}{\phi_*}\right)^{1/3}
    \right) + 2 \theta_n
\right\}
\right].
\label{dotpsiV}
\end{equation}
Here, in order to avoid clutter we have left $\phi$, which is given in
terms of~$\tau$ in (\ref{phitau}).

\vspace{\baselineskip}

Introducing the ratio
\begin{equation}
 s_{\boldsymbol{k}}^{(1)} \equiv
  \frac{q_{\boldsymbol{k}}^{(1)}}{q_{\boldsymbol{k}}^{(0)}}, 
\end{equation}
and substituting (\ref{qkzero}) for~$q_{\boldsymbol{k}}^{(0)}$, the
equation (\ref{maruA}) can be recast in the form 
\begin{equation}
 \left(\frac{i}{k \tau } - 1\right) s_{\boldsymbol{k}}''^{(1)} + 
 2 \left(-\frac{i}{k \tau^2} + \frac{1}{ \tau} + ik  \right)
 s_{\boldsymbol{k}}'^{(1)} \simeq 
 - \frac{ik}{\tau } 
 \frac{16}{H} \sum_{n \neq 0} \alpha_n^2 \frac{\psi_n
 \dot{\psi}_n}{\phi^2}
- \left(k^2 - i \frac{k}{\tau}\right) \sum_{n \neq
0}\frac{\dot{\psi}_n^2}{V}.
\label{sk1}
\end{equation}
Supposing, for instance, that the KK modes are suddenly excited
at time~$\tau_{\mathrm{exc}}$, then one can obtain $s_{\boldsymbol{k}}^{(1)}$ by
solving (\ref{sk1}) (with (\ref{phitau}), (\ref{8psipsi}), and
(\ref{dotpsiV})), with the initial conditions
\mbox{$s_{\boldsymbol{k}}^{(1)}(\tau_{\mathrm{exc}}) =
s_{\boldsymbol{k}}'^{(1)}(\tau_{\mathrm{exc}}) = 0$}. 

We choose the vacuum as
\begin{equation}
 a_{\boldsymbol{k}} |0 \rangle = 0, \quad ^{\forall} \, \boldsymbol{k},
 \label{eq:BSvac}
\end{equation}
hence the two point function of the field fluctuations is 
\begin{equation}
 \langle 0 | \delta \phi_{\boldsymbol{k}} \delta \phi_{\boldsymbol{k'}}
 | 0 \rangle = \frac{(2 \pi)^9}{a^2} \delta^{(3)} (\boldsymbol{k} +
 \boldsymbol{k'})
 \left| q_{\boldsymbol{k}}  \right|^2.
\label{4.35}
\end{equation}
Considering the KK mode effects to first order in $\eta_n$, then
\begin{equation}
 |q_{\boldsymbol{k}}|^2 \simeq 
 |q_{\boldsymbol{k}}^{(0)}|^2
\left\{ 1 + 2 \mathrm{Re} (s_{\boldsymbol{k}}^{(1)})  \right\}
 \simeq \frac{1}{(2 \pi)^6} \frac{1}{2k} \left(\frac{1}{k^2 \tau^2} + 1
				    \right)
\left\{ 1 + 2 \mathrm{Re} (s_{\boldsymbol{k}}^{(1)})  \right\}.
\label{4.36}
\end{equation}

\subsection{Curvature Perturbations}
\label{subsec:CurvPert}

Finally we use the $\delta
\mathcal{N}$-formalism~\cite{Starobinsky:1986fxa,Sasaki:1995aw,Wands:2000dp,Lyth:2004gb}
to compute the curvature perturbation which is expressed as 
\begin{equation}
 \zeta_{\boldsymbol{k}} \simeq \frac{\partial \mathcal{N}}{\partial
  \phi} \delta \phi_{\boldsymbol{k}}
\label{zetaNphi}
\end{equation}
at the leading order of the inflaton field fluctuation. 
Here, $\mathcal{N}$ is the e-folding number between an initial flat
hypersurface and a final uniform density hypersurface.
Let us repeat that, even though there are multiple degrees of freedom
involved in wrapped brane inflation, the KK modes are massive hence 
the curvature perturbations are sourced by the inflaton field fluctuations.
We take the final uniform density surface at some later time when
the wave mode is well outside the horizon so the separate
universe assumption is a good approximation, and also when the KK modes
have sufficiently damped away such that the inflationary universe can be
considered as nearly single-component.
Computing (\ref{zetaNphi}) at such time, one obtains
\begin{equation}
 \frac{\partial \mathcal{N}}{\partial \phi } \delta   \phi_{\boldsymbol{k}} 
 \simeq   - \frac{H}{\dot{\phi}} \delta \phi_{\boldsymbol{k}},
\label{4.38}
\end{equation}
up to corrections from the damped KK modes. 
Here $H/\dot{\phi}$ can be considered as a function of~$\phi$ since
now inflation is nearly single-field.

The (damped) KK corrections, i.e. corrections to the single-component
treatment, can be estimated from the Friedmann equation~(\ref{Fvarphi})
which can be written as 
\begin{equation}
 3 M_p^2 H^2 = V 
\left\{ 1 + \mathcal{O}(\epsilon) + \mathcal{O} (\eta_n \xi_n) \right\},
\end{equation}
and the equation of motion of~$\phi$,
\begin{equation}
 3 H \dot{\phi}  = -V'
\left\{ 1 + \mathcal{O}(\epsilon) + \mathcal{O} (\eta_n ) \right\} .
\label{4.40}
\end{equation}
From these equations, the KK corrections to the expression
(\ref{4.38}) can be estimated to be of size~$\sim \eta_n$ at the time
when evaluating its right hand side.
(Here, note that $\eta_n$ decreases in time as the KK oscillations are damped.)
On the other hand, the KK mode effects are also imprinted on the
inflaton field fluctuations~$\delta \phi_{\boldsymbol{k}}$. We will see
in the following sections that these effects on~$\delta
\phi_{\boldsymbol{k}}$ are at least of order~$\sim 
(\eta_{n} \xi_{n})|_{\mathrm{exc}}$ which are values at the KK
excitation. Therefore, by 
evaluating (\ref{4.38}) at times when the wave mode of interest has
exited the horizon, i.e.
\begin{equation}
 -k \tau \ll 1,
 \label{6star-1}
\end{equation}
and also when the KK modes are damped such that the parameter~$\eta_n$
is small enough to satisfy
\begin{equation}
 \eta_n \ll (\eta_{n} \xi_{n})|_{\mathrm{exc}} ,
\label{KKdamp}
\end{equation}
then one can safely ignore the KK modes at around the final uniform
density hypersurface but still capture the main KK effects
through the inflaton field fluctuations.

Combining (\ref{4.35}) and (\ref{zetaNphi}) - (\ref{4.40}), then at zeroth order
of~$\epsilon$ and $\eta_n$ (we stress that this $\eta_n$ is the value 
after the KK modes have damped away), one obtains the two point function 
of the curvature perturbations up to linear order in the
inflaton field fluctuation as
\begin{equation}
  \langle \zeta_{\boldsymbol{k}} \zeta_{\boldsymbol{k'}} \rangle 
\simeq
 (2 \pi)^9  \delta^{(3)} (\boldsymbol{k} + \boldsymbol{k'})
 \left( \frac{V}{M_p^2 V' a} \right)^2
 |q_{\boldsymbol{k}}|^2 ,
\label{4.43}
\end{equation}
where the right hand side should be evaluated at times when
(\ref{6star-1}) and (\ref{KKdamp}) are satisfied. 
Further using (\ref{esttau})\footnote{As we have stated in
Footnote~\ref{foot:appint}, the estimated errors of~$\tau$ in
(\ref{esttau}) are actually those parameters integrated over a finite
time period. However, since we fix $\tau \to 0^- $ as $a \to \infty$, 
the integration is from~$\tau$ to the future and thus 
errors in~$\tau$ at the final hypersurface are not sourced by 
KK modes at times when they were excited.} and the order~$\eta_{n\,
\mathrm{exc}}$ expression (\ref{4.36}), the two
point function can also be expressed as
\begin{equation}
 \langle \zeta_{\boldsymbol{k}} \zeta_{\boldsymbol{k'}} \rangle
 \simeq 
 (2 \pi)^3 \delta^{(3)} (\boldsymbol{k} + \boldsymbol{k'}) \frac{2
 \pi^2}{k^3} \times
\frac{V^3 }{12 \pi^2  M_p^6 V'^2 }
 \left(1 + k^2 \tau^2\right)
 \left\{1 + 2 \mathrm{Re} (s_{\boldsymbol{k}}^{(1)}) \right\}.
\label{zetafinal}
\end{equation}
Here we should remark that the expression~(\ref{zetafinal}) does not capture 
the large (super-Planckian) variation of $\phi$ during inflation, since 
in the previous subsection we have focused on the KK mode effects and
dropped $\mathcal{O} (\epsilon)$ 
corrections upon calculating $q_{\boldsymbol{k}}$.

Defining the power spectrum~$\mathcal{P}_\zeta (k)$ as 
\begin{equation}
 \langle \zeta_{\boldsymbol{k}} \zeta_{\boldsymbol{k'}} \rangle = (2
  \pi)^3 \delta^{(3)} (\boldsymbol{k} + \boldsymbol{k'}) \frac{2
 \pi^2}{k^3} \mathcal{P}_\zeta (k),
\end{equation}
one sees that effects from the excited KK modes are
represented by $\mathrm{Re} (s_{\boldsymbol{k}}^{(1)})$ evaluated at a 
fixed time~$\tau$ when the wave number range $k \sim k + \Delta k$ of
interest satisfies (\ref{6star-1}), i.e.,
\begin{equation}
 \frac{\mathcal{P}_\zeta (k)}{\mathcal{P}_{\zeta 0} (k)}  \simeq 1 +
2 \mathrm{Re} (s_{\boldsymbol{k}}^{(1)}) .
\label{PzetaPzeta0}
\end{equation}
Here, $\mathcal{P}_{\zeta 0} (k)$ denotes the curvature perturbation
power spectrum in the absence of the KK modes.

\section{Weak Resonance from Small KK Excitations}
\label{sec:smallKK}

We now analyze KK mode effects on the curvature perturbation spectrum 
by solving the evolution equation derived above.
In this section we consider cases where the condition
(\ref{smallpar}) holds and the inflationary dynamics are well
approximated by (\ref{star2}) - (\ref{star4}). 
Such small KK excitations give rise to weak parametric resonance,
sourcing oscillations on the perturbation spectrum.
In this small KK regime we can carry out analyses
analytically, enabling us to understand the particular oscillatory forms 
the KK modes source on the power spectrum,
cf. Figure~\ref{fig:schematic}. We will go beyond the small KK
approximations in Section~\ref{sec:largeKK} where we find strong resonance
sourcing sharp spikes, but the basic properties of the KK signals will
still be described by the analyses given in this section.

\subsection{Solving the Evolution Equation}

Recall from the discussions in Section~\ref{subsec:CurvPert} that
for small KK excitations~(\ref{smallpar}), their leading order effects on the
curvature perturbation spectrum is obtained by computing the asymptotic
value of $2 \mathrm{Re} (s_{\boldsymbol{k}}^{(1)})$.
Focusing on wave modes~$\boldsymbol{k}$ that are well inside the horizon
when the KK modes are excited, then 
the evolution equation for $s_{\boldsymbol{k}}^{(1)}$~(\ref{sk1})
prior to horizon exit (i.e. during $-k \tau \gg 1$) is approximated by
\begin{multline}
 -s_{\boldsymbol{k}}''^{(1)}(\tau) + 2 i k s_{\boldsymbol{k}}'^{(1)}(\tau) 
\\
\simeq 
  16 \cdot 3^{1/2} \frac{ik}{\tau} \left(
					   \frac{\tau}{\tau_*}\right)^3
 \sum_{n \neq 0} \alpha_n^3 
\frac{ M_p \widetilde{\psi}_{n*}^2 }{\phi^2 \phi_*}
 \sin \left\{
-4 \cdot 3^{1/2} \alpha_n \frac{M_p}{ \phi} 
\ln \left( \frac{\tau}{\tau_*} \left( \frac{\phi}{\phi_*}\right)^{1/3}
    \right) + 2 \theta_n
\right\}
\\
+ 
 2 k^2 \left(\frac{\tau}{\tau_*} \right)^3 \sum_{n \neq 0} \alpha_n^2
 \frac{\widetilde{\psi}_{n*}^2}{\phi \phi_*} 
 \left[
-1 
+  \sin \left\{
-4 \cdot 3^{1/2} \alpha_n \frac{M_p}{ \phi} 
\ln \left( \frac{\tau}{\tau_*} \left( \frac{\phi}{\phi_*}\right)^{1/3}
    \right) + 2 \theta_n + \frac{\pi}{2}
\right\}
\right],
\label{sk1inho}
\end{multline}
where $\phi$ is given in (\ref{phitau}). 
(Here one sees that the KK terms in the third line of (\ref{eq:qk}) is
irrelevant when inside the horizon.)
The homogeneous solution of this equation is
\begin{equation}
 s_{\boldsymbol{k},  \mathrm{h}}^{(1)}(\tau) = C_0 + C_1 e^{2 i k
  \tau} ,
\label{homo}
\end{equation}
denoting oscillations with frequency~$k/\pi$ and an offset. Since
(\ref{sk1inho}) is a linear equation, let us look into 
$ s_{\boldsymbol{k}}^{(1)}$ induced by each of the source terms in its
right hand side, and then add up the results at the end.
We suppose that the KK modes are excited at time~$\tau_{\mathrm{exc}}$,
and set the initial conditions as 
\begin{equation}
 s_{\boldsymbol{k}}^{(1)} (\tau_{\mathrm{exc}}) =
s_{\boldsymbol{k}}'^{(1)}(\tau_{\mathrm{exc}}) = 0 .
\label{inicexc}
\end{equation}
Hereafter we use the subscript ``exc'' to denote values at the KK excitation.

\subsubsection{Non-Oscillatory Source}

Upon dealing with the non-oscillatory source term (i.e. the one without
the sine in the right hand side of (\ref{sk1inho})), we ignore the
logarithmic $\tau$-dependence of $\phi$ (cf.~(\ref{phitau})) and solve the
equation
\begin{equation}
 -s_{\boldsymbol{k}}''^{(1)}(\tau) + 2 i k s_{\boldsymbol{k}}'^{(1)}(\tau)  =
 - 2 k^2  \left( \frac{\tau }{\tau_{\mathrm{exc}}} \right)^3 
 \sum_{n \neq 0 } \alpha_n^2
 \frac{\widetilde{\psi}_{n \, \mathrm{exc}}^2}{\phi_{\mathrm{exc}}^2 },
\label{eq5.4}
\end{equation}
where we have taken the arbitrary time~$\tau_*$ in
(\ref{sk1inho}) as at the KK excitation.
Solving (\ref{eq5.4}) with the initial conditions~(\ref{inicexc}), 
then the real part of $s_{\boldsymbol{k}}^{(1)}$ which is relevant for
the curvature perturbations is obtained as
\begin{equation}
 \mathrm{Re} (  s_{\boldsymbol{k}}^{(1)} ) \simeq
 \frac{1}{2} \sum_{n \neq 0} \alpha_n^2 \frac{\widetilde{\psi}_{n\,
 \mathrm{exc}}^2}{\phi_{\mathrm{exc}}^2} 
\left[ \left(\frac{\tau}{\tau_{\mathrm{exc}}}\right)^3
 - \cos \{ 2 k (\tau - \tau_{\mathrm{exc}}) \} \right].
\label{solfornonosc}
\end{equation}
This solution induced by the non-oscillatory source consists of an
oscillatory part with frequency $k / \pi$, and an decaying offset.
Since we are now focusing on wave modes inside the horizon, in the above
solution we have dropped oscillatory terms and (decaying) offset that
are suppressed by $\mathcal{O} (|k\tau |^{-1})$ compared to the shown terms.

\subsubsection{Oscillatory Source}

In order to analyze $s_{\boldsymbol{k}}^{(1)}$ induced by the
oscillatory sources, we focus on time intervals of order the KK-mode
oscillations (which is much shorter than the Hubble time due to $\xi_n
\ll 1$). 
Then we can ignore the time dependence of the source terms except for
that showing up explicitly in the sines, thus obtain
\begin{equation}\label{sk1eqsimp}
\begin{split}
 -s_{\boldsymbol{k}}''^{(1)}(\tau) + 2 i k s_{\boldsymbol{k}}'^{(1)}(\tau) 
 & \simeq
  16 \cdot 3^{1/2}  \frac{i k}{\tau_*} \sum_{n \neq 0} \alpha_n^3
 \frac{M_p \widetilde{\psi}_{n*}^2}{\phi_*^3} \sin\left[
-4 \cdot 3^{1/2} \alpha_n \frac{M_p}{\phi_*} 
  \ln  \left(\frac{\tau}{\tau_*}\right) + 2 \theta_n 
\right] \\
& \qquad  + 2 k^2 \sum_{n \neq 0} \alpha_n^2
 \frac{\widetilde{\psi}_{n*}^2}{\phi_*^2} 
\sin\left[
-4 \cdot 3^{1/2} \alpha_n \frac{M_p}{\phi_*} 
  \ln  \left(\frac{\tau}{\tau_*}\right) + 2 \theta_n + \frac{\pi}{2} 
\right]
\\
 &  \simeq
16 \cdot 3^{1/2} \frac{ik}{\tau_*} \sum_{n \neq 0} \tilde{\eta}_{n*} 
 \sin \left( \omega_{n*} \tau + \Theta_{n*}\right) 
\\
& \qquad - 8 \cdot 3^{1/2} \frac{k^2 }{\tau_*} \sum_{n \neq 0}  
 \frac{\tilde{\eta}_{n*}}{\omega_{n*}} 
 \sin \left( \omega_{n*} \tau + \Theta_{n*} + \frac{\pi}{2} \right) .
\end{split}
\end{equation}
Here, the parameters with the subscript~$*$ represent their typical values during
the time scale of interest at around some arbitrary time $\tau =
\tau_*$, and especially 
$\widetilde{\psi}_{n*}$ denotes the oscillation amplitude of the KK-mode~$\psi_n$. 
In the second line we have expanded the log term and introduced the parameters
\begin{equation}
 \tilde{\eta}_n \equiv \alpha_n^3 \frac{M_p \widetilde{\psi}_n^2}{\phi^3}, \quad
 \omega_n \equiv - 4 \cdot 3^{1/2} \alpha_n \frac{  M_p}{ \tau  \phi },
 \quad
 \Theta_n \equiv 4 \cdot 3^{1/2} \alpha_n \frac{M_p}{\phi } + 2 \theta_n.
\end{equation}
These parameters can be considered as constants during the time intervals of
the KK-mode oscillations, however at larger time scales (i.e. Hubble
time or more) they vary approximately as
\begin{equation}
 \tilde{\eta}_n \propto \sim \tau^3 , \qquad \omega_n \propto \sim
  \tau^{-1}. 
\end{equation}
From the homogeneous solution~(\ref{homo}),
one can expect parametric resonance to happen when the
oscillation frequency~$\omega_n/2 \pi$ becomes equal to the characteristic
frequency of the mode~$k/\pi$, i.e. $2 k \approx \omega_n$.
This is at a time
\begin{equation}
 -\tau_{\mathrm{res}} (k) = \frac{2\cdot 3^{1/2}}{k \xi_n} \gg
  \frac{1}{k} ,
\end{equation}
when the mode is well inside the Hubble horizon.

\vspace{\baselineskip}

We collectively study the oscillatory terms in (\ref{sk1eqsimp}) by
analyzing the following equation
\begin{equation}
 -s_{\boldsymbol{k}}''^{(1)}(\tau) + 2 i k
  s_{\boldsymbol{k}}'^{(1)}(\tau) 
 = i A \sin (\omega \tau + \Theta),
\label{tanjun}
\end{equation}
where $A$ is a complex constant, while $\omega$ and $\Theta$ are real
constants. The particular solution of (\ref{tanjun}) is
\begin{equation}
 s_{\boldsymbol{k},  p}^{(1)}(\tau) = \frac{A}{\omega^2 - 4 k^2} 
 \left\{ \frac{2k}{\omega } \cos (\omega \tau + \Theta) + i \sin (\omega
  \tau + \Theta   ) \right\}
\label{part}
\end{equation}
for $\omega^2 \neq 4 k^2$, and 
\begin{equation}
  \tilde{s}_{\boldsymbol{k}, p}^{(1)}(\tau) = \frac{A}{ 8 k^2} 
 \left\{2 i k \tau e^{i (2 k \tau + \Theta)} - \cos (2 k \tau + \Theta)
 \right\} 
\label{partamp+}
\end{equation}
for $\omega = 2 k$. 
The particular solutions (\ref{part}) and (\ref{partamp+}) connect in
the $\omega \to 2 k$ limit up to the homogeneous solution (\ref{homo}):
\begin{equation}
 \tilde{s}_{\boldsymbol{k}, p}^{(1)}(\tau)  =
 \lim_{\omega \to 2 k } \left\{
  s_{\boldsymbol{k}, p}^{(1)}(\tau) +
 \frac{A (\omega - 6k)}{16 k^2 (\omega - 2 k)} e^{i (2 k \tau + \Theta
 )}   \right\}.
\label{partmix}
\end{equation}

While the homogeneous solution~(\ref{homo}) sets the offset as well as
oscillations with frequency $k/\pi$, the particular
solution~(\ref{part}) gives oscillations with
frequency $\omega / 2 \pi$.
Here, the parameters such as $A$ and $\omega$ actually vary in
cosmological time scales in the original equation for~$s_{\boldsymbol{k}}^{(1)}$.
However, since their varying time scales are much longer than the oscillation
period~$\sim \omega^{-1}$ (and also than~$k^{-1}$), one can
consider~(\ref{part}) to be the solution at each cosmological
time scale. In other words, the particular solution denotes
oscillations with time dependent amplitude and frequency, in contrast to
the homogeneous solution giving oscillations with constant amplitude and
frequency.

As for the resonant solution~(\ref{partamp+}), the first term in the
right hand side represents parametric resonance. 
The degree of parametric amplification is determined by how long the
wave mode stays in the resonance band~$\omega \approx 2k$,
see discussions in Appendix~\ref{subsec:C.2}. 

\vspace{\baselineskip}

Detailed solutions for the equation~(\ref{tanjun}) are given in
Appendix~\ref{app:c} (e.g. (\ref{C.10}), (\ref{C.11}), and
(\ref{C.12}) for time dependent parameters),
but here let us give a rough description on 
how the real part of~$s_{\boldsymbol{k}}^{(1)}$ reacts to the
oscillating sources of~(\ref{sk1eqsimp}). 

Upon the KK excitation, if the KK-modes' oscillation frequencies 
are larger than the characteristic frequency of the wave mode, i.e. $2k
\ll \omega_{n\, \mathrm{exc}}$, then $\mathrm{Re} (s_{\boldsymbol{k}}^{(1)})$ is
insensitive to the KK oscillations and is well described by the homogeneous
solution~(\ref{homo}). The oscillation amplitude and offset are set by
the initial ``kick'' at the KK excitation, and are of
order~$\eta_{n\, \mathrm{exc}} \xi_{n\, \mathrm{exc}}$. 
Such wave modes do not experience parametric resonance and eventually
exit the horizon.

On the other hand, wave modes satisfying $2 k \gg \omega_{n\,
\mathrm{exc}}$ go through the resonance band before
exiting the horizon. When the KK modes are excited, oscillations
of~$\mathrm{Re} (s_{\boldsymbol{k}}^{(1)})$ with 
frequencies $k/\pi$ and $\omega_n / 2\pi $ both can be generated with
amplitude of order~$\eta_{n\, \mathrm{exc}} \xi_{n\,
\mathrm{exc}}$. However the former keeps a constant amplitude
(corresponding to the homogeneous solution~(\ref{homo})), while the
latter decays as $\tau^{3}$ as the KK mode oscillations
are damped (corresponding to the particular solution~(\ref{part})). 
The offset is also of order $\eta_{n\, \mathrm{exc}} \xi_{n\, \mathrm{exc}}$.

The wave mode undergoes parametric resonance with the KK 
oscillations while $2 k \approx \omega_n$, during which
the oscillation amplitude of~$\mathrm{Re} (s_{\boldsymbol{k}}^{(1)})$ is amplified. 
At this time $\mathrm{Re} (s_{\boldsymbol{k}}^{(1)})$ oscillates only with
frequency~$k /\pi$. 
The wave mode stays in the resonance band for 
$\Delta \tau \sim -  \xi_{n\, \mathrm{exc}}^{1/2} \tau_{\mathrm{res}} $
(cf.~(\ref{res_time})), and by the time it leaves the resonance band,
the oscillation amplitude of $\mathrm{Re} (s_{\boldsymbol{k}}^{(1)})$ is
parametrically amplified to~$\sim \eta_{n\, \mathrm{exc}}
\xi_{n\, \mathrm{exc}}^{1/2} (k_n / k)^3$. 
The amplification is peaked at a wave number~$\sim k_n$ 
experiencing resonance right when the KK modes are excited.
For larger~$k$, the amplification is less significant since such wave
modes enter the resonance band
at later times when the KK oscillations have damped away. We also note
that the $k/\pi$ oscillation of $\mathrm{Re} (s_{\boldsymbol{k}}^{(1)})$ with 
constant amplitude~$\eta_{n\, \mathrm{exc}} \xi_{n\,
\mathrm{exc}}$ (denoted by the homogeneous solution) survives
through the resonance band. This can be the dominant oscillations for 
large~$k$ modes that are not so amplified at the resonance band.

After leaving the resonance band (i.e. when $2 k \ll
\omega_n$), then $\mathrm{Re} (s_{\boldsymbol{k}}^{(1)})$ becomes insensitive to 
the source terms and maintains its oscillation with frequency~$k/\pi$ and 
amplitude at the resonance band, until the wave mode exits the horizon.

\subsubsection{After Horizon Exit}
\label{subsubsec:AHE}

The evolution of $s_{\boldsymbol{k}}^{(1)}$ after horizon exit can be
analyzed in a similar fashion, by studying the equation~(\ref{sk1}) for 
$-k \tau \ll 1$. (In such case, the second source term $\dot{\psi}_n^2 /
V $ in (\ref{sk1}) is much smaller than the first term and thus can be ignored.) 
Analyzing the super-horizon version of (\ref{sk1eqsimp}) by neglecting
the time-dependence of the parameters such as~$\omega_n$, then one can
easily check that the solution outside the horizon 
consists of a constant term and decaying terms.
Hence the oscillation of $\mathrm{Re} (s_{\boldsymbol{k}}^{(1)})$
becomes frozen as the wave mode crosses the horizon, i.e. at $-k \tau \approx 1$. 
One can also check that the induced constant term for wave modes that
are super-horizon at the KK excitation is tiny compared to 
the typical amplitude of $\mathrm{Re} (s_{\boldsymbol{k}}^{(1)})$
for wave modes inside the horizon at~$\tau_{\mathrm{exc}}$.

\subsection{Approximate Solutions}

Now let us present approximate solutions to the evolution equation
(\ref{sk1}) with initial conditions (\ref{inicexc}), 
based on the discussions above and in Appendix~\ref{app:c}. 
The results will be written in terms of the following parameters:
\begin{equation}
 \tilde{\eta}_{n\, \mathrm{exc}} = 
\left. \alpha_n^3 \frac{M_p   \tilde{\psi}_n^2}{\phi^3} \right|_{\mathrm{exc}},
\quad
\xi_{n\, \mathrm{exc}} = \left. \frac{\phi}{\alpha_n M_p}
			 \right|_{\mathrm{exc}}, 
\quad
 k_n \equiv -\frac{2 \cdot 3^{1/2}}{\tau_{\mathrm{exc}} \xi_{n\,
 \mathrm{exc}}},
\label{etaxiexc}
\end{equation}
where $\tilde{\psi}_n$ is the oscillation amplitude of the KK mode~$\psi_n$, 
and $ k_n$ denotes the wave number in the resonance band when the KK modes
are excited, i.e. $2 k_n = \omega_{n\, \mathrm{exc}}$.
Furthermore, since $\omega_n \propto \tau^{-1}$,
the time when the wave number~$k$ enters the resonance band can be
estimated as 
\begin{equation}
 \tau_{ \mathrm{res}}(k) = 
 \frac{\left. \left(\tau \omega_n\right) \right|_{\mathrm{exc}}}{2k} 
 =   - \frac{2 \cdot 3^{1/2}}{k \xi_{n\, \mathrm{exc}}}.
\label{tau_res}
\end{equation}
The time range a certain wave mode stays in the resonance
band is derived in~(\ref{lambda}), which can now be rewritten as
\begin{equation}
 \tau_f (k) -  \tau_{\mathrm{res}} (k)  \simeq
  -  \xi_{n\, \mathrm{exc}}^{1/2} \tau_{\mathrm{res}} (k).
\label{res_time}
\end{equation}

\vspace{\baselineskip}

The approximate solutions for $\mathrm{Re} (s_{\boldsymbol{k}}^{(1)})$
are obtained by adding up the contributions from the oscillatory sources  
(i.e. (\ref{C.10}), (\ref{C.11}), and (\ref{C.12}) for (\ref{sk1eqsimp}))
and non-oscillatory sources (i.e. (\ref{solfornonosc})).
Here we show fluctuations induced by a single KK mode~$n$,
but the full expression is simply a sum of the
following expressions over all excited KK modes.

For wave modes that are sub-horizon at KK excitation but do not cross the
resonance band, i.e. $-1/\tau_{\mathrm{exc}} < k \leq k_n$, one finds
\begin{multline}
 \mathrm{Re} (s_{\boldsymbol{k}}^{(1)}(\tau)) \simeq
 \tilde{\eta}_{n\,  \mathrm{exc}}\,   \xi_{n\, \mathrm{exc}} 
\left[
\left(2 \cos 2 \theta_n - \frac{1}{2} \right) \cos \left\{ 2 k (\tau -
 \tau_{\mathrm{exc}})  \right\} 
 - 2 \cos 2 \theta_n + \frac{1}{2}
 \left(\frac{\tau}{\tau_{\mathrm{exc}}}\right)^3
\right]
\\
 \qquad  \mathrm{for}  \, \, \, \, 
\tau_{\mathrm{exc}} \leq \tau < -\frac{1}{k} .
\label{eq102}
\end{multline}

For modes that do cross the resonance band, i.e. $k_n \leq k$, the
solution before crossing the resonance band is:
\begin{multline}
 \mathrm{Re} (s_{\boldsymbol{k}}^{(1)}(\tau)) \simeq
  \frac{3}{2} \tilde{\eta}_{n\,  \mathrm{exc}} \, \xi_{n\, \mathrm{exc}} 
 \left(\frac{\tau}{\tau_{\mathrm{exc}}} \right)^3
\cos \left\{
-\frac{4 \cdot 3^{1/2}}{\xi_{n\, \mathrm{exc}}} \ln
 \left(\frac{\tau}{\tau_{\mathrm{exc}}}  \right) + 2 \theta_n
\right\}
 \\ \qquad 
 +  \tilde{\eta}_{n\,  \mathrm{exc}}\,   \xi_{n\, \mathrm{exc}} 
\left[
\frac{1}{2}\left(\cos 2 \theta_n - 1  \right) \cos \left\{ 2 k (\tau -
 \tau_{\mathrm{exc}})  \right\} 
 - 2 \cos 2 \theta_n + \frac{1}{2} \left(\frac{\tau}{\tau_{\mathrm{exc}}}\right)^3
\right]
\\
  \mathrm{for} \, \, \, \, 
 \tau_{\mathrm{exc}} \leq \tau \leq \tau_{\mathrm{res}}(k),
\label{eq103}
\end{multline}
and after crossing the resonance band:
\begin{multline}
 \mathrm{Re} (s_{\boldsymbol{k}}^{(1)}(\tau)) \simeq
 3^{3/2} \tilde{\eta}_{n\,  \mathrm{exc}} \, \xi_{n\, \mathrm{exc}}^{1/2}
 \left(\frac{k_n}{k}\right)^3
\cos \left\{
2 k \tau  + \frac{4 \cdot 3^{1/2}}{\xi_{n\, \mathrm{exc}}}
\left( 1 + \ln \left(\frac{k}{k_n} \right) \right)
+ 2 \theta_n
\right\}
 \\ \qquad 
 +  \tilde{\eta}_{n\,  \mathrm{exc}}\,   \xi_{n\, \mathrm{exc}} 
\left[
\frac{1}{2}\left(\cos 2 \theta_n - 1  \right) \cos \left\{ 2 k (\tau -
 \tau_{\mathrm{exc}})  \right\} 
 - 2 \cos 2 \theta_n + \frac{1}{2} \left(\frac{\tau}{\tau_{\mathrm{exc}}}\right)^3
\right]
\\
 \mathrm{for}  \, \, \, \, 
\tau_{\mathrm{res}}(k)  \leq \tau  <  - \frac{1}{k},
\label{eq104}
\end{multline}
where for simplicity we have supposed that the parametric amplification
happens suddenly at $\tau = \tau_{\mathrm{res}}(k)$.
Upon obtaining the first line of (\ref{eq104}), we have used
$\tilde{\eta}_{\mathrm{res}} \simeq \tilde{\eta}_{\mathrm{exc}} (k_n / k)^3 $,
which follows from $\tilde{\eta}_n \propto \sim \tau^3$,
(\ref{etaxiexc}), and (\ref{tau_res}).  

In the above results we have dropped sub-leading
contributions to oscillations with frequency~$k /\pi $ that would follow
from simply adding the results (\ref{C.10}), (\ref{C.11}), and
(\ref{C.12}), induced by the two oscillatory source terms
in~(\ref{sk1eqsimp}).\footnote{In particular, the dropped terms are  
$ - 4^{-1} 3^{-1/2} k \tau_{\mathrm{exc}} \tilde{\eta}_{n\, \mathrm{exc}} \xi_{n\,
\mathrm{exc}}^2 \sin 2 \theta_n \sin \{2 k (\tau -
\tau_{\mathrm{exc}})\}$ in (\ref{eq102}), and 
$- 4 \cdot 3^{1/2} k^{-1} \tau_{\mathrm{exc}}^{-1}   \tilde{\eta}_{n\,
\mathrm{exc}}\sin2 \theta_n \sin \{2 k (\tau - \tau_{\mathrm{exc}})\}  $
in (\ref{eq103}) and (\ref{eq104}). They can become comparable to the
leading $k/\pi$ oscillatory terms for $k \approx k_n $, however,
for such wave modes the parametrically amplified term (i.e. the first
term in the right hand side of (\ref{eq104})) is basically dominant anyway.} 
We should also remark that since we have collected leading contributions
from each source term, the above results can become inaccurate when the
main terms in the expressions exactly cancel each other (for e.g., when $2 \cos
2 \theta_n = 1/2$ in (\ref{eq102})). 

When the wave mode exits the horizon at around $\tau = -1/k$, then
$ \mathrm{Re} (s_{\boldsymbol{k}}^{(1)})$ freezes out for both
(\ref{eq102}) and (\ref{eq104}).

\vspace{\baselineskip}

\begin{figure}[htbp]
 \begin{minipage}{.50\linewidth}
  \begin{center}
 \includegraphics[width=\linewidth]{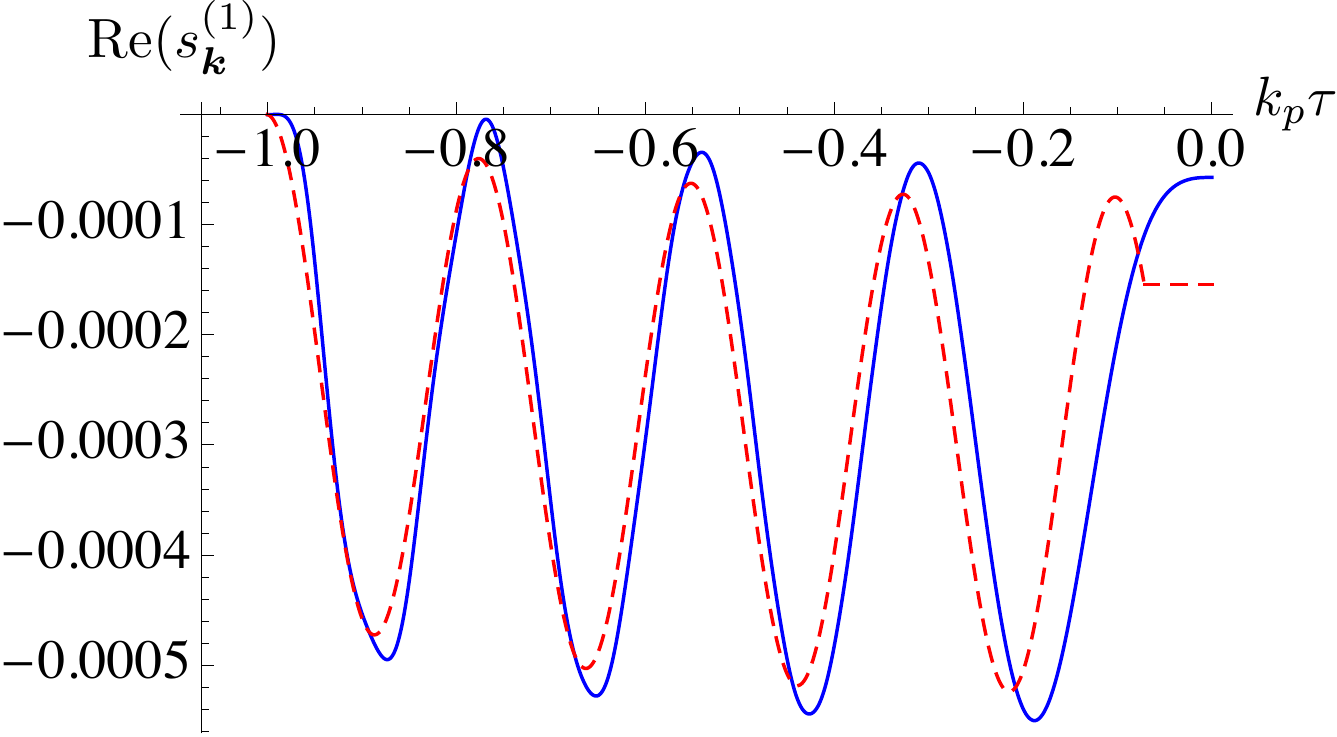}
  \end{center}
 \end{minipage} 
 \begin{minipage}{.50\linewidth}
  \begin{center}
 \includegraphics[width=\linewidth]{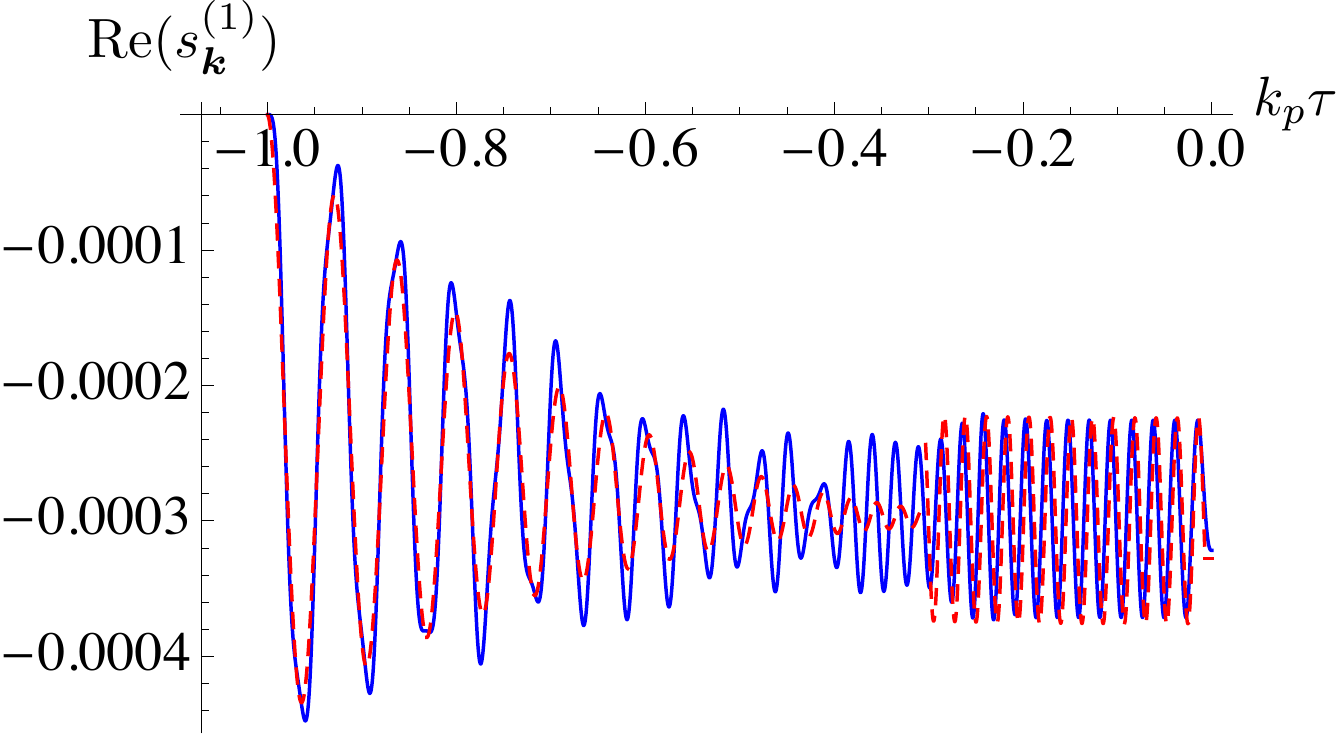}
  \end{center}
 \end{minipage} 
  \caption{Time evolution of $\mathrm{Re} (s_{\boldsymbol{k}}^{(1)})$ in
 terms of normalized time~$k_p \tau$. The KK mode is excited at
 $k_p \tau \approx -1$. Blue solid lines show numerically
 computed results, while the red dashed lines are the analytic
 estimations (\ref{eq102}) - (\ref{eq104}). Right (Left): Case for a
 wave mode that does (not) cross the resonance band.}
  \label{fig:sk-tau}
\end{figure}

In Figure~\ref{fig:sk-tau} we plot the time evolution of 
$\mathrm{Re} (s_{\boldsymbol{k}}^{(1)})$, comparing the approximate
solutions with numerically computed results. 
Here the inflationary parameters are chosen as $\phi_{\mathrm{exc}}
\approx 8.2 M_p$ and $\mu \approx 0.0016 M_p$, such that 
the KK modes are excited at about 50~e-foldings before
the end of inflation, and 
the COBE normalization value
$\mathcal{P}_{\zeta} \approx 2.4 \times 10^{-9}$~\cite{Komatsu:2010fb}
is realized at the wave number~$k_p$ which exits the horizon at the KK
excitation.
We have assumed the KK mode $n=1$ with $\alpha_1 = 100$ to be excited as
$\tilde{\psi}_{n\, \mathrm{exc}} = 0.001 M_p$ and 
$\theta_1 = 0$ (i.e. $\dot{\psi}_{n\, \mathrm{exc}} \approx 0$). This
set of parameters is translated into the small parameters $\xi_{1\,
\mathrm{exc}} \approx 0.082$ and $\tilde{\eta}_{1\, \mathrm{exc}}
\approx 0.0018$, and realizes the resonance peak at around
$k_1  \approx 42  k_p$. 
We show two plots for $k \approx 14 k_p$, $140 k_p$, each representing 
wave modes that do not/do cross the resonance band, respectively.
The blue solid lines are obtained by numerically 
solving (\ref{sk1}) (with (\ref{phitau}), (\ref{8psipsi}), and
(\ref{dotpsiV})), while the red 
dashed lines show the analytic estimations (\ref{eq102}) -
(\ref{eq104}). 

One sees that the analytic estimations well describe the numerical
results, except for at around the time~$\tau_{\mathrm{res}}$ in the
right figure, where we have made the simplifying assumption that the
amplification happens instantaneously.
It should be noted that parametric amplification actually happens
roughly as $ \propto (\tau- \tau_{\mathrm{res}})$, cf.~(\ref{exsolres}). 

We also note that in the analytic estimations, we have simply frozen the
fluctuations at $\tau =-1 / k $, which also give
rise to errors for estimating the asymptotic values
$ \mathrm{Re} (s_{\boldsymbol{k}}^{(1)})|_{\tau \to 0^-}$. However, such
errors only lead to shifts in the phase of the oscillations in the
$\mathrm{Re} (s_{\boldsymbol{k}}^{(1)})$ spectrum in $k$-space, as we
will see in the next subsection.

\subsection{Curvature Perturbation Spectrum}

The effects from the KK-mode oscillations on the curvature perturbation
spectrum are captured by the asymptotic value
$2 \mathrm{Re} (s_{\boldsymbol{k}}^{(1)})|_{\tau \to 0^-} $,
cf.~(\ref{PzetaPzeta0}). 
We obtain this analytically by freezing the expressions
(\ref{eq102}) and (\ref{eq104}) at horizon exit $\tau = -1/k$.

For $ k \leq - 1/\tau_{\mathrm{exc}}$:
\begin{equation}
  \mathrm{Re} (s_{\boldsymbol{k}}^{(1)}) |_{\tau \to 0^-} \simeq 0.
\label{eq105}
\end{equation}
For $-1/\tau_{\mathrm{exc}} \leq k \leq k_n$:
\begin{equation}
 \mathrm{Re} (s_{\boldsymbol{k}}^{(1)}) |_{\tau \to 0^-}
\simeq
 \tilde{\eta}_{n\,  \mathrm{exc}}\,   \xi_{n\, \mathrm{exc}} 
\left[
\left(2 \cos 2 \theta_n - \frac{1}{2} \right) \cos 
 \left( 2 k \tau_{\mathrm{exc}} + 2  \right)
 - 2 \cos 2 \theta_n 
 -  \frac{1}{2 (k \tau_{\mathrm{exc}})^3}
\right].
\label{eq106}
\end{equation}
For $k_n \leq k$:
\begin{multline}
 \mathrm{Re} (s_{\boldsymbol{k}}^{(1)}) |_{\tau \to 0^-} \simeq
 3^{3/2} \tilde{\eta}_{n\,  \mathrm{exc}} \, \xi_{n\, \mathrm{exc}}^{1/2}
 \left(\frac{k_n}{k}\right)^3
\cos \left\{
 \frac{4 \cdot 3^{1/2}}{\xi_{n\, \mathrm{exc}}}
\left( 1 + \ln \left(\frac{k}{k_n} \right) \right)
-2  + 2 \theta_n
\right\}
 \\
 +  \tilde{\eta}_{n\,  \mathrm{exc}}\,   \xi_{n\, \mathrm{exc}} 
\left[
\frac{1}{2} \left(\cos 2 \theta_n - 1  \right) \cos 
 \left(2 k\tau_{\mathrm{exc}} + 2  \right)
 - 2 \cos 2 \theta_n 
\right].
\label{eq107}
\end{multline}
Here in (\ref{eq107}) we have dropped a contribution originating from 
the term $(\tau / \tau_{\mathrm{exc}})^3$ in (\ref{eq104}) which
asymptotically becomes tiny.
As stated in the previous subsection, the actual 
$\mathrm{Re} (s_{\boldsymbol{k}}^{(1)}) $ spectrum is obtained by adding 
the above expressions over all excited KK-modes~$n$. 

\vspace{\baselineskip}

The resonant peak rises at around $k = k_n$, dividing wave modes by
whether or not they went through the resonance band.
Here, since we have adopted the simplifying assumption of
instantaneous resonant amplification, the above approximate expressions give a 
peak that rises suddenly at~$k = k_n$. 
Actually, the resonant oscillations arise sharply but within a finite wave number
range, reflecting the finite time interval that a wave mode
stays in the resonance band.
As is shown in the first term in~(\ref{exsolres}), when in the resonance band,
$s_{\boldsymbol{k}}^{(1)}$ is parametrically amplified roughly proportional
to $(\tau - \tau_{\mathrm{res}})$.
The rising region~$\Delta k$ around~$k_n$ consists of wave modes for
which the KK excitation at $\tau_{\mathrm{exc}}$ happens to be 
within the range $\Delta \tau \sim -\xi_{n\, \mathrm{exc}}^{1/2}
\tau_{\mathrm{res}} (k)  $ (\ref{res_time})
around their resonance times~$\tau_{\mathrm{res}} (k) $. 
One can estimate that the resonant peak in the $k$-space
rises sharply within a wave number interval of 
\begin{equation}
 \Delta k \sim \xi_{n\, \mathrm{exc}}^{1/2} k_n,
\label{res_k_region}
\end{equation}
around $ k = k_n$. Thus actually the resonant oscillations peak at wave
numbers slightly larger than~$k_n$.
One can also see that in the rising region~(\ref{res_k_region}), effects
from parametric resonance (i.e. the first line of (\ref{eq107})) is
dominant over other effects (i.e. second line of (\ref{eq107})).

\vspace{\baselineskip}

\begin{figure}[htbp]
  \begin{center}
  \begin{center}
  \includegraphics[width=0.85\linewidth]{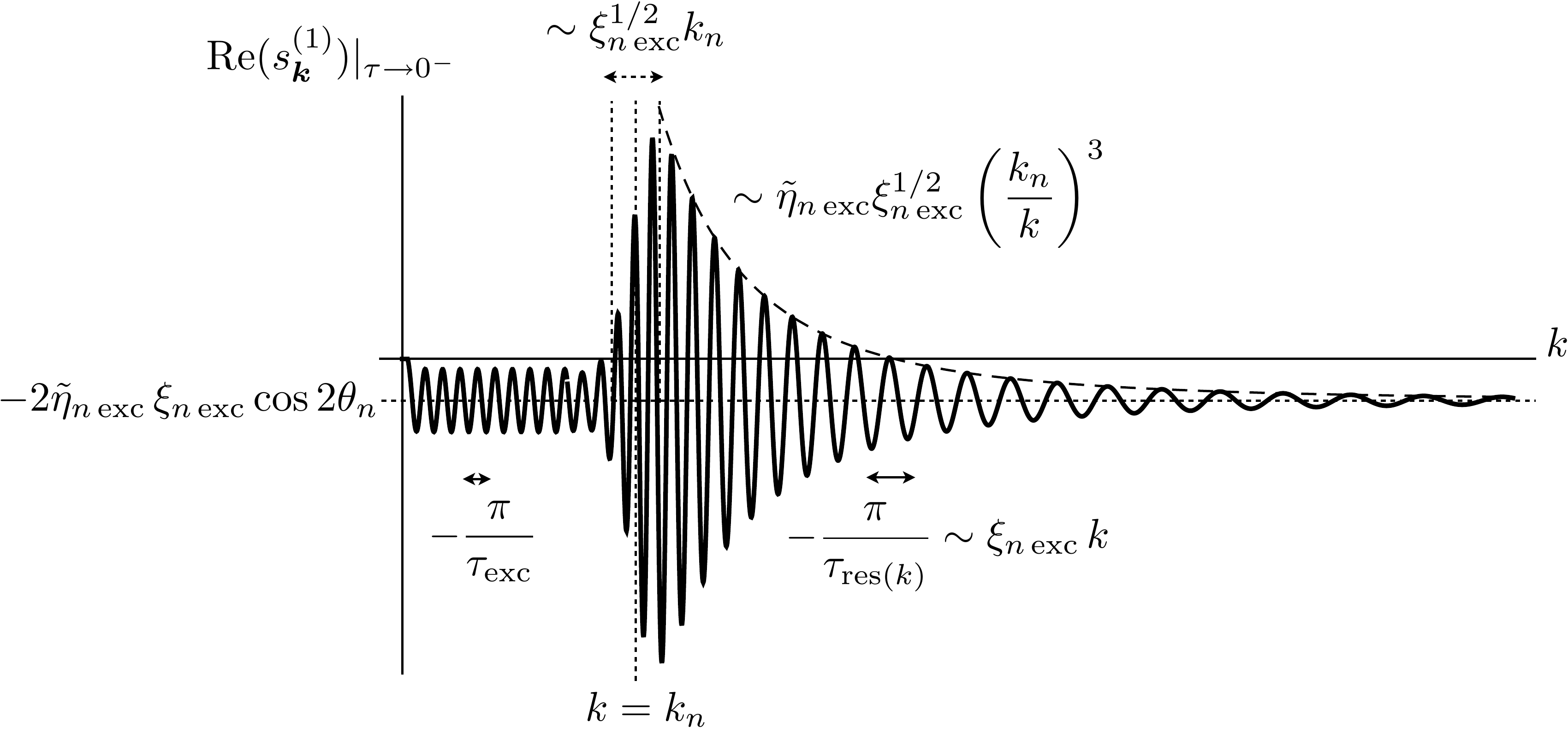}
  \end{center}
  \caption{Schematic of the oscillations on the curvature
   perturbation spectrum induced by an excited KK-mode with $\theta_n
   \approx 0$, i.e. $\dot{\psi}_{n\, \mathrm{exc}} \approx 0$. 
   The envelope of the resonant oscillations is sharper and more
   asymmetric for smaller $\xi_{n\, \mathrm{exc}}$, i.e. for heavier KK
   modes.}
  \label{fig:schematic}
  \end{center}
\end{figure}

In Figure~\ref{fig:schematic} we illustrate a typical oscillation form
induced on the curvature perturbation spectrum from an excited KK-mode
with $\theta_n  \approx 0$, i.e. $\dot{\psi}_{n\, \mathrm{exc}} \approx 0$.
Wave modes that are sub-horizon at the KK excitation obtain
oscillations about the offset $ -2 \tilde{\eta}_{n\,  \mathrm{exc}}\,
\xi_{n\, \mathrm{exc}} \cos 2 \theta_n $. For modes in the range $
-\tau_{\mathrm{exc}}^{-1} \leq k \lesssim k_n$ which do not cross the
resonance band, the oscillation period is
a constant~$-\pi/\tau_{\mathrm{exc}}$,
and the oscillation amplitude is comparable to the offset. 

The oscillation amplitude obtains a peak of 
$\sim \tilde{\eta}_{n\, \mathrm{exc}} \xi_{n\, \mathrm{exc}}^{1/2}$
at the wave number $k \approx k_n$, beyond which the amplitude decays proportionally to
$\propto k^{-3}$, with $k$-dependent oscillation period $-\pi /
\tau_{\mathrm{res}} (k) \sim \xi_{n\, \mathrm{exc}} k$. 
Here one clearly sees that the induced oscillations are determined
basically by the two parameters $\tilde{\eta}_{n\, \mathrm{exc}}$ and
$\xi_{n\, \mathrm{exc}}$. 

Let us further remark that for nonzero~$\theta_n$, oscillations with
period~$-\pi / \tau_{\mathrm{exc}}$ is dominant also at wave
numbers sufficiently larger than~$k_n$, as can be seen from the second
line of (\ref{eq107}).
(Such oscillations at large~$k$ do exist
even for $\theta_n = 0$, though they are suppressed. See discussions
below (\ref{eq104}).) 
On the other hand, oscillations in the range $-\tau_{\mathrm{exc}}^{-1}
\leq k \lesssim k_n$ are suppressed for $\cos 2 \theta_n \approx
1/4$. 

For smaller $\xi_{n\, \mathrm{exc}}$, i.e. heavier KK modes, the resonant
peak rises more sharply, and thus the envelope of the oscillations takes
a more asymmetric form in $k$-space.

\vspace{\baselineskip}

\begin{figure}[htbp]
 \begin{minipage}{.49\linewidth}
  \begin{center}
 \includegraphics[width=\linewidth]{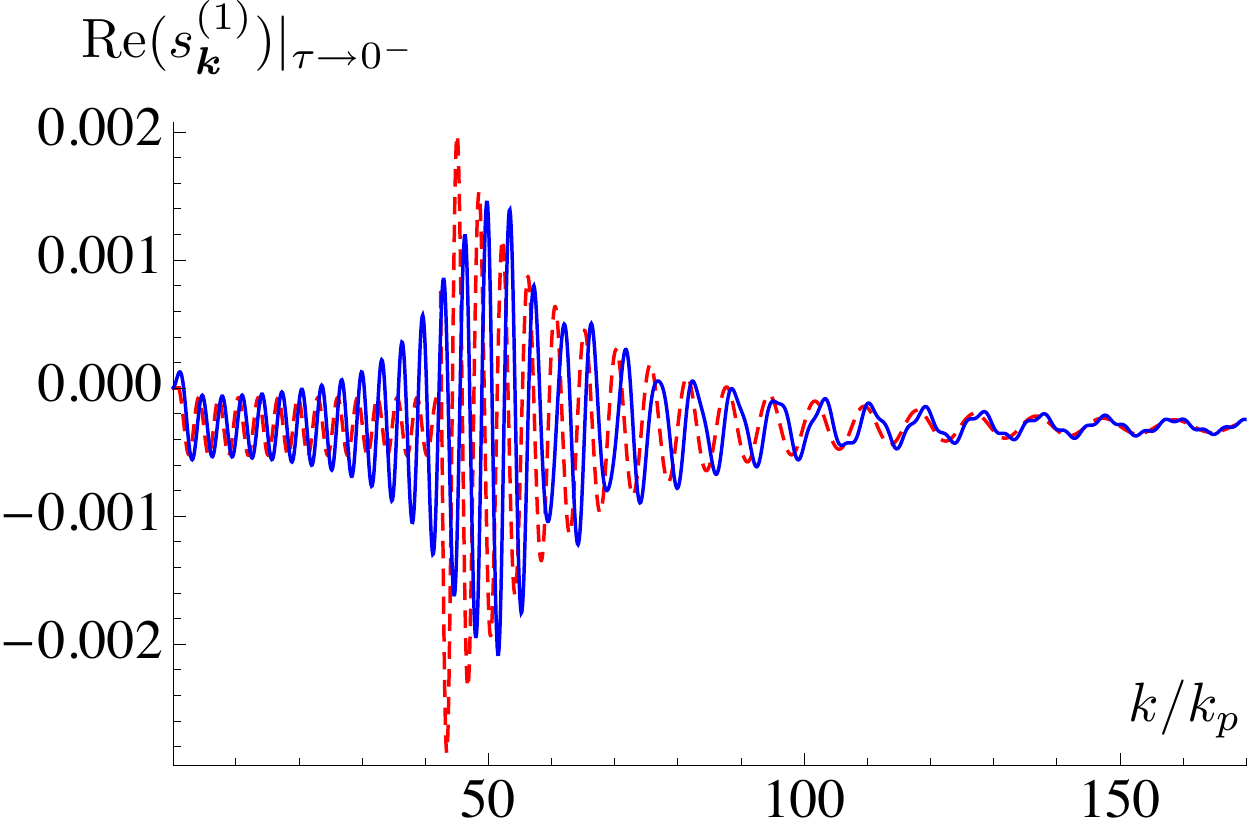}
  \end{center}
 \end{minipage} 
 \begin{minipage}{0.01\linewidth} 
  \begin{center}
  \end{center}
 \end{minipage} 
 \begin{minipage}{.49\linewidth}
  \begin{center}
 \includegraphics[width=\linewidth]{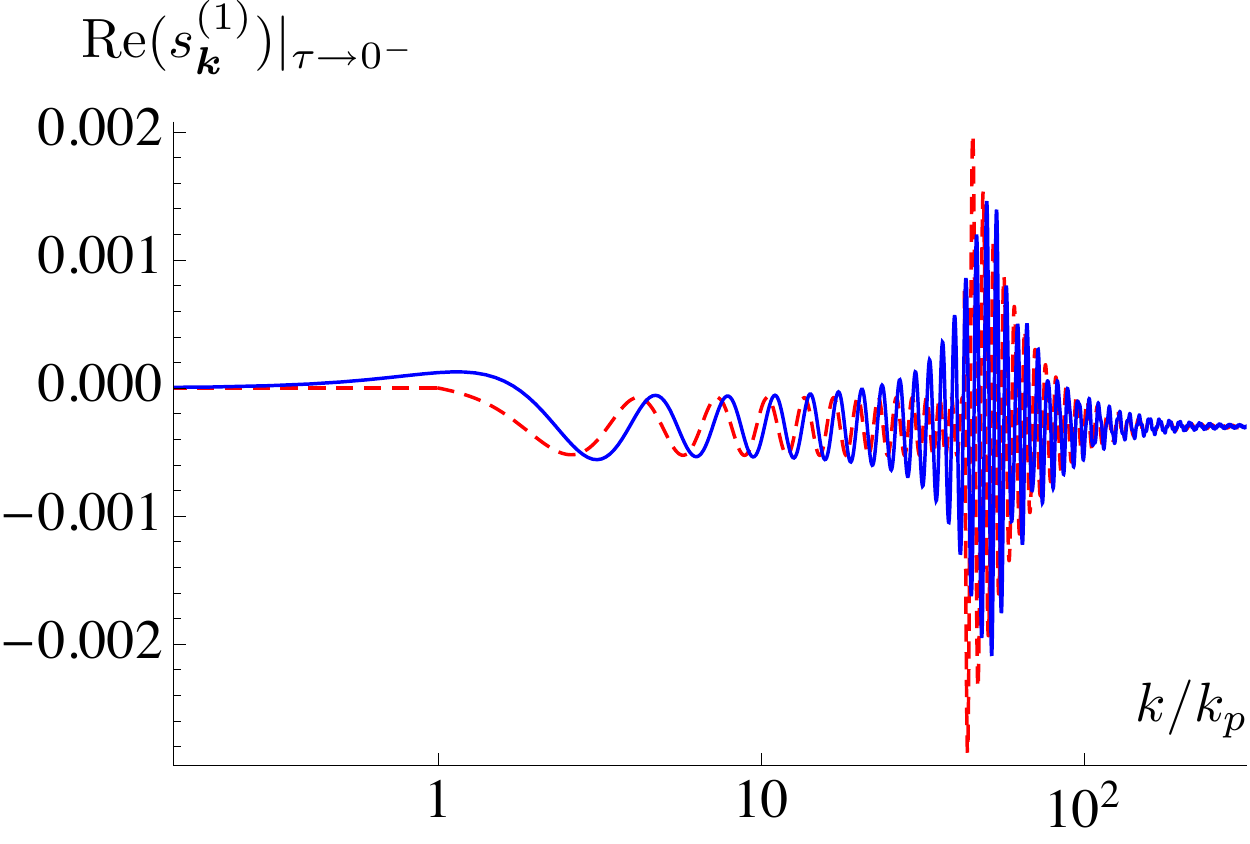}
  \end{center}
 \end{minipage} 
  \caption{Asymptotic values of $\mathrm{Re}
 (s_{\boldsymbol{k}}^{(1)})$, denoting the KK induced oscillations on
 the curvature perturbation spectrum $\frac{1}{2}
 (\mathcal{P}_{\zeta}(k) /  \mathcal{P}_{\zeta0}(k) - 1)$, 
 as a function of wave number in linear (left figure) and log (right) scales.
 Blue solid lines show numerically computed  results, while the red
 dashed lines are the analytic estimations (\ref{eq105}) - (\ref{eq107}).}
 \label{fig:sk-k}
\end{figure}

We have also numerically computed the $\mathrm{Re}
(s_{\boldsymbol{k}}^{(1)})$ spectrum as a function of the wave
number, for the parameter set chosen in the previous subsection.
The results are shown in Figure~\ref{fig:sk-k}, where the blue solid
lines show numerically calculated
values of $\mathrm{Re} (s_{\boldsymbol{k}}^{(1)})$ at a time when all the
displayed wave modes are well outside the horizon (and also when the KK mode has
sufficiently damped away), and the red dashed lines denote the
analytic estimations (\ref{eq105}) - (\ref{eq107}). 
One sees that the analytic result captures the overall behavior of the
spectrum, though overestimates the resonant peak due to the
approximation of the instantaneous parametric amplification. 
One can also see that as the dominant oscillation with period $- \pi /
\tau_{\mathrm{res}} (k)$ decays away for large~$ k $, tiny oscillations
with period $- \pi / \tau_{\mathrm{exc}}$ show up. Such
oscillations are suppressed from $\theta_1 = 0$, but are present at
sub-leading orders beyond the approximate estimation of (\ref{eq107}). 

\vspace{\baselineskip}

Before ending this subsection, let us repeat that the particular forms of
the oscillations generated by the KK modes can be attributed to a few
parameters, which are $\tilde{\eta}_{n\, \mathrm{exc}}$ representing the
excited KK amplitudes, 
and $\xi_{n\, \mathrm{exc}}$ setting the effective masses of the KK modes. In
addition to these two parameters, the time of KK 
excitation~$\tau_{\mathrm{exc}}$ determines where in $k$-space the
oscillations rise.

\subsection{Effects from Multiple KK Modes}

\begin{figure}[htbp]
  \begin{center}
  \begin{center}
  \includegraphics[width=0.48\linewidth]{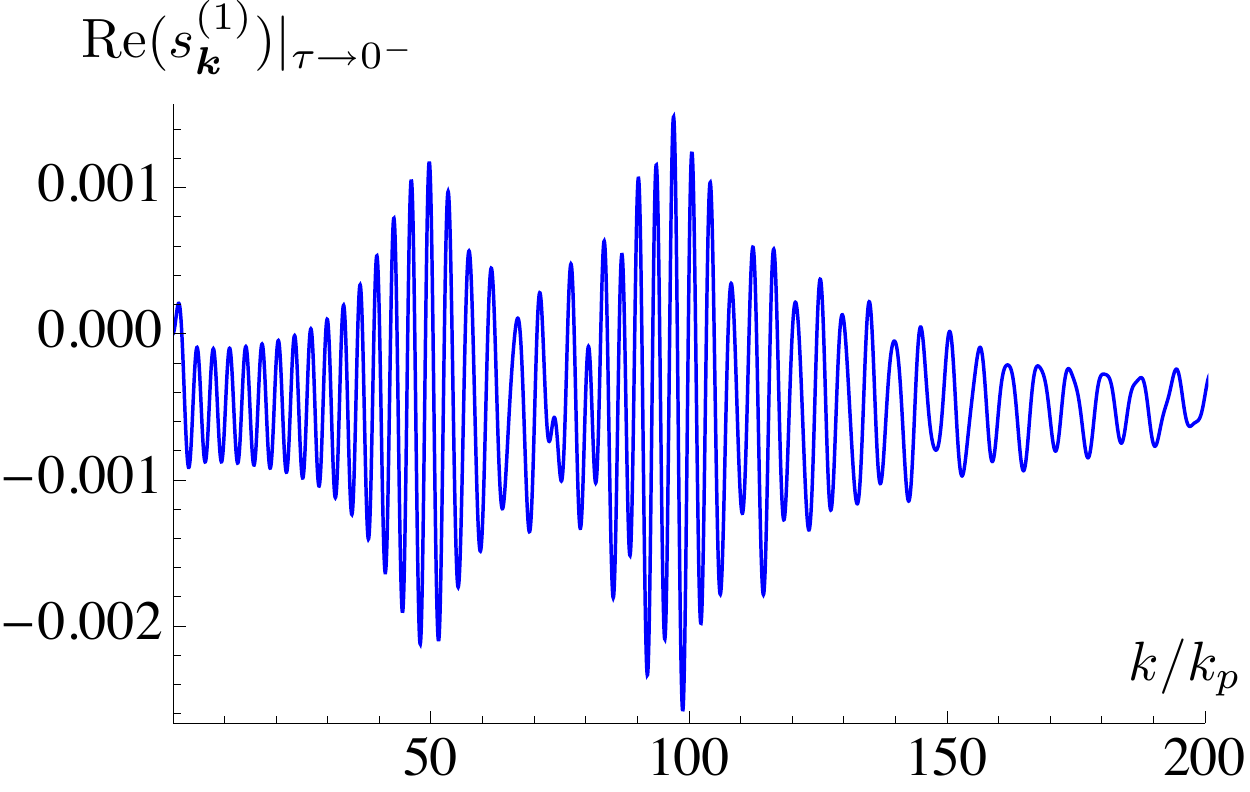}
  \end{center}
  \caption{$\mathrm{Re} (s_{\boldsymbol{k}}^{(1)})$ spectrum in
   $k$-space when two KK modes are excited.}
  \label{fig:two-KK}
  \end{center}
\end{figure}

As we have stated above, for small KK excitations, effects from multiple
KK modes simply add up. 
As an example, in Figure~\ref{fig:two-KK} we show the numerically
computed $\mathrm{Re} (s_{\boldsymbol{k}}^{(1)})$ spectrum for the case
where two KK modes $n =1$, $2$ are excited. 
The parameters are the same as in the previous subsections, except for
that the $n=2$ mode is further excited with $\alpha_2 = 200$,
$\tilde{\psi}_{2\, \mathrm{exc}} = 0.0004 M_p$, $\theta_2 = 0$.
(Note that $\alpha_n \propto n$, $\xi_{n\, \mathrm{exc}} \propto
n^{-1}$, $\tilde{\eta}_{n\, \mathrm{exc}} \propto n^3 \tilde{\psi}_{n\,
\mathrm{exc}}^2$, and $k_n \propto n$.) 
The plot shows superimposed oscillations from the two KK modes, which are
peaked at around $k_1 \approx 42 k_p$ and $k_2  \approx 85  k_p$. 

Let us also note that effects from multiple KK modes can cancel each
other, given that the excited KK amplitudes obey
specific hierarchies such that different $n$ modes produce $\mathrm{Re}
(s_{\boldsymbol{k}}^{(1)})$ with similar oscillation amplitudes/offsets, 
and further if the phases~$\theta_n$ take appropriate values. 
When focusing on a certain wave number~$k$ and considering
the time evolution of~$\mathrm{Re} (s_{\boldsymbol{k}}^{(1)})$, its 
oscillation amplitude can be suppressed (instead of amplified) when
going through multiple 
resonance bands, if the phases~$\theta_n$ of the KK modes are
suitably distributed, cf. first line of~(\ref{eq104}).
However, the resonant oscillations of the $\mathrm{Re}
(s_{\boldsymbol{k}}^{(1)})$~spectrum in $k$-space
(i.e. first line of (\ref{eq107})) cannot be cancelled entirely,
since $\xi_{n\, \mathrm{exc}} \propto n^{-1}$ and thus 
different $n$~modes source different oscillation periods.

As we have seen explicitly, the KK tower generates oscillations on the
curvature perturbation spectrum with specific patterns characterized by
the integers~$n$. It would be very interesting to investigate such
specific patterns in cosmological observables, which can allow us to
extract information about the extra dimensional space.

\section{Strong Resonance from Large KK Excitations}
\label{sec:largeKK}

In this section we study cases with large KK excitations beyond the 
condition~(\ref{smallpar}), i.e., when (\ref{star2}) - (\ref{star4}) are
no longer good approximations of the inflationary dynamics.
We will see that when the KK mode amplitudes are as large as
$\tilde{\eta}_{n\, \mathrm{exc}} \xi_{n\, \mathrm{exc}} \gtrsim 0.1$,
the inflaton field fluctuations are strongly amplified through
parametric resonance, generating sharp spikes in the curvature
perturbation spectrum.

\subsection{Curvature Perturbation Spectrum}
\label{subsec:CPS}

\begin{figure}[p]
 \begin{minipage}{.48\linewidth}
  \begin{center}
 \includegraphics[width=\linewidth]{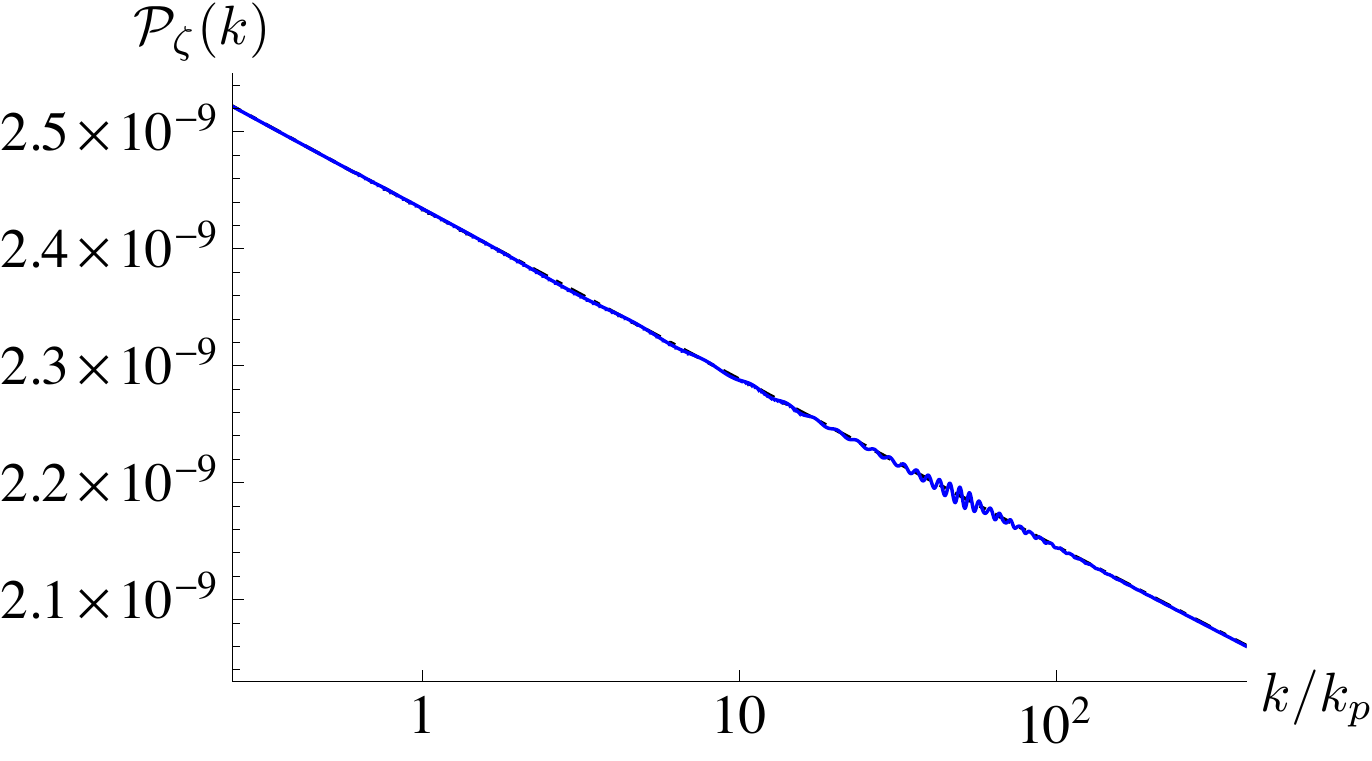}
  \end{center}
 \end{minipage} 
 \begin{minipage}{0.01\linewidth} 
  \begin{center}
  \end{center}
 \end{minipage} 
 \begin{minipage}{.48\linewidth}
  \begin{center}
 \includegraphics[width=\linewidth]{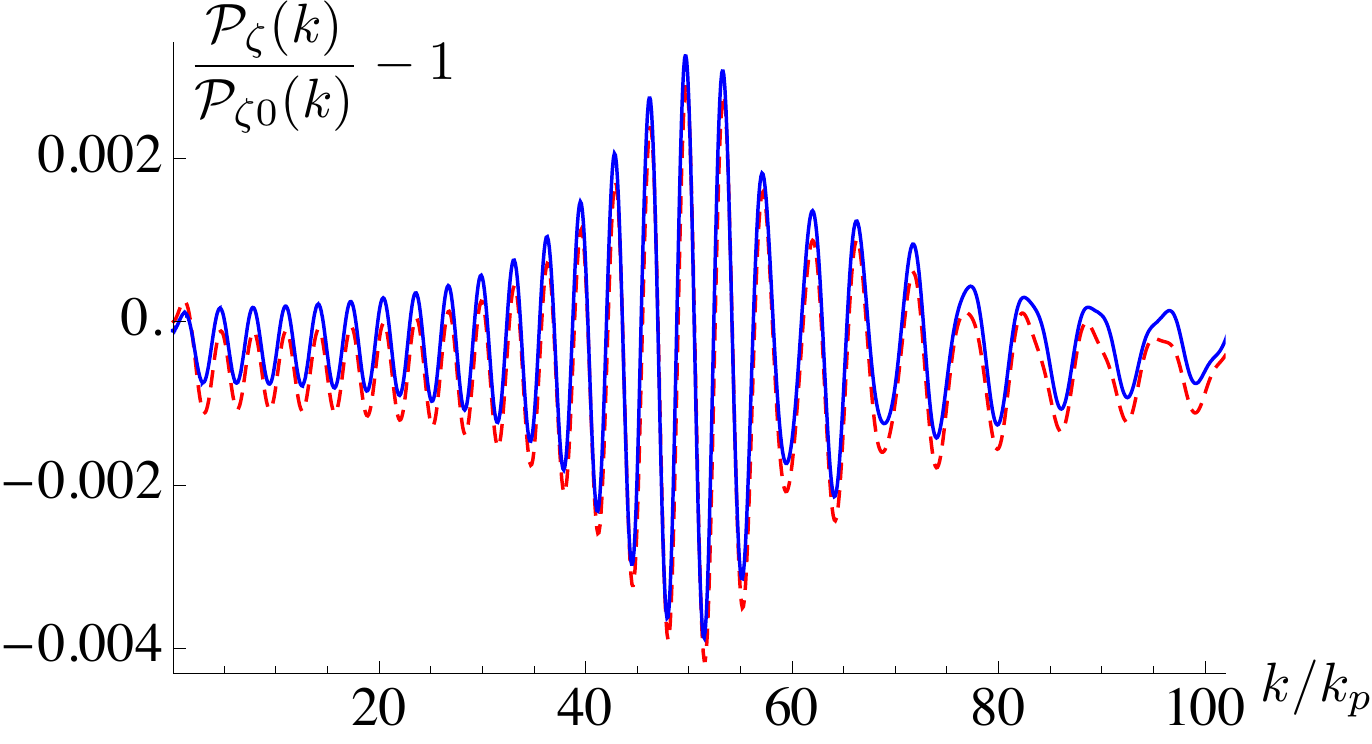}
  \end{center}
 \end{minipage} 
  \caption{Curvature perturbation spectrum with KK mode
 $\tilde{\psi}_{1\, \mathrm{exc}} = 0.001 M_p$ 
 ($\tilde{\eta}_{1\, \mathrm{exc}} \approx 0.0018$).}
  \label{fig:KK1000}
\end{figure}

\begin{figure}[htbp]
 \begin{minipage}{.48\linewidth}
  \begin{center}
 \includegraphics[width=\linewidth]{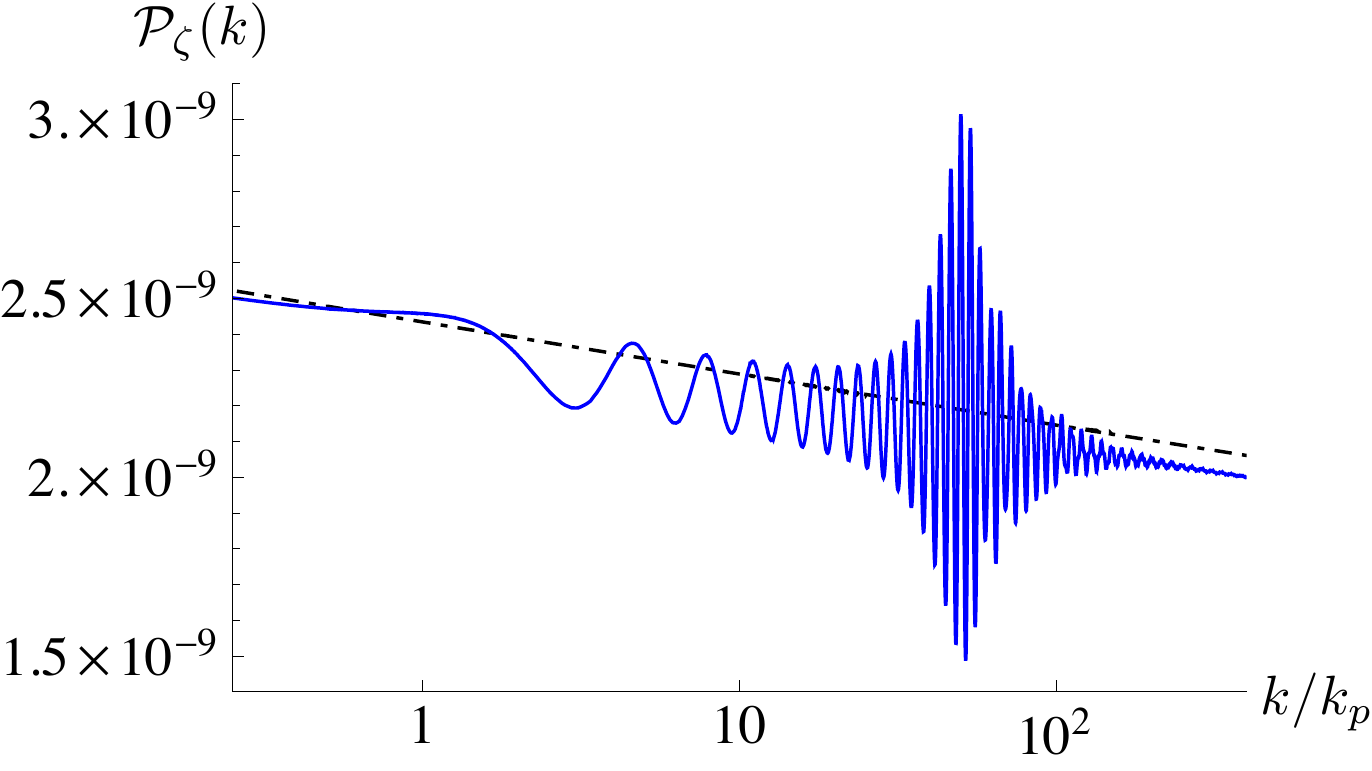}
  \end{center}
 \end{minipage} 
 \begin{minipage}{0.01\linewidth} 
  \begin{center}
  \end{center}
 \end{minipage} 
 \begin{minipage}{.48\linewidth}
  \begin{center}
 \includegraphics[width=\linewidth]{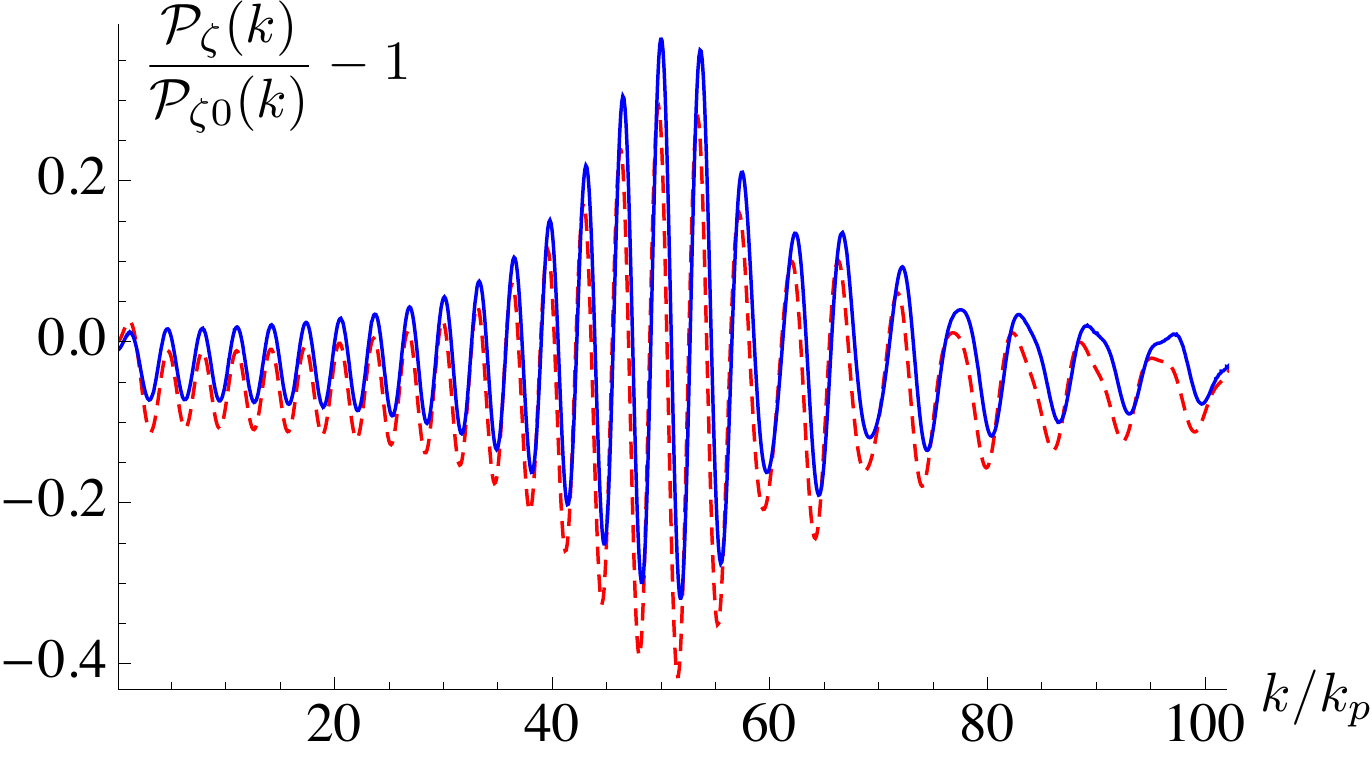}
  \end{center}
 \end{minipage} 
  \caption{Curvature perturbation spectrum with KK mode
 $\tilde{\psi}_{1\, \mathrm{exc}} = 0.01 M_p$ 
 ($\tilde{\eta}_{1\, \mathrm{exc}} \approx 0.18$).}
  \label{fig:KK100}
\end{figure}

\begin{figure}[htbp]
 \begin{minipage}{.48\linewidth}
  \begin{center}
 \includegraphics[width=\linewidth]{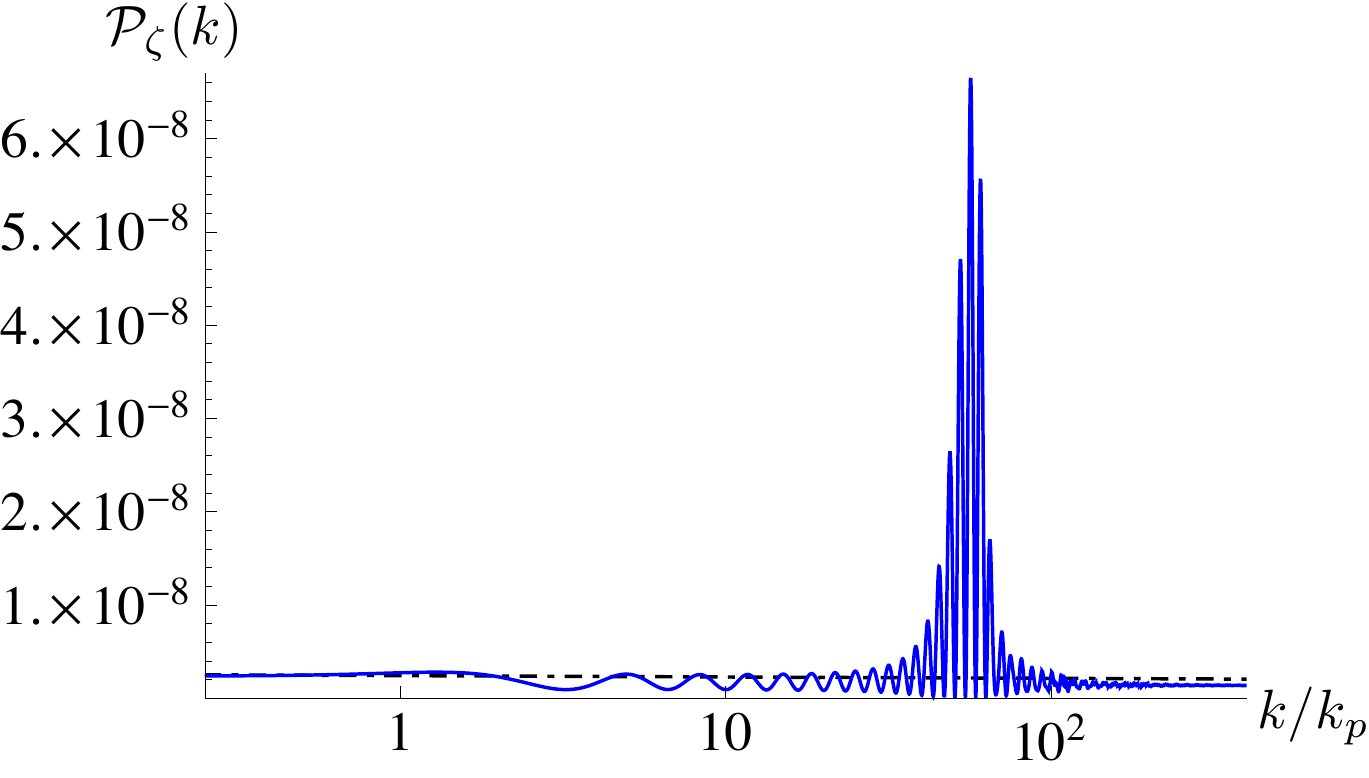}
  \end{center}
 \end{minipage} 
 \begin{minipage}{0.01\linewidth} 
  \begin{center}
  \end{center}
 \end{minipage} 
 \begin{minipage}{.48\linewidth}
  \begin{center}
 \includegraphics[width=\linewidth]{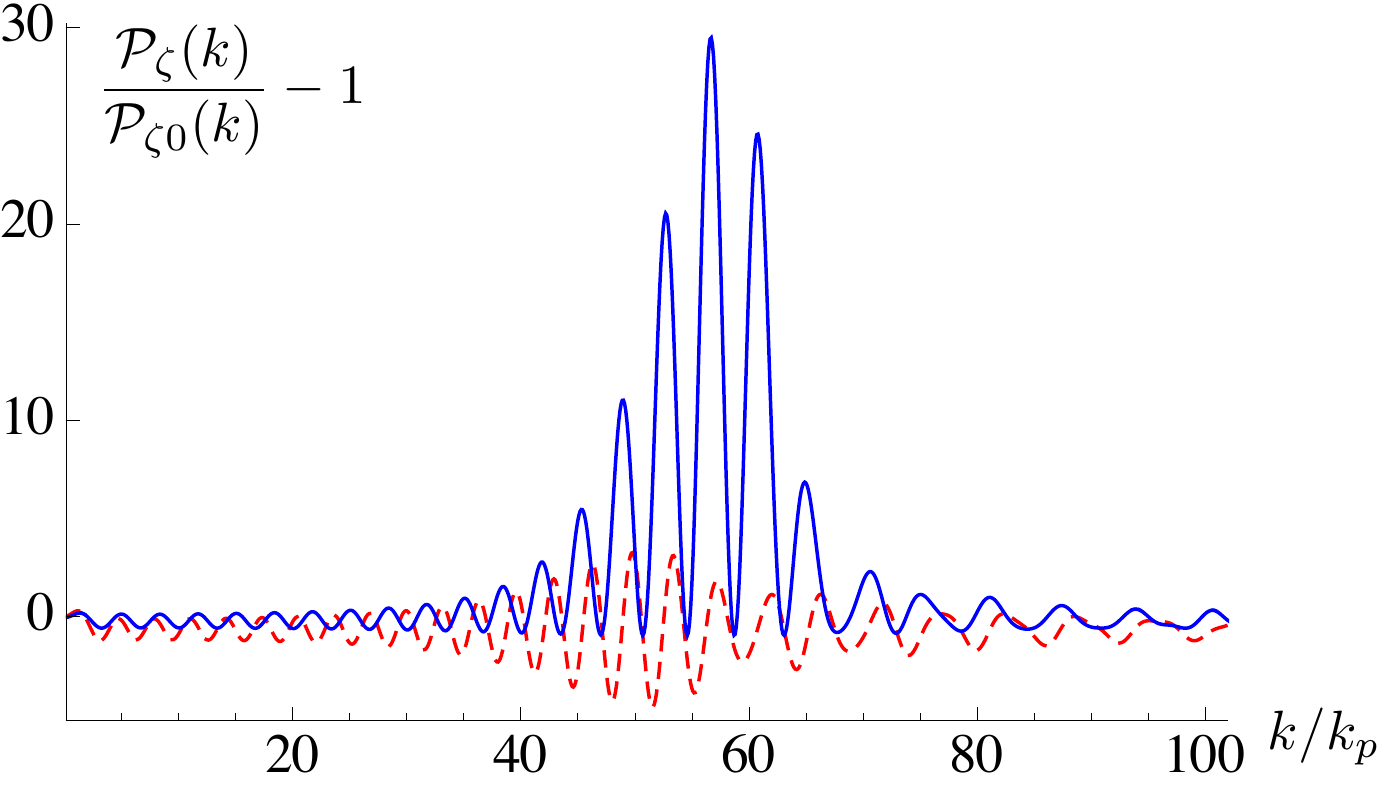}
  \end{center}
 \end{minipage} 
  \caption{Curvature perturbation spectrum with KK mode
 $\tilde{\psi}_{1\, \mathrm{exc}} \approx  0.033 M_p$
 ($\tilde{\eta}_{1\, \mathrm{exc}} \approx 2.0$).}
  \label{fig:KK30}
\end{figure}

Let us first show how the resulting curvature perturbation spectrum
changes when the KK mode is excited with larger amplitude.
In Figures~\ref{fig:KK1000}, \ref{fig:KK100}, and \ref{fig:KK30}, we
show the results for the case where the $n=1$ KK mode is
excited at about 50 e-foldings before the end of inflation. 
The parameters are chosen similarly as in the previous section, i.e.
$\phi_{\mathrm{exc}} \approx 8.2 M_p$, $\mu \approx 0.0016 M_p$,
$\alpha_1 = 100$, but with various KK mode amplitudes~$\psi_{1\,
\mathrm{exc}}$. We again use $k_p$ to denote the wave number that exits the
horizon at around the time of KK excitation.
(The following parameters are also the same as in the previous section:
$\epsilon_{\mathrm{exc}} \approx 0.015$, $\xi_{1\, \mathrm{exc}} \approx
0.082$, and $k_1 \approx 42 k_p$, though $k_1$ defined in
(\ref{etaxiexc}) no longer denotes the position of the resonant peak for
strong resonance.) 
The excited KK mode amplitudes are taken as 
$\tilde{\psi}_{1\, \mathrm{exc}} = 0.001 M_p$  
($\tilde{\eta}_{1\, \mathrm{exc}} \approx 0.0018$), 
$\tilde{\psi}_{1\, \mathrm{exc}} = 0.01 M_p$ 
($\tilde{\eta}_{1\, \mathrm{exc}} \approx 0.18$), and
$\tilde{\psi}_{1\, \mathrm{exc}} \approx  0.033 M_p$
($\tilde{\eta}_{1\, \mathrm{exc}} \approx 2.0$), 
respectively, 
for Figures~\ref{fig:KK1000}, \ref{fig:KK100}, and \ref{fig:KK30},
with vanishing initial velocity~$\dot{\psi}_{1\, \mathrm{exc}} = 0$.
Upon obtaining the power spectrum, we have numerically solved
the full set of equations of motion for the homogeneous background
(\ref{Fvarphi}), (\ref{EoMphi}), and (\ref{EoMvarphi}), 
as well as for the inflaton field
fluctuation~$q_{\boldsymbol{k}}$~(\ref{4747}).\footnote{If the initial
conditions of~$q_{\boldsymbol{k}}$ were set by the Hankel 
type solution~(\ref{qkzero}), it would contain errors of
$\mathcal{O}(\epsilon)$ (in the absence of KK-modes). Since we are
dealing with a non-canonical inflaton, such errors modify the sound
speed and induce artificial oscillations of $\mathcal{O}(\epsilon)$ on
the resulting perturbation spectrum, which can be confused with the KK
effects. Therefore in our numerical computations we have set the initial
conditions using the WKB-type solution: 
\begin{equation}
 q_{\boldsymbol{k}} = - \frac{1}{(2 \pi)^3} \frac{1}{(2 \varpi_k
  (\tau))^{1/2}}
 \exp \left\{
-i \int^\tau \varpi_k (\tau') d\tau'
\label{WKB-type}
\right\}
\end{equation}
where
\begin{equation}
 \varpi_k (\tau) \equiv
\left[
 \frac{1}{B_{\phi \phi}}
 \left\{ 2 L_A G_{\phi \phi}^A k^2 - B_{\phi \phi} \frac{a''}{a}
 + \left(-M_{\phi \phi} + \dot{C}_{\phi \phi}\right)  a^2 + 
 \left( -\dot{B}_{\phi \phi} + 3 C_{\phi \phi}  \right) a'\right\}
\right]^{1/2}.
\end{equation}
This is a solution to the equation of motion~(\ref{4747}) while the
wave modes are well inside the horizon, within errors (in the absence of
KK-modes) of $\mathcal{O} (\epsilon^2 \frac{aH}{k}, \,
(\frac{aH}{k})^4)$, providing more 
accurate initial conditions than~(\ref{qkzero}).
One can also check that (\ref{WKB-type}) satisfies (\ref{eq:commu2})
within errors of $\mathcal{O} (\epsilon)$, and approaches
$q_{\boldsymbol{k}} \propto e^{-i k \tau}$ in the asymptotic past.
(The $\mathcal{O} (\epsilon)$ error for satisfying (\ref{eq:commu2}) can
affect the overall amplitude of the perturbations, but does not directly
lead to sourcing artificial oscillations in the spectrum.)} Note that
even when the KK modes are largely excited, the power spectrum 
can be obtained by evaluating~(\ref{4.43}) when the wave modes have exited the
horizon~(\ref{6star-1}) and the KK mode oscillations are damped away such that
$\eta_n$ is sufficiently smaller than unity, satisfying for e.g.
(\ref{KKdamp}). This gives the power spectrum 
within errors of $\mathcal{O}(\epsilon, \eta_n)$ upon evaluation.
We have assumed the KK mode to be excited suddenly to~$\tilde{\psi}_{n\,
\mathrm{exc}}$, which sources unphysical jumps at~$t_{\mathrm{exc}}$
in quantities such as the Hubble parameter.
However, we expect this treatment to be valid for discussing the KK
signals since it is the parametric resonance following
the KK excitation that sources sharp features in the perturbation
spectrum.

In the figures on the left side, the blue solid lines denote the power
spectrum~$\mathcal{P}_{\zeta}$ of the curvature perturbations, shown
with log scale for the 
wave number. The black dot-dashed lines represent the
spectrum~$\mathcal{P}_{\zeta0}$ in the absence of the KK mode. On the
right side, we show in blue solid lines the ratio $(\mathcal{P}_{\zeta} -
\mathcal{P}_{\zeta 0}) /  \mathcal{P}_{\zeta 0}$ denoting the relative
amplitude of the KK mode effects, with linear scale for the wave number.
For comparison, we have also plotted the red dashed lines which are 
the asymptotic values of $2 \mathrm{Re} (s_{\boldsymbol{k}}^{(1)}) $,
obtained by solving (\ref{sk1}), with (\ref{phitau}), (\ref{8psipsi}), and
(\ref{dotpsiV}). As we have discussed in the previous section, 
$2 \mathrm{Re} (s_{\boldsymbol{k}}^{(1)}) $  represents
$(\mathcal{P}_{\zeta} -\mathcal{P}_{\zeta 0}) /  \mathcal{P}_{\zeta 0}$
in the small KK regime $\tilde{\eta}_{n\, \mathrm{exc}} \ll 1$
(cf.~(\ref{PzetaPzeta0})), as is seen in
Figure~\ref{fig:KK1000}.\footnote{The slight difference in the center of
the oscillations between $2 \mathrm{Re} (s_{\boldsymbol{k}}^{(1)}) $ and 
$(\mathcal{P}_{\zeta} -\mathcal{P}_{\zeta 0}) /  \mathcal{P}_{\zeta 0}$
arise due to the sub-leading KK contributions that were dropped in the
$s_{\boldsymbol{k}}^{(1)} $ computations. For example, the $O(\eta_n
\xi_n)$ corrections to the conformal time~(\ref{esttau}) slightly affects
the expansion of the universe and thus modifies the offset of the
perturbation spectrum. Such KK mode effects on the expansion history
also source the deviation of $\mathcal{P}_\zeta$ from
$\mathcal{P}_{\zeta 0}$ for wave modes that are super-horizon at the KK
excitation, as can be seen in the figures on the left side.} 
However, $2 \mathrm{Re} (s_{\boldsymbol{k}}^{(1)}) $ fails in describing
the KK effects as $\tilde{\eta}_{n\, \mathrm{exc}} $ increases and the
small KK approximations break down.
As the KK amplitude increases, the resonant oscillations
around $k = k_1$ grow asymmetrically towards the positive side
of~$\mathcal{P}_{\zeta 0}$, and eventually form sharp spikes on the
power spectrum. 

The strong parametric resonance significantly enhances the inflaton
field fluctuations, and the resulting spikes in the perturbation spectrum
grows non-linearly with respect to~$\tilde{\eta}_{n\, \mathrm{exc}}$.
A slight increase in the KK amplitude enhances the spike by orders of
magnitude, for e.g., 
$\tilde{\psi}_{n\, \mathrm{exc}} = 0.05 M_p$ sources
$(\mathcal{P}_{\zeta} -\mathcal{P}_{\zeta 0}) /  \mathcal{P}_{\zeta 0}$
as large as $\sim 3000$ at its peak. 
It should also be noted that the peak of the oscillations/spikes is shifted
slightly towards larger~$k$ for strong parametric resonance, as can be
seen in Figure~\ref{fig:KK30}. 
While the amplitude of the oscillations on $\mathcal{P}_\zeta$ are
significantly enhanced by the strong resonance, the oscillation period
follows similar behavior as for the weak resonance. 
Moreover, the oscillations at wave numbers away from the peak
(e.g. $k \lesssim 40 k_p$ and $k \gtrsim 70 k_p$ in
Figure~\ref{fig:KK30}) are not so different from the 
weak resonant calculations represented by~$2 \mathrm{Re}
(s_{\boldsymbol{k}}^{(1)}) $. 
Here we also remark that when the resonance is very strong, the spectrum can
obtain multiple peaks, although the one at the wave number slightly
larger than~$k_1$ is the largest.

The resonant spikes grow significantly as the KK amplitude is increased,
until $\tilde{\psi}_{n\, \mathrm{exc}}$ becomes so large that the field
dynamics are completely modified and/or the higher order KK interactions
are important.
For the Lagrangian~(\ref{Lphi}) with the set of parameters chosen here,
the KK mode no longer oscillates 
when $\tilde{\psi}_{n\, \mathrm{exc}} \gtrsim 0.06$ (or
$\tilde{\eta}_{n\, \mathrm{exc}} \gtrsim 6$). In addition, $B_{\phi \phi}$ becomes
negative and thus the inflaton field fluctuation~$q$ is tachyonic,
cf.~(\ref{S2q}). 
However we should remark that in such case the product of 
$\tilde{\eta}_{n\, \mathrm{exc}}$ and $ \xi_{n\, \mathrm{exc}} $ is as
large as
$\tilde{\eta}_{n\, \mathrm{exc}} \xi_{n\,
\mathrm{exc}} \gtrsim 0.5$ and the condition~(\ref{3.23}) is almost
broken. Hence the cubic or higher order KK terms in the Lagrangian can
become important, and may prevent $q$ from becoming tachyonic.
It would be very interesting to explore this regime, where the higher
order KK terms may contribute to source even stronger resonant features
in the primordial spectrum.


\subsection{Effective Frequency}

The behavior of the strong resonance can be better understood by
studying how the oscillation frequency of $q_{\boldsymbol{k}}$ resonates
with that of the KK modes. For this purpose, let us redefine the field fluctuation as
\begin{equation}
 u_{\boldsymbol{k}} \equiv (a B_{\phi \phi})^{1/2} 
 q_{\boldsymbol{k}},
\label{defuk}
\end{equation}
so that its equation of motion takes the form
\begin{equation}
 0 = \ddot{ u_{\boldsymbol{k}}} + f_{k\, \mathrm{eff}}^2  \,
  u_{\boldsymbol{k}}, 
\end{equation}
where $f_{k\, \mathrm{eff}}^2$ denotes the effective frequency
squared defined as
\begin{equation}
f_{k\, \mathrm{eff}}^2 (t)  \equiv 
 \frac{2 L_A G^A_{\phi \phi}}{B_{\phi \phi}} \frac{k^2}{a^2} 
 - \frac{M_{\phi \phi}}{B_{\phi \phi}} + 
 \frac{\left(a^{3} C_{\phi \phi}\right)^{\cdot }}{a^{3} B_{\phi \phi} }
 -  \frac{\left(a^{3/2} B_{\phi \phi}^{1/2}\right)^{\cdot \cdot}}{a^{3/2} B_{\phi
 \phi}^{1/2} } .
\label{eq:fkeff2}
\end{equation}
The oscillating KK modes force the frequency $f_{k\, \mathrm{eff}}^2$ itself to
oscillate in time, and when $f_{k\, \mathrm{eff}}^2$ becomes similar to the
KK mode frequency~$\sim m_{\mathrm{KK}}$, the field
fluctuation~$u_{\boldsymbol{k}}$ experiences parametric resonance. 
The weak resonance discussed in
Section~\ref{sec:smallKK} corresponds to tiny oscillations of $f_{k\,
\mathrm{eff}}^2$, and the strong resonance happens when 
$f_{k\, \mathrm{eff}}^2$ oscillates considerably.

\begin{figure}[htbp]
 \begin{minipage}{.48\linewidth}
  \begin{center}
 \includegraphics[width=\linewidth]{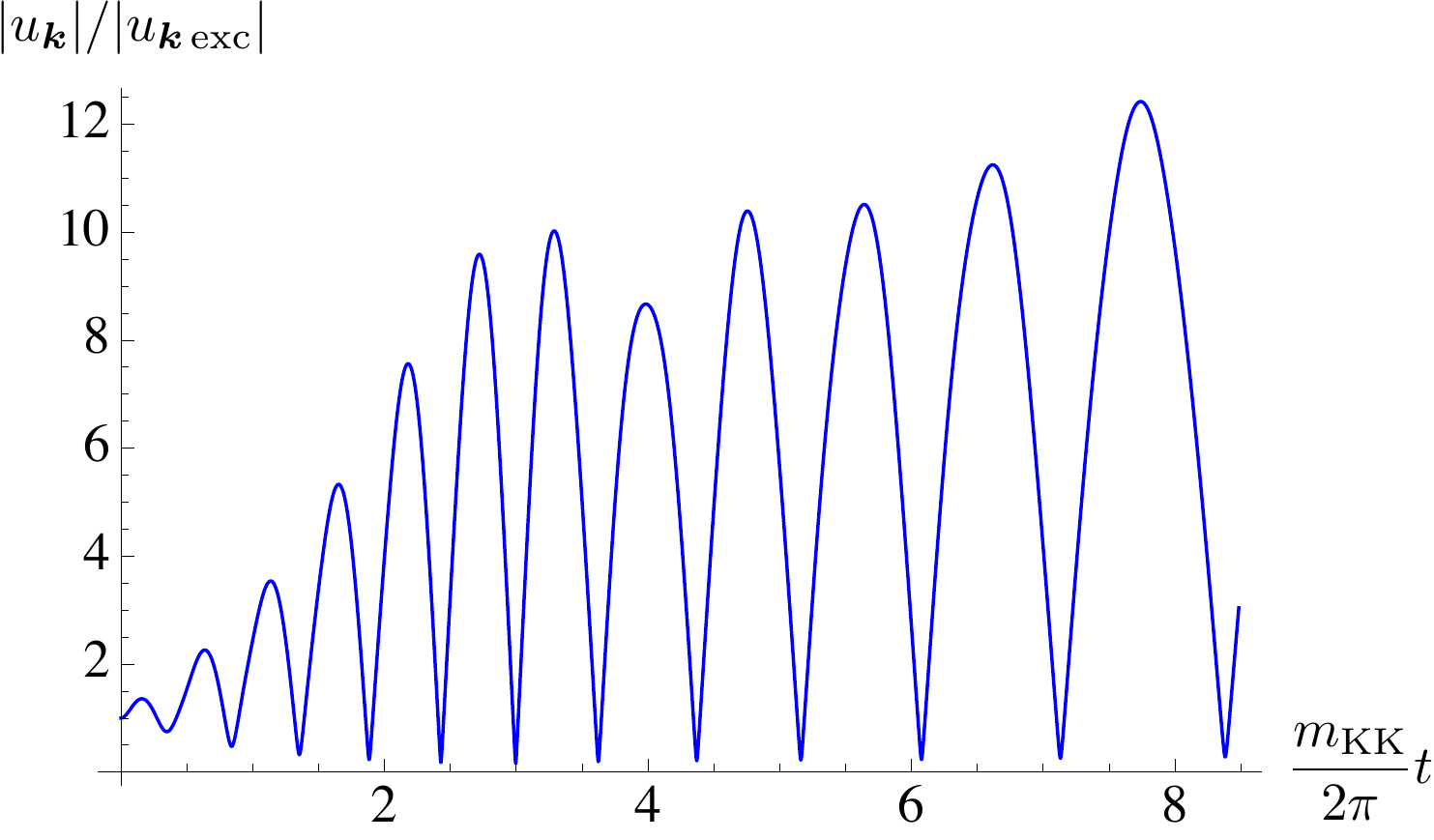}
  \end{center}
 \end{minipage} 
 \begin{minipage}{0.01\linewidth} 
  \begin{center}
  \end{center}
 \end{minipage} 
 \begin{minipage}{.48\linewidth}
  \begin{center}
 \includegraphics[width=\linewidth]{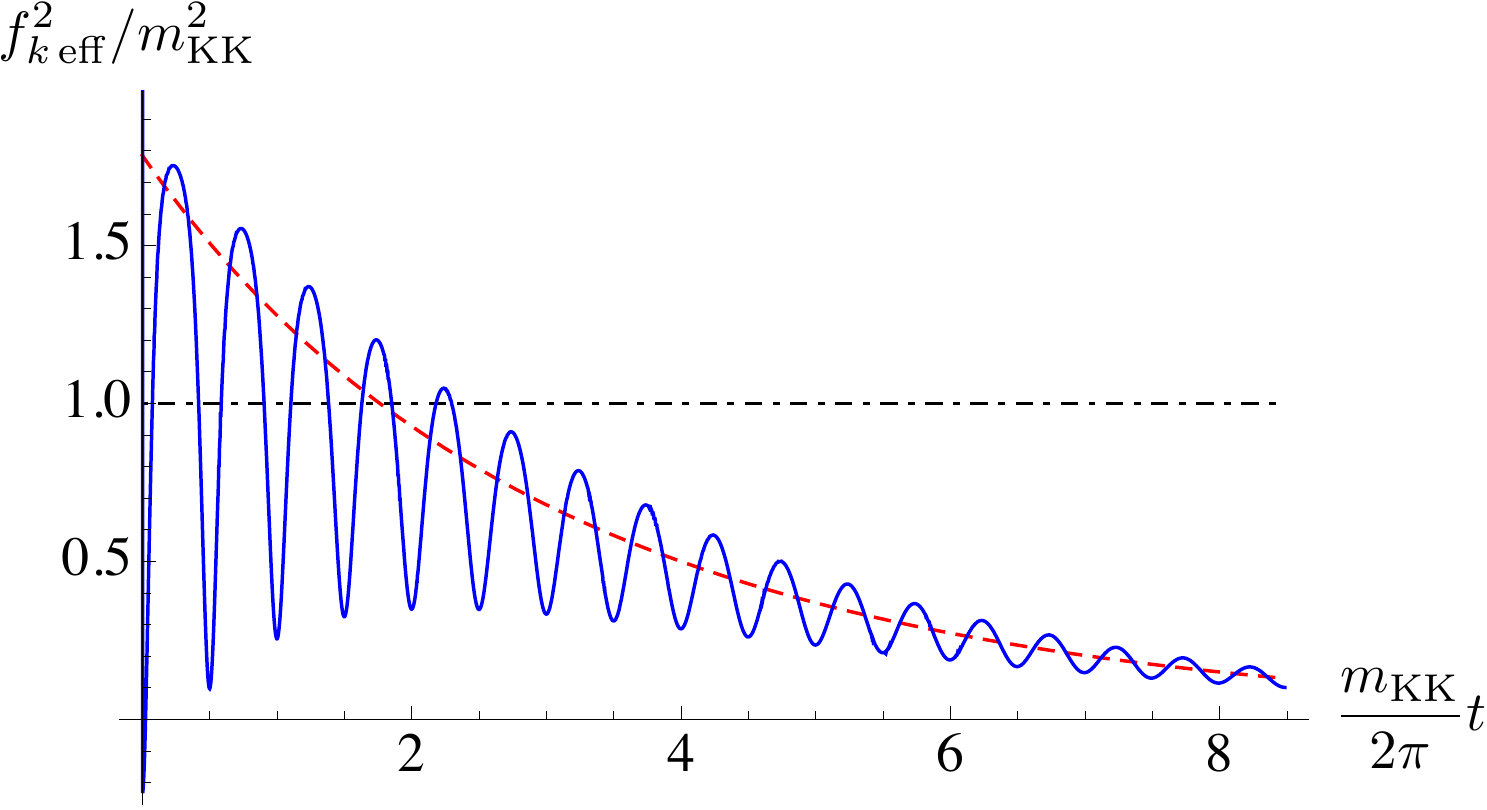}
  \end{center}
 \end{minipage} 
  \caption{Time evolution of the inflaton field
 fluctuation~$u_{\boldsymbol{k}}$ (left figure) and its effective frequency
 squared (right figure), for $\tilde{\psi}_{1\, \mathrm{exc}} \approx
 0.033 M_p$ and $k \approx 57 k_p$. The time is shown in units of the
 oscillation period of the KK mode $2 \pi / m_{\mathrm{KK}}$, and the KK
 mode is excited at $t=0$.} 
  \label{fig:fkeff2}
\end{figure}

In Figure~\ref{fig:fkeff2} we plot the time evolution of the field
fluctuation~$u_{\boldsymbol{k}}$ and its effective frequency 
$f_{k\, \mathrm{eff}}^2$ for the case of 
$\tilde{\psi}_{1\, \mathrm{exc}} \approx  0.033 M_p$ and
$k \approx 57 k_p$. This wave number is where the resonant spike is
peaked at in Figure~\ref{fig:KK30}.
The time is shown in units of the KK mode oscillation
period $2\pi / m_{\mathrm{KK}} \simeq \pi \phi_{\mathrm{exc}} / \alpha_1
V_{\mathrm{exc}}^{1/2}$,\footnote{Now we are beyond the small KK
approximation, but the KK mode $\psi_1$ still oscillates with period
$\simeq 2\pi / m_{\mathrm{KK}} \simeq \pi \phi_{\mathrm{exc}} / \alpha_1 
V_{\mathrm{exc}}^{1/2}$.} therefore is equivalent to the number of
oscillations of~$\psi_1$. Here, recall that the KK modes
affect $f_{k\, \mathrm{eff}}^2$ through quadratic terms
(see e.g.~(\ref{Bphiphi})), hence $f_{k\, \mathrm{eff}}^2$ oscillates
with half the period of the KK mode. The KK mode is excited at $t=0$ in
the figures. In the left figure we show the growth of the fluctuation
$|u_{\boldsymbol{k}}|$ relative to its value at the KK
excitation~$|u_{\boldsymbol{k}\, \mathrm{exc}}|$. 
In the right figure, the blue solid line denotes the 
ratio~$f_{k\, \mathrm{eff}}^2 / m_{\mathrm{KK}}^2$ between the effective
frequency of $u_{\boldsymbol{k}}$ and the KK mass. 
Moreover, the red dashed line shows $k^2 / a^2 m_{\mathrm{KK}}^2$, while the
black dot-dashed line at unity represents the KK mass squared.
We remark that the displayed wave mode~$k \approx 57 k_p$ is inside the
horizon in the plotted time range.

In the absence of the KK modes, the effective frequency~$f_{k\,
\mathrm{eff}}^2 $ traces $k^2  /a^2$ when the wave mode is inside the
horizon. This is still the case for small KK excitations triggering 
weak resonance discussed in Section~\ref{sec:smallKK}.
However when the KK modes are largely excited, 
$f_{k\, \mathrm{eff}}^2 $ considerably oscillates and can even become
negative. From the figures, one sees that $u_{\boldsymbol{k}}$ is
enhanced by an order of magnitude soon after the KK mode excitation,
while $f_{k\, \mathrm{eff}}^2 $ wildly oscillates
around~$m_{\mathrm{KK}}^2$. 
The fluctuation~$u_{\boldsymbol{k}}$ initially oscillates with twice the
frequency of the KK mode (i.e. the period matches with the tics on the
time axis), and is enhanced at each oscillation.
As $f_{k\, \mathrm{eff}}^2 $ becomes smaller than $m_{\mathrm{KK}}^2$,
parametric resonance ceases and $|u_{\boldsymbol{k}}|$ starts to
increase slowly in time as $\propto \sim 
a^{1/2}$. (Note that, in the absence of the KK modes,
$|u_{\boldsymbol{k}}|$ scales proportional to $a^{1/2}$ when inside the
horizon, and as $a^{3/2}$ after horizon exit, as can be seen from its
definition (\ref{defuk}) and (\ref{qkzero}).)
Let us also remark that the parametric resonance of the inflaton field
fluctuation shares common features with
preheating~\cite{Kofman:1994rk,Kofman:1997yn}.
We have studied that parametric resonance happens for $u_{\boldsymbol{k}}$ 
when its effective frequency~$f_{k\, \mathrm{eff}}^2$
oscillates with frequency similar to itself, hence in this sense,
both the weak and strong resonance is analogous to the narrow resonance
in preheating.

In Figures~\ref{fig:KK1000} - \ref{fig:KK30} we have seen that the
resonant peak in the perturbation spectrum slightly shift towards
larger~$k$ for stronger resonance. This can be understood from the
oscillating $f_{k\, \mathrm{eff}}^2 $ mostly taking values that are 
smaller than $k^2  /a^2$. In other words, the characteristic frequency of
$u_{\boldsymbol{k}}$ effectively becomes smaller than $k / a$, and
therefore the resonant features shift towards (slightly) larger wave numbers.

The oscillations of the effective frequency~$f_{k\, \mathrm{eff}}^2 $
are quickly suppressed as the KK mode oscillations damp away. 
The wave modes that enter the resonance band while the oscillations are
large experience strong parametric resonance, and moreover, such resonant
effects depend non-linearly on the KK amplitude (squared). 
This is why the strong enhancement of the curvature perturbations can be
localized to a narrow wave number range, forming spikes in the
perturbation spectrum.

\subsection{Condition for Strong Resonance}

For the set of parameters chosen in Section~\ref{subsec:CPS}, 
the oscillation of the effective frequency~$f_{k\, \mathrm{eff}}^2
$~(\ref{eq:fkeff2}) is dominated by its components
\begin{equation}
  \frac{2 L_A G^A_{\phi \phi}}{B_{\phi \phi}} \frac{k^2}{a^2}
 + \frac{1}{4} \left(\frac{\dot{B_{\phi \phi}}}{B_{\phi \phi}}\right)^2 -
 \frac{1}{2} \frac{\ddot{B_{\phi \phi}}}{B_{\phi \phi}},
\label{rough6.7}
\end{equation}
and furthermore, by the same terms that were important for weak resonance,
i.e., the terms explicitly written in (\ref{Bphiphi}), (\ref{4949}), and
(\ref{LAGAphiphi}). 
Recall that these terms originate from the three kinetic
couplings~(\ref{threekings}) in the inflaton Lagrangian.  
Here we focus on those terms and quantitatively discuss the strong parametric
resonance. We note that our aim here is to give a rough estimate on when
strong resonance happens, and in this subsection we use some identities
without proof.

As a rule of thumb, we consider strong resonance to happen for wave
modes in the resonance band if $f_{k\, \mathrm{eff}}^2 $ oscillates with
amplitude as large as $k^2 / a^2$. 
This happens if, for e.g., $B_{\phi \phi}$ in the denominator of the
first term in (\ref{rough6.7}) oscillates substantially.
Using $\psi_n \simeq \tilde{\psi}_n \cos (2 \alpha_n V^{1/2} t / \phi +
\cdots)$ and 
$\dot{\psi}_n \simeq -2 \alpha_n V^{1/2}  \tilde{\psi}_n \sin (2
\alpha_n V^{1/2} t / \phi + \cdots)$, then one finds from
(\ref{Bphiphi}) that $B_{\phi \phi}$ oscillates with amplitude of
order unity or larger for 
(for simplicity we consider a single KK mode excitation),
\begin{equation}
 0.1 \lesssim \alpha_n^2 \frac{\tilde{\psi}_{n\,
  \mathrm{exc}}^2}{\phi_{\mathrm{exc}}^2} = 
  \tilde{\eta}_{n\, \mathrm{exc}} \xi_{n\, \mathrm{exc}}.
\label{srcond}
\end{equation}
One can further check that the other terms in (\ref{rough6.7}) also 
oscillate $f_{k\, \mathrm{eff}}^2 $ with amplitude $ \sim k^2 / a^2$
for similar KK amplitudes while in the resonance band. Therefore, we can consider
(\ref{srcond}) as the rough indicator for strong resonance to happen. 

A wave mode undergoes strong resonance if it approaches the resonance
band (i.e. $k^2 / a^2 \sim m_{\mathrm{KK}}^2$) while (\ref{srcond})
holds. Denoting the largest (smallest) wave number that experiences
strong resonance by $k_{\mathrm{max}}$ ($k_{\mathrm{min}}$), then from
$\tilde{\eta}_n \xi_n \propto \sim a^{-3}$ one can estimate the ratio
\begin{equation}
 \frac{k_{\mathrm{max}}}{k_{\mathrm{min}}} \sim
 \left(10 \tilde{\eta}_{n\, \mathrm{exc}} \xi_{n\, \mathrm{exc}}\right)^{1/3}.
\end{equation}
Therefore, in the regime~(\ref{3.23}) where one can neglect the cubic or
higher order KK terms in the Lagrangian, the wave interval over which 
strongly resonant features arise is quite narrow. Moreover, since the 
strong resonant amplification non-linearly depends on the KK amplitudes, the
resonant features can further have hierarchical structures.

\section{Conclusions}
\label{sec:Conc}

Wrapped brane inflation necessarily possesses infinite KK degrees of
freedom in addition to the zero-mode inflaton, and 
the brane's Nambu-Goto action naturally gives kinetic couplings among
them. In this paper, we have investigated resonant amplification of the curvature
perturbations triggered by the KK oscillations during inflation.
We found that the resonant signals in the perturbation
spectrum can be localized to narrow wave number ranges, in contrast to
previous studies where signals were normally distributed over a rather
wide $k$-range.
In the effective four dimensional theory, the zero-mode inflaton
couples to the heavy KK modes which are excited and oscillates during
inflation. Small KK excitations lead to weak resonance, sourcing oscillatory
features in the perturbation spectrum (cf. Figure~\ref{fig:schematic},
\ref{fig:KK1000}) whose oscillation amplitude is proportional to the
excited KK amplitude squared. 
For larger KK excitations, the resonant amplification becomes extremely
efficient with non-linear dependence on the KK amplitude. 
We saw that such strong resonant effects damp more rapidly than the KK oscillations
during inflation, therefore leave spiky features localized to a narrow
wave number range in the perturbation spectrum (cf. Figure~\ref{fig:KK30}).
Both weak and strong resonant effects were mainly sourced by the
couplings between the inflaton kinetic terms and the oscillating KK modes.

Motivated by the string theory construction of~\cite{Silverstein:2008sg}, 
we have supposed the brane tension to be the main driving force for
inflation and considered only the Nambu-Goto action. However we note
that, even if there are further contributions to the brane potential, 
the Nambu-Goto action gives the kinetic terms that contain the important
interactions with the KK modes. Therefore we expect the resonant
effects studied in this paper to be rather generic features for
inflation driven by extended sources. 
From an effective theory point of view, it would also be
interesting to systematically examine resonant effects arising from
kinetic couplings in general. 

We have restricted ourselves to KK modes that are not
excited too largely and thus studied up to quadratic KK interactions in 
the Lagrangian. Some region of the strong resonance was
close to the regime where the higher order KK terms can become
important, thus it is of interest to analyze effects from higher order
interactions. Especially when going beyond the strong resonance
regime we have studied, the tower of KK interactions may further enhance
the resonant signals. 

The wrapped brane model provides an example where the kinetic couplings
between the inflaton and the heavy KK modes produce sharp features in
the curvature perturbation spectrum, and to that end we have focused on the
evolution of the perturbations under the KK oscillations. 
Although the KK modes can be excited as the wrapped inflaton brane passes by
structures or localized sources along the inflationary
trajectory,\footnote{See also~\cite{Mukohyama:1997bd} which discusses KK
excitations due to oscillations in the compactification scale of the
extra dimension.}
we have not yet specified the excitation mechanism in detail.
It would be very interesting to study the whole picture including
dynamical excitation of the KK modes. One can imagine that
the KK excitation is triggered as the inflaton comes across some point(s)
in field space. Then the KK excitation itself may become inhomogeneous
and give further contributions to the curvature perturbations. 
It could also be that the KK modes are repeatedly excited during inflation,
if, for example, the trajectory of the wrapped brane is a repeated circuit as in
monodromy models~\cite{Silverstein:2008sg,McAllister:2008hb}. 
In such case, the spikes/oscillations would repeatedly show up in the
perturbation spectrum with the interval corresponding to the size of the
inflaton circuit.
Let us further mention that, if the KK excitations are triggered by
another brane along the inflationary trajectory in string theory models,
then as the branes encounter each other, 
strings stretched between them can become massless and
be produced~\cite{Green:2009ds}. Such effects can slow
down the inflaton brane, and may further generate oscillations/bumps in
the perturbation spectrum by temporarily affecting the inflaton 
dynamics, and also by the produced strings (or particles in the
effective theory) rescattering off the inflaton condensate, see
e.g. \cite{Romano:2008rr,Barnaby:2009mc,Barnaby:2009dd,Barnaby:2010ke}. 
However, these signals are present for wide ranges of the wave number,
thus can be distinguished from the sharp spikes sourced by the KK oscillations.
We also note that, when constructing explicit models for wrapped brane
inflation, the fate of the excited KK modes need to be taken into
account, especially in relation to the moduli
problem~\cite{Coughlan:1983ci,Banks:1993en,de Carlos:1993jw}. 
(See also~\cite{Dufaux:2008br} for discussions in this direction.)

Interestingly, the resonant signals in the curvature perturbations are
characterized by the microphysics of the KK modes, which is tied to the 
properties of the internal manifold. The height of the spikes/oscillations are
basically determined by the excited amplitude of the KK modes, while the
width and oscillation period of the features in $k$-space are set by the KK
mass corresponding to the size of the wrapped cycle. The positions of the
resonant signals in $k$-space tell us when during inflation the KK
modes were excited (possibly denoting the positions of localized sources
in the internal manifold), and furthermore, the KK tower generates signals
periodic in wave number~$k$, cf. Figure~\ref{fig:two-KK}. All these
features open up possibilities for probing the extra dimensional space
through cosmological experiments. It will also be interesting to study
how well we may detect and measure these resonant signals through
observations of the cosmic microwave background (CMB) and large scale
structure.  The oscillatory features from weak resonance may be
somewhat smoothed out by the CMB transfer function (as is discussed in,
e.g.~\cite{Chen:2012ja}),   
but the spikes from strong resonance are expected to leave distinct imprints in
the CMB temperature anisotropy. We leave to future work a detailed
analysis of the observational consequences of weak/strong resonance
during inflation. We should also mention that the resonant signals can
show up in the non-Gaussian signals as well, as discussed
in~\cite{Flauger:2009ab,Flauger:2010ja,Behbahani:2011it}. 
It is quite possible that the strong resonance which sources spikes in
the power spectrum also leave violent marks on the higher order
correlation functions.

\section*{Acknowledgements}

We thank Masahiro Nakashima for many discussions and initial
collaboration. TK would also like to thank Dick Bond, Jonathan
Braden, A.~Emir~G\"umr\"uk\c{c}\"uo\u{g}lu, Amir Hajian, Shinji
Mukohyama, Paniez Paykari, Sohrab Rahvar, Christophe Ringeval, Ryo
Saito, Alexei A. Starobinsky, and Yi Wang for helpful discussions.


\appendix

\section{Primordial Density Perturbations from Multi-Field Inflation
 with Various Kinetic Terms}
\label{sec:app}

In this appendix we consider density perturbations from multi-field
inflation with an action containing various forms of kinetic terms.
We make use of the $\delta
\mathcal{N}$-formalism~\cite{Starobinsky:1986fxa,Sasaki:1995aw,Wands:2000dp,Lyth:2004gb}
for obtaining the density perturbations, so here we simply derive the
action of the field fluctuations up to quadratic order. The calculations
are an extension of the previous works discussing inflationary models
with multi-fields and/or non-canonical kinetic terms,
e.g. \cite{Maldacena:2002vr,Seery:2005wm,Seery:2005gb,Chen:2006nt,Langlois:2008wt}. 
We also refer the reader to~\cite{Langlois:2008qf,Arroja:2008yy} which
discuss similar Lagrangians as in this appendix.

The action we consider is of the following form:
\begin{equation}
 S = \int d^4 x \sqrt{-g} \left\{
\frac{M_p^2}{2}  R + L \left(\varphi^I, \, X^A \right)
\right\},
\label{SinAppendix}
\end{equation}
where $\varphi^I$ denotes the fields (labeled by~$I$), 
and $X^A$ the kinetic terms,
\begin{equation}
 X^A \equiv -G^A_{IJ} g^{\mu \nu} \partial_\mu \varphi^I \partial_\nu \varphi^J.
\end{equation}
$G^A_{IJ}$ is the field space metric, where $A$ labels different
metrics, i.e. different forms of kinetic terms. 
We impose $G^A_{IJ} = G^A_{JI}$, and consider $G^A_{IJ}$
to be constants (in other words, independent of $\varphi^I$).
Note that in this appendix we label fields by $I, J, \cdots$, and field metrics by
$A, B, \cdots$. 
Moreover, we express partial derivatives in terms of $\varphi^I $ and
$X^A$ as, respectively,
\begin{equation}
 L_I \equiv \frac{\partial L}{\partial \varphi^I}, \quad
 L_A \equiv \frac{\partial L}{\partial X^A}.
\end{equation}

Then one can derive the energy-momentum tensor,
\begin{equation}
 T_{\mu\nu}  = 
g_{\mu\nu} L - 2 \frac{\partial L}{\partial g^{\mu\nu}} 
  = g_{\mu\nu} L + 2 L_A G^A_{IJ} \partial_\mu \varphi^I \partial_\nu
 \varphi^J,
\end{equation}
as well as the equation of motion of $\varphi^I$,
\begin{equation}
 L_I + \frac{2}{\sqrt{-g}}
\partial_\mu \left(
\sqrt{-g} L_A G^A_{IJ} g^{\mu\nu} \partial_\nu \varphi^J 
\right)
= 0.
\label{varphiEoM}
\end{equation}

\subsection{Homogeneous Background}

First let us derive the equations of motion of the homogeneous
background. We fix the background metric to a flat FRW:
\begin{equation}
 ds^2 = -dt^2 + a^2 (t) \delta_{ij} dx^i dx^j,
\end{equation}
where $i, j = 1,2,3$ run over the spatial directions. 

Then the Einstein equations are
\begin{equation}
 3 M_p^2 H^2 = -L + 2 L_A G^A_{IJ} 
\dot{\varphi}^I \dot{\varphi}^J ,
\label{eins7}
\end{equation}
\begin{equation}
  -M_p^2 (2 \dot{H} + 3 H^2) = L,
\label{eins8}
\end{equation}
where an overdot denotes a time derivative, and $ H = \dot{a} / a$.

The equation of motion of $\varphi^I$ (\ref{varphiEoM}) now takes the
form
\begin{equation}
 L_I - 2 \left(L_A G^A_{IJ} \dot{\varphi}^J \right)^{\cdot} - 6 H L_A
  G^A_{IJ} \dot{\varphi}^J = 0.
\end{equation}

\subsection{Field Fluctuations}

We then study fluctuations around the homogeneous background. We
adopt the ADM formalism,
\begin{equation}
 ds^2 = -N^2 dt^2 + h_{ij} (dx^i + N^i dt) (dx^j + N^j dt),
\end{equation}
under which the action (\ref{SinAppendix}) is rewritten as (we take $N >0$): 
\begin{equation}
 S = \int dt d^3 x N \sqrt{h}
 \left[
\frac{M_p^2}{2}\left\{
R^{(3)} + \frac{1}{N^2} \left( E_{ij} E^{ij} - E^2\right) 
\right\} + L
\right].
\end{equation}
Here $h \equiv \det (h_{ij})$, and $R^{(3)}$ is the scalar curvature on the $t
= \mathrm{const.}$ spatial hypersurface.
The symmetric tensor~$E_{ij}$ is defined as 
\begin{equation}
 E_{ij} \equiv \frac{1}{2} \left( \dot{h}_{ij} - \nabla_i N_j - \nabla_j
			   N_i\right),
\end{equation}
where $\nabla_i$ is a derivative associated with $h_{ij}$, and $E =
h^{ij} E_{ij }$. We also note
that the indices $i,j$ are raised and lowered by $h^{ij}$ and
$h_{ij}$, respectively. $X^A$ is now expressed as
\begin{equation}
 X^A = -G^A_{IJ} \left\{
-\frac{1}{N^2} \dot{\varphi}^I \dot{\varphi}^J + 
 \left( h^{ij} - \frac{N^i N^j}{N^2}\right) \partial_i \varphi^I
 \partial_j \varphi^J
 + 2 \frac{N^i}{N^2} \dot{\varphi}^I \partial_i \varphi^J
\right\}.
\end{equation}

\vspace{\baselineskip}

Hereafter we take uniform curvature slicing, such that $t =
\mathrm{const.}$ slices have vanishing Ricci curvature, thus
\begin{equation}
 h_{ij} = a(t)^2 \delta{ij}, \qquad
 R^{(3)}=0, \qquad
 \nabla_i \to \partial_i,
\end{equation}
and the action becomes
\begin{equation}
 S = \int dt d^3x \sqrt{h} \left[
\frac{M_p^2}{2N}
\left( E_{ij} E^{ij} - E^2 \right) + N L
\right].
\label{Sunicurv}
\end{equation}
We will focus on the field fluctuations $Q^I$ defined as 
\begin{equation}
 \varphi^I = \varphi^I_0 + Q^I
\end{equation}
where $\varphi_0^I$ is the homogeneous classical background, 
and later on convert them into the curvature perturbations using the $\delta
\mathcal{N}$-formalism (\ref{deltaNform}).

The lapse~$N$ and shift~$N^i$ are Lagrange multipliers in the
action~(\ref{Sunicurv}), hence their equations of motion can be used as
constraints, which are, respectively,
\begin{equation}
 N^2 L = \frac{M_p^2}{2} \left(E_{ij} E^{ij} - E^2\right) + 2 L_A
  G^A_{IJ} v^I v^J,
\label{equA}
\end{equation}
\begin{equation}
 M_p^2 \partial_j \left\{
\frac{1}{N}\left( E^{ij} - E h^{ij}\right)
\right\}
 = \frac{2}{N}L_A h^{ij} G^A_{IJ} v^I \partial_j \varphi^J.
\label{equB}
\end{equation}
Here we have defined
\begin{equation}
 v^I \equiv \dot{\varphi}^I - N^i \partial_i \varphi^I.
\end{equation}
Let us rewrite the lapse and shift as
\begin{equation}
\begin{split}
 N & = 1 + \alpha, \\
 N_i & = \partial_i \psi + \widetilde{N}_i,
\end{split}
\end{equation}
where $\widetilde{N}_i$ is the incompressible part, i.e.,
\begin{equation}
 h^{ij} \partial_j \widetilde{N}_i = 0.
\end{equation}
We further expand the functions in terms of the field fluctuations~$Q$, 
\begin{equation}
\begin{split}
 \alpha & = \alpha_1 + \alpha_2 + \cdots , \\
 \psi &= \psi_1 + \psi_2 + \cdots , \\
 \widetilde{N}_i & = \widetilde{N}_i^{(1)} + \widetilde{N}_i^{(2)} +
 \cdots ,
\end{split}
\end{equation}
where the numbers represent orders of~$Q$, e.g. $\alpha_n =
\mathcal{O}(Q^n)$. 
We require the constraint equations to hold order by order.

In order to expand the action up to second order in~$Q$, 
one needs to consider the lapse and shift only up to the first order,
i.e. $\alpha_1$, $\psi_1$, and $\widetilde{N}_i^{(1)}$
(cf.~\cite{Maldacena:2002vr,Chen:2006nt}). 
The constraint equation (\ref{equA}) at its zeroth order reproduce the
Friedmann equation~(\ref{eins7}), while its first order is
\begin{multline}
 \frac{2 M_p^2 H}{a^2} \delta^{ij} \partial_i \partial_j \psi_1
= L_I^{(0)} Q^I + L_A^{(0)}G^A_{IJ} \left(-2 \alpha_1
\dot{\varphi}_0^I \dot{\varphi}_0^J + 2 \dot{\varphi}_0^I \dot{Q}^J
 \right)
 + 2 \alpha_1 L^{(0)}  \\
 - 4 L_A^{(0)} G^A_{IJ}\dot{Q}^I \dot{\varphi}_0^J 
 - 2 \left( L_{AI}^{(0)} Q^I + L_{AB}^{(0)}X_1^B \right)
 G^A_{IJ}\dot{\varphi}_0^I \dot{\varphi}_0^J,
\end{multline}
where we have used (\ref{eins7}) upon obtaining this form. The zeroth
order of (\ref{equB}) is trivial, and the first order is
\begin{equation}
 2 \partial_i \left( M_p^2 H \alpha_1 - L_A^{(0)} G^A_{IJ}
	       \dot{\varphi}^I_0 Q^J \right)
 = \frac{M_p^2}{2} h^{jk} \partial_j \partial_k \widetilde{N}_i^{(1)}.
 \label{Bat1}
\end{equation}
Since $h^{ij} \partial_i ( h^{kl} \partial_k \partial_l
\widetilde{N}_j^{(1)}) = 0$, after choosing proper boundary conditions,
one arrives at
\begin{equation}
 h^{ij} \partial_i \partial_j \widetilde{N}_k^{(1)} = 0,
\label{B11}
\end{equation}
\begin{equation}
 M_p^2 H \alpha_1 = L_A^{(0)} G^A_{IJ}\dot{\varphi}_0^I Q^J.
\label{B12}
\end{equation}
We also note that the evolution equation (\ref{eins8}) is
obtained by combining the zeroth order of (\ref{equA}) (i.e. the Friedmann
equation) and the zeroth order equation of motion of $\varphi^I$.

Using the zeroth order of (\ref{equA}), 
the zeroth order equation of motion of $\varphi^I$, and the first order of
(\ref{equB}) (i.e. (\ref{B11}) and (\ref{B12})), one can expand the 
action up to second order in~$Q$, 
\begin{equation}
\begin{split}
 S \triangleq 
 \int d^3 x dt \, &  a^3   \Biggl[ 
-3 M_p^2 H^2 + L^{(0)} 
  - L_A^{(0)} G^A_{IJ}  \frac{1}{a^2} \delta^{ij}
\partial_i Q^I  \partial_j Q^J
\\
 & 
 + \Biggl\{
 \frac{1}{2} L_{IJ}^{(0)} 
+ \frac{2}{M_p^2}\frac{1}{a^3} \left(
 \frac{a^3}{H}L_A^{(0)} L_B^{(0)} G^A_{IK} G^B_{JL} \dot{\varphi}_0^K
 \dot{\varphi}_0^L
\right)^{\cdot} 
\\
&
- \frac{1}{M_p^2 H}L_A^{(0)} \left( L_{BI}^{(0)}  G_{JK}^A  +
 L_{BJ}^{(0)}  G_{IK}^A  \right) X_0^B  \dot{\varphi}_0^K 
\\
& + \frac{2}{M_p^4 H^2} L_{AB}^{(0)}  L_C^{(0)} L_D^{(0)} G_{I
 K}^C G_{JL}^D X_0^A X_0^B \dot{\varphi}_0^K \dot{\varphi}_0^L
\Biggr\} Q^I Q^J
 \\
 & +  \left(
L_A^{(0)} G^A_{IJ} + 2 L_{AB}^{(0)} G_{IK}^A G_{JL}^B \dot{\varphi}_0^K
\dot{\varphi}_0^L 
\right) \dot{Q}^I \dot{Q}^J
\\
 & + \left( 2 L_{AI}^{(0)} G^A_{KJ} \dot{\varphi}_0^K 
 - \frac{4}{M_p^2 H}L_{AB}^{(0)} L_C^{(0)} G_{JK}^A  G_{IL}^C
 X_0^B  \dot{\varphi}_0^K \dot{\varphi}_0^L \right) Q^I \dot{Q}^J 
 + \mathcal{O} (Q)^3
\Biggr],
\label{2ndSQ}
\end{split}
\end{equation}
where ``$\triangleq$'' is used to denote that we have dropped 
total derivatives in the integrand.

Hereafter we omit the sub(super)script~$0$ denoting the homogeneous
background. 
Further introducing 
\begin{equation}
\begin{split}
 M_{IJ} & \equiv L_{IJ} + \frac{4}{M_p^2 a^3}
 \left(\frac{a^3}{H}L_A L_B G^A_{IK} G^B_{JL} \dot{\varphi}^K
 \dot{\varphi}^L  \right)^{\cdot}
 \\
 & \quad - \frac{2}{M_p^2 H}L_A \left( L_{BI}  G_{JK}^A  +
 L_{BJ} G_{IK}^A  \right) X^B  \dot{\varphi}^K 
\\
 & \quad  + \frac{4}{M_p^4 H^2} L_{AB}  L_C L_D G_{I
 K}^C G_{JL}^D X^A X^B \dot{\varphi}^K \dot{\varphi}^L, 
 \\
 B_{IJ} & \equiv 2 L_A G^A_{IJ} + 4 L_{AB} G_{IK}^A G_{JL}^B
 \dot{\varphi}^K \dot{\varphi}^L, \\
 C_{IJ} & \equiv 2 L_{AI} G^A_{KJ} \dot{\varphi}^K
 - \frac{4}{M_p^2 H}L_{AB} L_C G_{JK}^A  G_{IL}^C
 X^B  \dot{\varphi}^K \dot{\varphi}^L,
\label{MBC}
\end{split} 
\end{equation}
(note that $M_{IJ} = M_{JI}$ and $B_{IJ} = B_{JI}$, but $C_{IJ}$ is not
necessarily symmetric), then the second order action for $Q$ can be
rewritten in the following form: 
\begin{multline}
 S_2 \triangleq \int dt d^3 x \, a^3
 \Biggl[
 -\frac{1}{a^2} L_A G^A_{IJ} \partial Q^I \partial Q^J
 + \frac{1}{2} \left\{M_{IJ} - \frac{1}{a^3} \left(a^3 C_{IJ}
					     \right)^{\cdot}  \right\}
 Q^I Q^J
 \\ 
 + \frac{1}{2} B_{IJ} \dot{Q}^I \dot{Q}^J
 + \frac{1}{2} (C_{IJ} - C_{JI}) Q^I \dot{Q}^J
\Biggr],
 \label{S2Q}
\end{multline}
where we use the abbreviated expressions
\begin{equation}
  \partial X \partial Y \equiv \delta^{ij} \partial_i X \partial_j Y,
   \qquad
 \partial^2 X \equiv \delta^{ij} \partial_i \partial_j X.
\end{equation}
The equation of motion for $Q^I$ can be obtained from (\ref{S2Q}) as 
\begin{equation}
0 =  \frac{2}{a^2}L_A G^A_{JI} \partial^2 Q^J + M_{JI} Q^J 
 + C_{IJ} \dot{Q}^J 
 - \frac{1}{a^3}  \left\{
a^3 \left(B_{JI} \dot{Q}^J + C_{JI} Q^J \right)
\right\}^{\cdot}.
 \label{EoMQ}
\end{equation}

The results can also be expressed in terms of the conformal time
\begin{equation}
 d t = a \, d\tau.
\end{equation}
Redefining the field as 
\begin{equation}
 q^I \equiv a\, Q^I,
\end{equation}
and further Fourier expanding $q^I$,
\begin{equation}
 q^I (\tau , \boldsymbol{x}) = \frac{1}{ (2 \pi)^3} 
 \int d^3 k\,  e^{-i \boldsymbol{k \cdot x}}
 \,  q_{\boldsymbol{k}}^I (\tau), 
\end{equation}
then the equation of motion (\ref{EoMQ}) can be rewritten as
\begin{multline}
0 =  B_{JI} q_{\boldsymbol{k}}''^{J} + 
\left( C_{JI} - C_{IJ} + \dot{B}_{JI} \right) a q_{\boldsymbol{k}}'^J
 \\
 + \left\{
 2 L_A G^A_{JI} k^2 - B_{JI} \frac{a''}{a} 
 + \left( -M_{JI} + \dot{C}_{JI}  \right) a^2
 + \left( -\dot{B}_{JI} + 2 C_{JI} + C_{IJ} \right) a'
\right\} q_{\boldsymbol{k}}^I .
 \label{EoMq}
\end{multline}
Here $k = |\boldsymbol{k} |$, and a prime denotes a derivative with
respect to the conformal time~$\tau$. 

\vspace{\baselineskip}

These are the main results of this appendix, which can be used to obtain
the field fluctuations after choosing appropriate initial conditions (which
corresponds to choosing the vacuum state).
Then one can calculate the resulting curvature perturbations using the $\delta
\mathcal{N}$-formalism~\cite{Starobinsky:1986fxa,Sasaki:1995aw,Wands:2000dp,Lyth:2004gb},
\begin{equation}
 \zeta = \mathcal{N}_I Q^I + \frac{1}{2} \mathcal{N}_{IJ} Q^I Q^J +
  \cdots,
 \label{deltaNform}
\end{equation}
where $\mathcal{N}$ is the number of e-folds between an initial flat
hypersurface and a final uniform density hypersurface.
The right hand side of (\ref{deltaNform}) can be computed at any time
after the separate universe picture becomes a good approximation,
provided that there are no isocurvature perturbations sourcing
further~$\delta \mathcal{N}$. 
The computations of the curvature perturbations using (\ref{EoMq})
are carried out in detail in Section~\ref{sec:CPcalc},
focusing on the specific action of (\ref{Lphi}).

\section{Homogeneous Functions in the Field Fluctuation Action}
\label{app:B}

Here we write down the forms of the homogeneous functions defined
in Appendix~\ref{sec:app} (at e.g. (\ref{MBC})), when applied to the
action~(\ref{Lphi}). They show up in the second order action of the
field fluctuation~(\ref{S2q}). 

Expressing the kinetic terms as
\begin{equation}
 X \equiv  -(\partial \phi)^2, \qquad
 \widetilde{X} \equiv -\sum_{n \neq 0} (\partial \psi_n)^2, \qquad 
 X_n \equiv   -(\partial \phi \cdot \partial \psi_n ) 
 \quad (\mathrm{for}\, \, n \neq 0),
\end{equation}
then the Lagrangian (\ref{Lphi}) can be written as  $ \mathcal{L} =
\sqrt{-g} L $ where
\begin{multline}
L = -V \left( \frac{1}{\gamma} + 2 \gamma \sum_{n \neq 0} \alpha_n^2
	 \frac{\psi_n^2}{\phi^2} \right)
 + \gamma \left(
\frac{1}{2} \widetilde{X} - \frac{1}{6} \sum_{n \neq 0}
\frac{\psi_n^2}{\phi^2} X + \frac{1}{3} \sum_{n \neq 0}
\frac{\psi_n}{\phi}X_n
\right)
 \\
+ 
\frac{\gamma^3}{2 V} \sum_{n \neq 0}^{ } \left( X_n^2  
+ \frac{1}{9} \frac{\psi_n^2 }{\phi^2 } X^2
- \frac{2}{3} \frac{\psi_n}{\phi }X X_n 
\right),
\end{multline}
with
\begin{equation}
 \gamma = \left(1 - \frac{X}{V}\right)^{-1/2}.
\end{equation}
With the above~$L$, the functions in the fluctuation equations are
expressed as follows: 
\begin{equation}
 L_A G_{\phi\phi}^A = \frac{\partial L}{\partial X} ,
\label{eqB4}
\end{equation}
\begin{equation}
 B_{\phi \phi} = 2 \frac{\partial L}{\partial X}
 + 4  \frac{\partial^2 L}{\partial X^2}\dot{\phi}^2  
+ 4  \sum_{n \neq 0}
\frac{\partial^2 L}{\partial X_n \partial X} \dot{\phi} \dot{\psi_n}
+ \sum_{m, n \neq 0}
\frac{\partial^2 L}{\partial X_m \partial X_n} \dot{\psi_m}
\dot{\psi_n},
\label{eqB5}
\end{equation}
\begin{equation}
\begin{split}
 M_{\phi \phi} & = \frac{\partial^2 L}{\partial \phi^2} + 
 \frac{4 }{M_p^2 a^3} 
 \left\{ \frac{a^3}{H} \left(  
 \frac{\partial L}{\partial X} \dot{\phi} + \frac{1}{2} \sum_{n \neq 0} 
 \frac{\partial L}{\partial X_n} \dot{\psi}_n
\right)^2 \right\}^{\cdot}
\\
& \quad -\frac{4}{M_p^2 H} 
\left(\frac{\partial L}{\partial X}\dot{\phi } + \frac{1}{2}
  \sum_{n \neq 0}\frac{\partial L}{\partial X_n}\dot{\psi}_n \right)
 \cdot \sum_{ X^A} \frac{\partial^2 L}{\partial X^A \partial \phi } X^A
\\
& \quad + \frac{4}{M_p^4 H^2} 
\left(\frac{\partial L}{\partial X}\dot{\phi } + \frac{1}{2}
  \sum_{n \neq 0}\frac{\partial L}{\partial X_n}\dot{\psi}_n \right)^2
 \cdot \sum_{ X^A,\,  X^B} \frac{\partial^2 L}{\partial X^A \partial X^B
 } X^A X^B,
\label{eqB6}
\end{split}
\end{equation}
\begin{equation}
\begin{split}
 C_{\phi \phi}  &= 2 \frac{\partial^2 L}{\partial \phi \partial
  X}\dot{\phi} + 
  \sum_{n \neq 0} \frac{\partial^2 L}{\partial \phi
  \partial X_n} \dot{\psi}_n
\\
& \quad -\frac{4}{M_p^2 H} 
 \left(\frac{\partial L}{\partial X}\dot{\phi } + \frac{1}{2}
  \sum_{n \neq 0}\frac{\partial L}{\partial X_n}\dot{\psi}_n \right)
\cdot \sum_{X^A} \left\{
\frac{\partial^2 L }{\partial X^A \partial X} X^A  \dot{\phi } +
\frac{1}{2} \sum_{m\neq 0}\frac{\partial^2 L }{\partial X^A \partial
 X_m} X^A  \dot{\psi}_m 
\right\},
\label{eqB7}
\end{split}
\end{equation}
where the sum $\sum_{X^A}$ runs over $X$, $\widetilde{X}$, and $X_n$
($n\neq 0$).

\section{Computation of Weak Resonance from Oscillatory Sources}
\label{app:c}

In this appendix we give detailed computations of weak resonance from
small KK excitations. We focus on the equation~(\ref{sk1eqsimp}) and
study the inflaton field fluctuations induced by oscillatory source
terms. Solutions of the evolution equation of the form~(\ref{tanjun}):
\begin{equation}
 -s_{\boldsymbol{k}}''^{(1)}(\tau) + 2 i k
  s_{\boldsymbol{k}}'^{(1)}(\tau) 
 = i A \sin (\omega \tau + \Theta)
\label{eq153}
\end{equation}
are analyzed, where $k$, $\omega$, and $\Theta$ are real
(here $k$ and $\omega$ are positive), while $A$ is a complex number. 
We start our discussions by supposing that these parameters are constants.
We arbitrarily set the initial conditions at a certain time~$\tau_*$ as
\begin{equation}
 s_{\boldsymbol{k}}^{(1)}(\tau_*) = s_{*}, \qquad
 s_{\boldsymbol{k}}'^{(1)}(\tau_*) = s_{*}'.
\label{sstar}
\end{equation}
The homogeneous solution of (\ref{eq153}) is (\ref{homo}), and the
particular solutions are given in (\ref{part}) and (\ref{partamp+}).

\subsection{Solution for $2k \neq \omega$}
\label{subsec:C.1}

When $2k \neq \omega$,
the solution of (\ref{eq153}) with initial conditions (\ref{sstar}) is
\begin{multline}
 s_{\boldsymbol{k}}^{(1)}(\tau) = \frac{A}{\omega^2 - 4 k^2}
 \Biggl[
\frac{2k}{\omega} \cos (\omega \tau + \Theta)
 + i \sin (\omega \tau + \Theta) 
- \left( \frac{2k}{\omega} - \frac{\omega}{2k}\right)
 \cos  (\omega \tau_* + \Theta)
\\
- \left\{
 i \sin (\omega \tau_* + \Theta)
  +  \frac{\omega}{2k} \cos (\omega \tau_* + \Theta)
\right\}
 e^{2 i k (\tau - \tau_*)}
\Biggr]
 + s_*
 + \frac{i}{2k}s_*'
\left\{
1 -  e^{2 i k (\tau - \tau_*)} 
\right\}.
\label{skexpsol}
\end{multline}
Especially when $s_* = s_*' = 0$ and further $2 k \gg \omega$, then the
solution is\footnote{Here we use $x = \mathcal{O}(y)$ for a complex
variable~$x$ to denote $|x| \lesssim |y|$.}
\begin{multline}
 s_{\boldsymbol{k}}^{(1)}(\tau) = 
-\frac{A}{2 k \omega}
 \Biggl[
 \cos (\omega \tau + \Theta) 
 + i \frac{\omega}{2 k}\sin (\omega \tau + \Theta)
- \cos (\omega    \tau_* + \Theta) 
 \\
 - i \frac{\omega}{2 k}\sin (\omega \tau_* + \Theta) e^{2
 i k (\tau - \tau_*)}  + \mathcal{O} \left(\frac{\omega^2 }{k^2}  \right)
\Biggr],
\label{eq156}
\end{multline}
while $s_* = s_*' = 0$ and $2 k \ll \omega$ give
\begin{equation}
 s_{\boldsymbol{k}}^{(1)}(\tau) = 
 \frac{A}{2 k \omega }
 \Biggl[  
 \cos (\omega \tau_* +  \Theta) 
 \left\{ 1 - e^{2 i k ( \tau - \tau_*)} \right\} 
+ \mathcal{O} \left( \frac{k}{\omega  }\right)
\Biggr].
\label{eq157}
\end{equation}
One sees from (\ref{eq156}) and (\ref{eq157}) that the system 
basically chooses the smaller of the two frequencies $k/\pi$ and $\omega/2 \pi$.
This can also be understood as follows: 
In the case $2 k \gg \omega$, the particular solution (\ref{part}) mainly
determines the oscillatory behavior, while for $2 k \ll \omega$
the homogeneous solution (\ref{homo}) dominates.
Therefore, when further taking into account
the time variation of the parameters such as~$A$, then
(\ref{eq156}) shows oscillations with a time-varying amplitude,
while (\ref{eq157}) gives a constant oscillation amplitude. 

However, it should also be noted that the story is more complicated when
focusing only on the real or imaginary part
of~$s_{\boldsymbol{k}}^{(1)}$. 
Especially for the $2 k \gg \omega$ case~(\ref{eq156}), the amplitude of
oscillations with frequency~$\omega/2 \pi$ for
$\mathrm{Re} (s_{\boldsymbol{k}}^{(1)})$ 
changes drastically depending on whether $A$ is real or imaginary.

\subsection{Solution for $2 k = \omega$}
\label{subsec:C.2}

For the resonant case $2 k = \omega$, the solution for arbitrary $s_*$
and $s_*'$ is
\begin{multline}
 s_{\boldsymbol{k}}^{(1)}(\tau) = 
\frac{A}{8 k^2} \Biggl[
 2 i k (\tau - \tau_*)  e^{i (2 k \tau + \Theta)}
-  \cos (2 k \tau + \Theta) 
+ \left\{
 2 - e^{2 i k (\tau - \tau_*)}
\right\}
 \cos (2 k \tau_* + \Theta)
\Biggr]
\\
 + s_* + \frac{i}{2k}s_*'
  \left\{
1 - e^{2 i k (\tau - \tau_*)} 
\right\}.
\label{exsolres}
\end{multline}
The term $2 i k (\tau - \tau_*) e^{i (2 k \tau + \Theta)}$ in the first
line represents parametric resonance, and 
since we have in mind wave modes well inside the horizon (i.e. $-k \tau
\gg 1$), one naively expects the term to dominate over the others in the $[ \, \,
]$~parentheses. This can be verified as follows:
Going back to the original equation shown in the first and second lines
of~(\ref{sk1eqsimp}), one should recall that parametric resonance
happens while the characteristic oscillation with phase~$2 k \tau$ is
synchronized with that of the KK-induced source term with~$
\omega_{*} \tau_* \ln (\tau / \tau_*) $ (where we have explicitly
written~$\omega_*$ to indicate $\omega$ at $\tau_*$).
Fixing the time~$\tau_*$ to be when $2 k = \omega_*$, then one can
estimate the time scale for the resonant solution~(\ref{exsolres}) to be
valid by calculating how long $2 k (\tau - \tau_*) \approx  2 k \tau_*
\ln (\tau / \tau_*)$ lasts. 
Considering the wave mode~$k$ to exit the resonance band when the two
phases are misaligned by half-period~$\pi$, one can compute this time~$\tau_f$ as
\begin{equation}
 \tau_f - \tau_* =    \tau_* \ln \frac{\tau_f}{\tau_*} - \frac{\pi}{ 2 k}. 
\end{equation}
Since $-k \tau_* \gg 1$ one obtains
\begin{equation}
 k (\tau_f - \tau_*) \simeq 
 \left(  - \pi k \tau_*\right)^{1/2} ,
\label{lambda}
\end{equation}
which shows that the resonant term $2 i k (\tau - \tau_*) e^{i (2
k \tau + \Theta)}$ does become much larger than the others in the
$[ \, \, ]$~parentheses of (\ref{exsolres}) by the time the wave mode
leaves the resonance band.

\subsection{Time Dependent Parameters}

We now extend the above discussions and 
consider the parameters $A$, $\omega$, and $\Theta$ in~(\ref{eq153}) to
evolve in time, as in the original equation~(\ref{sk1inho}) from
oscillating KK modes. 
The parameters are considered to vary with time scales much larger
than the oscillation periods $\pi / k$ and $2 \pi / \omega$.
Moreover, we suppose $\omega$ to monotonically increase in
time~$\tau$, and $A$ to decrease.
Here we provide rough expressions that track the time
evolution of~$s_{\boldsymbol{k}}^{(1)}$, from the KK excitation 
when we set the initial conditions
\begin{equation}
 s_{\boldsymbol{k}}^{(1)}(\tau_{\mathrm{exc}}) = 0, \qquad
 s_{\boldsymbol{k}}'^{(1)}(\tau_{\mathrm{exc}}) = 0,
\label{inicondexc}
\end{equation}
until some time $\tau $ ($< -1 /k$).
In other words, we focus on wave modes that are well inside the horizon
upon the KK excitations, and study their sub-horizon evolutions.
For the wave mode that is in the resonance band at the KK excitations, we
refer to its wave number as~$k_n$, i.e., $2
k_n = \omega(\tau_{\mathrm{exc}})$.

Keeping in mind the discussions in Section~\ref{subsec:C.1},
we obtain rough approximations for~$s_{\boldsymbol{k}}^{(1)} (\tau)$ by 
collecting the leading contributions
from the above solutions, under the following procedure:
For the oscillations with frequency~$k/\pi$, we treat them as the
homogeneous solution~(\ref{homo}) and consider their oscillation
amplitudes as frozen at the initial value. The offset of the
oscillations are treated similarly. Thus for wave modes $-1 /
\tau_{\mathrm{exc}} \leq k \leq k_n$ that do not
cross the resonance band, $2 k < \omega$ holds throughout and 
one sees from (\ref{eq157}) that the time evolution
of~$s_{\boldsymbol{k}}^{(1)}$ is approximated by
\begin{equation}
 s_{\boldsymbol{k}}^{(1)}(\tau) \simeq
\frac{A_{\mathrm{exc}}}{2 k \omega_{\mathrm{exc}} }
\cos (\omega_{\mathrm{exc}} \tau_{\mathrm{exc}} + \Theta_{\mathrm{exc}})
\left\{ 1 - e^{2 i k (\tau - \tau_{\mathrm{exc}})}  \right\} 
 \qquad  \mathrm{for}  \, \, \, \, 
\tau_{\mathrm{exc}} \leq \tau < -\frac{1}{k} .
\label{C.10}
\end{equation}
Here, note that no matter $A_{\mathrm{exc}}$ is real or imaginary, the
oscillatory term~$e^{2 k i (\tau - \tau_{\mathrm{exc}})}$ gives similar
oscillation amplitudes to both the real and imaginary parts
of~$s_{\boldsymbol{k}}^{(1)}$. 
Therefore we have neglected the $\mathcal{O}(k/\omega)$ terms in the
parentheses in (\ref{eq157}), which are sub-leading anyway.

On the other hand, oscillations with frequency~$\omega/2 \pi$ 
arise from the particular solutions (\ref{part}) and
(\ref{partamp+}), with time-dependent oscillation amplitude and frequency. 
Thus for wave modes that experience parametric resonance, i.e. $ k \geq
k_n$, the approximate expression before crossing the
resonance band (i.e. $2 k > \omega$) is obtained from (\ref{eq156}) as
\begin{multline} 
  s_{\boldsymbol{k}}^{(1)}(\tau) \simeq
  - \frac{A}{2 k \omega} \left\{   
 \cos (\omega \tau + \Theta) + i \frac{\omega}{2 k}  \sin (\omega \tau +
 \Theta)  \right\}
\\
+ \frac{A_{\mathrm{exc}}}{2 k \omega_{\mathrm{exc}} } 
 \cos (\omega_{\mathrm{exc}} \tau_{\mathrm{exc}} + \Theta_{\mathrm{exc}}
 ) 
 +i  \frac{A_{\mathrm{exc}}}{4 k^2} \sin (\omega_{\mathrm{exc}}
 \tau_{\mathrm{exc}} + \Theta_{\mathrm{exc}}) 
 e^{2 i k (\tau - \tau_{\mathrm{exc}})}
\\
 \qquad  \mathrm{for}  \, \, \, \, 
\tau_{\mathrm{exc}} \leq \tau \leq \tau_{\mathrm{res}},
\label{C.11}
\end{multline} 
where we define $\tau_{\mathrm{res}}$ as when $2 k =
\omega (\tau_{\mathrm{res}})$. (We also denote parameters 
at~$\tau_{\mathrm{res}}$ with the subscript~``res''.)
Focusing on~$\mathrm{Re}(s_{\boldsymbol{k}}^{(1)})$, it should be noted
that for real~$A$ the oscillation with frequency~$\omega / 2\pi$ dominates over
that with~$k /\pi$, but for imaginary~$A$ the oscillations with
different frequencies can have comparable amplitudes.

Upon obtaining the expression for $s_{\boldsymbol{k}}^{(1)}$ after crossing the
resonance band (i.e. $ 2 k < \omega$), we succeed the offset and the
oscillatory term with frequency~$k/\pi$ from the ``homogeneous solution
part'' in (\ref{C.11}) (i.e. the second line), but replace the
``particular solution part'' (i.e. the first line) by the resonant term
in~(\ref{exsolres}) combined with (\ref{lambda}).
Here for simplicity we treat the parametric amplification to happen
instantaneously. 
Since the amplified term oscillates with frequency~$k/\pi$ upon
leaving the resonance band, from then it can be considered as a
homogeneous solution with constant amplitude. 
We ignore the terms in (\ref{eq157}) which may arise
after leaving the resonance band, since their amplitudes are
smaller than the oscillation amplitude of the resonant term.
Therefore we obtain 
\begin{multline}
   s_{\boldsymbol{k}}^{(1)}(\tau) \simeq
  - \frac{ A_{\mathrm{res}}}{4 k^2} \left(-\pi k
				   \tau_{\mathrm{res}}\right)^{1/2}
 e^{i (2 k \tau + \Theta_{\mathrm{res}})}
 \\
 + \frac{A_{\mathrm{exc}}}{2 k \omega_{\mathrm{exc}} } 
\cos (\omega_{\mathrm{exc}} \tau_{\mathrm{exc}} + \Theta_{\mathrm{exc}})
 +i  \frac{A_{\mathrm{exc}}}{4 k^2} 
 \sin (\omega_{\mathrm{exc}} \tau_{\mathrm{exc}} +
 \Theta_{\mathrm{exc}})
 e^{2 i  k (\tau - \tau_{\mathrm{exc}})}
\\
 \qquad  \mathrm{for}  \, \, \, \, 
\tau_{\mathrm{res}} \leq \tau < -\frac{1}{k}.
\label{C.12}
\end{multline}
Here we have chosen the phase in the first line of the
right hand side such that it smoothly connects with that of the first
line of~(\ref{C.11}), when ignoring the time variation of the parameters
such as~$\omega$. 
(However, we should also remark that this choice of phase
is just a simplified procedure taken to smoothly connect the oscillation
phase through the assumed instantaneous parametric amplification.)

\vspace{\baselineskip}

Thus we have obtained rough approximations for~$s_{\boldsymbol{k}}^{(1)}$
(\ref{C.10}), (\ref{C.11}), and (\ref{C.12}).
It is shown in Section~\ref{sec:smallKK} that these expressions capture the
overall behavior of $s_{\boldsymbol{k}}^{(1)}$ when inside the horizon.

\end{document}